\documentclass[]{aastex631}

\hypersetup{pdfauthor={Name}}

\begin{document}

\title{The ALMA-ATOMS Survey: Exploring Protostellar Outflows in HC$_3$N}

\author[0009-0003-6633-525X]{Ariful Hoque}
\affiliation{S. N. Bose National Centre for Basic Sciences, Block-JD, Sector-III, Salt Lake City, Kolkata 700106, India}

\author[0000-0003-0295-6586]{Tapas Baug}
\affiliation{S. N. Bose National Centre for Basic Sciences, Block-JD, Sector-III, Salt Lake City, Kolkata 700106, India}

\author[0000-0001-6725-0483]{Lokesh K. Dewangan}
\affiliation{Physical Research Laboratory, Navrangpura, Ahmedabad 380009, India}

\author[0000-0002-5809-4834]{Mika Juvela}
\affiliation{Department of Physics, P.O. Box 64, FI- 00014, University of Helsinki, Finland}

\author[0000-0001-5917-5751]{Anandmayee Tej}
\affiliation{Indian Institute of Space Science and Technology, Thiruvananthapuram 695 547, Kerala, India}

\author[0000-0002-6622-8396]{Paul F. Goldsmith}
\affiliation{Jet Propulsion Laboratory, California Institute of Technology, 4800 Oak Grove Drive, Pasadena, CA 91109, USA}

\author[0000-0002-8586-6721]{Pablo Garc\'ia}
\affiliation{Chinese Academy of Sciences South America Center for Astronomy, National Astronomical Observatories, CAS, Beijing 100101, China}
\affiliation{Instituto de Astronom\'ia, Universidad Cat\'olica del Norte, Av. Angamos 0610, Antofagasta, Chile}

\author[0000-0003-2300-8200]{Amelia M. Stutz}
\affiliation{Departamento de Astronomía, Universidad de Concepción, Casilla 160-C, 4030000 Concepción, Chile}

\author[0000-0002-5286-2564]{Tie Liu}
\affiliation{Shanghai Astronomical Observatory, Chinese Academy of Sciences, 80 Nandan Road, Shanghai 200030, China}
\affiliation{Key Laboratory for Research in Galaxies and Cosmology, Shanghai Astronomical Observatory, Chinese Academy of Sciences, 80 Nandan Road, Shanghai 200030, China}

\author[0000-0002-3179-6334]{Chang Won Lee}
\affiliation{University of Science and Technology, Korea (UST), 217 Gajeong-ro, Yuseong-gu, Daejeon 34113, Republic of Korea}
\affiliation{Korea Astronomy and Space Science Institute, 776 Daedeokdae-ro, Yuseong-gu, Daejeon 34055, Republic of Korea}

\author[0000-0001-5950-1932]{Fengwei Xu}
\affiliation{I. Physikalisches Institut, Universität zu Köln, Zülpicher Str. 77, D-50937 Köln, Germany}
\affiliation{Kavli Institute for Astronomy and Astrophysics, Peking University, 5 Yiheyuan Road, Haidian District, Beijing 100871, China}
\affiliation{Department of Astronomy, School of Physics, Peking University, Beijing, 100871, China}

\author[0000-0002-7125-7685]{Patricio Sanhueza}
\affiliation{Department of Astronomy, School of Science, The University of Tokyo, 7-3-1 Hongo, Bunkyo, Tokyo 113-0033, Japan}

\author[0000-0001-8812-8460]{N. K. Bhadari}
\affiliation{Kavli Institute for Astronomy and Astrophysics, Peking University, 5 Yiheyuan Road, Haidian District, Beijing 100871, China}

\author[0000-0002-8149-8546]{K. Tatematsu}
\affiliation{Nobeyama Radio Observatory, National Astronomical Observatory of Japan, Nobeyama, Minamimaki, Minamisaku, Nagano 384-1305, Japan}
\affiliation{Astronomical Science Program, The Graduate University for Advanced Studies, SOKENDAI, 2-21-1 Osawa, Mitaka, Tokyo 181-8588, Japan}

\author[0000-0001-8315-4248]{Xunchuan Liu}
\affiliation{Shanghai Astronomical Observatory, Chinese Academy of Sciences, 80 Nandan Road, Shanghai 200030, China}

\author[0000-0003-3343-9645]{Hong-Li Liu}
\affiliation{Department of Astronomy, Yunnan University, Kunming, 650091, China}

\author[0000-0002-1086-7922]{Yong Zhang}
\affiliation{School of Physics and Astronomy, Sun Yat-sen University, 2 Daxue Road, Tangjia, Zhuhai, Guangdong Province, China}

\author[0000-0002-4154-4309]{Xindi Tang}
\affiliation{XingJiang Astronomical Observatory, Chinese Academy of Sciences(CAS), Urumqi 830011, PR China}

\author[0000-0003-1649-7958]{Guido Garay}
\affiliation{Departamento de Astronomía, Universidad de Chile, Las Condes, Santiago 7550000, Chile}
\affiliation{Chinese Academy of Sciences South America Center for Astronomy, National Astronomical Observatories, CAS, Beijing 100101, China}

\author[0000-0002-7237-3856]{Ke Wang}
\affiliation{Kavli Institute for Astronomy and Astrophysics, Peking University, 5 Yiheyuan Road, Haidian District, Beijing 100871, China}

\author[0000-0002-9836-0279]{Siju Zhang}
\affiliation{Departamento de Astronom\'{i}a, Universidad de Chile, Las Condes, 7591245 Santiago, Chile}

\author[0000-0001-7598-9026]{L. Viktor Tóth}
\affiliation{University of Debrecen, Institute of Physics, H-4032 Debrecen, Bem tér 1.}

\author{Hafiz Nazeer}
\affiliation{Indian Institute of Space Science and Technology, Thiruvananthapuram 695 547, Kerala, India}

\author{Jihye Hwang}
\affiliation{Korea Astronomy and Space Science Institute, 776 Daedeokdae-ro, Yuseong-gu, Daejeon 34055, Republic of Korea}

\author[0000-0003-1602-6849]{Prasanta Gorai}
\affiliation{Department of Space, Earth \& Environment, Chalmers University of Technology, SE-412 93 Gothenburg, Sweden}

\author[0000-0002-9574-8454]{Leonardo Bronfman}
\affiliation{Departamento de Astronomía, Universidad de Chile, Las Condes, Santiago 7550000, Chile}

\author[0000-0002-3658-0516]{Swagat Ranjan Das}
\affiliation{Departamento de Astronomía, Universidad de Chile, Las Condes, Santiago 7550000, Chile}

\author[0000-0001-5508-6575]{Tirthendu Sinha}
\affiliation{S. N. Bose National Centre for Basic Sciences, Block-JD, Sector-III, Salt Lake City, Kolkata 700106, India}

\begin{abstract}
We present the first systematic study of bipolar outflows using HC$_3$N as a tracer in a sample of 146 massive star-forming regions from ALMA-ATOMS survey. Protostellar outflows arise at the initial stage of star formation as a consequence of active accretion.
In general, these outflows play a pivotal role in regulating the star formation processes by injecting energetic material in the parent molecular clouds.
In such process, lower velocity components of outflows contain a significant portion of the energy. 
However, extraction of those component is difficult as the corresponding gas is often mixed with that of the ambient cloud.
In our sample, we identified 44 bipolar outflows and one explosive outflow in HC$_3$N (J=11--10). The host clumps of these outflows are found to be at different evolutionary stages, suggesting that outflows in HC$_3$N are detectable in different stages of star formation. Also, the non-correlation of HC$_3$N outflows with clump evolutionary stages suggests that HC$_3$N is an unbiased tracer of outflows. Analyses revealed that HC$_3$N performs slightly better in detecting low-velocity components of outflows than traditionally employed 
tracers like SiO. The derived outflow parameters (i.e outflow mass, momentum, and energy) show moderate correlations with clump mass and luminosity. 
Our analysis of outflow opening angles and position-velocity diagrams across the outflow lobes show that, HC$_3$N is not only a good tracer of low-velocity outflows, but can also detect high-velocity collimated outflows. Overall, this study indicates that HC$_3$N can be used as a complementary outflow tracer along with the traditionally known outflow tracers, particularly in the detection of the low-velocity components of outflows.
\end{abstract}


\section{Introduction} \label{sec:intro}
Protostellar outflows are ubiquitous phenomena at the initial stages of star formation. Highly-collimated bipolar jets originate as a consequence of active accretion in the central protostar. Less-collimated bipolar outflows are traced as ambient gas driven by shocks from these highly-collimated energetic jets. These bipolar jets and outflows also play a crucial role in dispersing the excess angular momentum and therefore allow the accretion process of the central protostar to continue.
Since their discovery in 1980 \citep[][]{snell1980}, molecular outflows have been studied extensively in active star-forming regions \citep[see, e.g.,][and references therein]{bachiller1996, arce2010, bally2016}. Being an energetic mass ejection phenomenon, outflows serve as one of the primary feedback mechanisms in star formation. A significant amount of cloud mass is entrained from the central protostar by protostellar jets and thereby controls the star formation process. Powerful feedback from outflows can inject enough turbulence into the natal clump to support it against further collapse \citep[][]{nakamura2007, carroll2009, baug2021}, and thus, regulates star formation in molecular clouds.
The formation mechanism of outflows is still a topic of debate. There are two main scenarios proposed as the origin of outflows -- the `X-wind' model \citep[][]{shu1994} and the `disk wind' model \citep[][]{pelletier1992}. In both models, a bipolar flow of high-velocity matter
is expected from the central protostar, 
which forms an elongated shocked region that can be detected observationally. Traditionally, outflows are traced with molecules  that show abundance enhancement in  the shocked environment (e.g., CO, SiO, SO, CH$_3$OH).
Among the commonly used outflow tracers, carbon monoxide (CO) and silicon monoxide (SiO) are most often used to infer the morphology and kinematics of the outflowing material \citep[see e.g.,][]{arce2010, maud2015, baug2020, li2020, guerra2023, towner2024}. CO mainly traces the envelopes of the outflow (or low-velocity, wide-angle outflow) that originate from %because of 
the interaction between the ambient medium and the high velocity jets launched from the forming star.
In contrast, SiO traces more collimated (or high-velocity, narrow-angle) outflow components that originate under high-velocity shock propagation \citep[][]{lee2002, tafalla2015, dutta2024}. %However, 
Tracers including HCO$^+$, HCN, SO, HNCO, H$_2$CO, and CH$_3$OH are also widely used to characterize the outflowing gas \citep[][]{palau2017, holdship2019, baug2021, Xie2023, izumi2024}.

In general, the low-velocity components of outflows contain the major fraction of the outflow mass \citep{machida2014}. Therefore, they inject most of the mechanical energy into the host cloud. Low-J CO transitions are capable of detecting low-velocity components of outflows \citep[][and references therein]{dunham2014, LiuCF2025}. However, CO is ubiquitous in the interstellar medium. Thus, CO outflows, specifically the low-velocity components, are often contaminated by the ambient gas close to the local standard of rest velocity (V$_{\text{lsr}}$) of the host cloud.

Cyanoacetylene (HC$_3$N) is generally considered as a tracer of dense molecular gas and is observed mainly toward dense cores (including hot molecular cores) in active star-forming regions \citep[][]{bergin1996, Cordiner2012, Taniguchi2017, taniguchi2019a}. \citet[][]{Bachiller1997} were the first to observe an enhancement in the abundance of HC$_3$N in shocked regions. Later, shock enhancement of HC$_3$N was observed in several studies \citep[][]{mendoza2018,taniguchi2018a, taniguchi2019b, lu2021, wang2022}.
In shocked environment, HC$_3$N and its precursors could be released from grain surfaces (the formation pathway is discussed in detail in the next paragraph), potentially contributing to the ability of HC$_3$N to trace the low-velocity components of outflows.
Being relatively less abundant compared to CO in the interstellar medium, low-velocity outflows traced by HC$_3$N are less contaminated by the ambient medium. This suggests that HC$_3$N could be a useful tracer for low-velocity components of outflows.

Several chemical models suggested that the main formation pathway for HC$_3$N in the interstellar medium is through the neutral-neutral reaction between acetylene (C$_2$H$_2$) and cyano radical (CN) \citep[C$_2$H$_2$ + CN$\rightarrow$ HC$_3$N + H;][]{Fukuzawa1997, meier2005, chapman2009}. \citet[][]{Fukuzawa1997} and \citet[][]{meier2005} suggested that the abundance of HC$_3$N is enhanced in shocked environment because of the release of HC$_3$N and its precursor from grain surfaces. \citet[][]{Hassel2008} proposed another gas-phase model where methane (CH$_4$) undergoes a series of reactions resulting in the formation of C$_2$H$_2$, which reacts with CN to produce HC$_3$N. Later, \citet[][]{mendoza2018} proposed that the enhancement of HC$_3$N is possible by orders of magnitude in the protostellar environment because of a two-step process resulting from the passage of the shock. In the first step, the abundance of HC$_3$N may increase by sputtering of grains, and further in the second step by the efficient reaction of C$_2$H$_2$ with CN. 
Indeed, \citet[][]{beltran2004,taniguchi2018b,Yu2019, Zinchenko2021}{}{} detected extended wings in the spectra of HC$_3$N molecule and inferred the origin of those extended wings to be outflowing gas. \citet{lu2021} also reported the detection of HC$_3$N outflows associated with several star-forming cores in the Central Molecular Zone (CMZ).

In this paper, we present observational evidence of HC$_3$N as an outflow tracer with a catalog of 45 outflows identified using HC$_3$N (J=11--10) in a sample of 146 massive star forming clumps of the ALMA three-millimeter observations of massive star-forming regions (ATOMS) survey \citep[][]{liu2020}{}{}. The primary emphases of the paper are -- whether HC$_3$N can be used as an outflow tracer in all kinds of star-forming clumps, how similar or different are HC$_3$N outflows in comparison to known outflow tracers like SiO, and how the derived HC$_3$N outflow parameters depend on the mass and luminosity of the host clump. 
The paper is organized in the following manner. In Section \ref{sec:data}, we briefly describe the observational details. The identification of outflows and outflow host cores, estimation of HC$_3$N column density and several other outflow parameters are presented in Section \ref{sec:results}. In Section \ref{sec:discussion}, we discuss the implication of HC$_3$N as an outflow tracer, a comparison of HC$_3$N outflow with commonly known outflow tracer, and the dependency of outflow parameters on clump properties. Finally, we present a summary of this work in Section \ref{sec:summary}.

\section{Observations and Data} \label{sec:data}
In our present study, we utilize ALMA data from the ATOMS survey (Project ID: 2019.1.00685.S; PI: Tie Liu). The correlator setup includes 8 spectral windows (SPWs), where SPWs 1-6 have a high spectral resolution ($\sim 0.2 - 0.4$ km s$^{-1}$) and SPWs 7 and 8 have a comparatively lower spectral resolution ($\sim$ 1.6 km s$^{-1}$). For our analysis, we used 12-m array data that have an angular resolution of $\sim$1.2$''$–1.9$''$ and sensitivity of $\sim$3–10 mJy beam$^{-1}$ per 0.122 MHz channel. For the identification of outflows, we choose the HC$_3$N (J=11--10) transition from SPW 8. We also utilized the H$^{13}$CO$^+$ (J=1--0) transition from SPW 2 to trace the dense gas. The ATOMS 3 mm continuum images have a rms noise of $\sim$0.2 mJy beam$^{-1}$ for most of the sources. Data reduction was performed using the CASA software package version 5.6 \citep[][]{McMullin2007}. More details of the observations and data reduction can be found elsewhere in \citet[][]{liu2020, HLiu2021, HLiu2022}.

\section{Results} \label{sec:results}
\subsection{Identification of Outflows} \label{subsec:identification}
\begin{figure}[ht!]
\centering
    \includegraphics[width=0.32\textwidth]{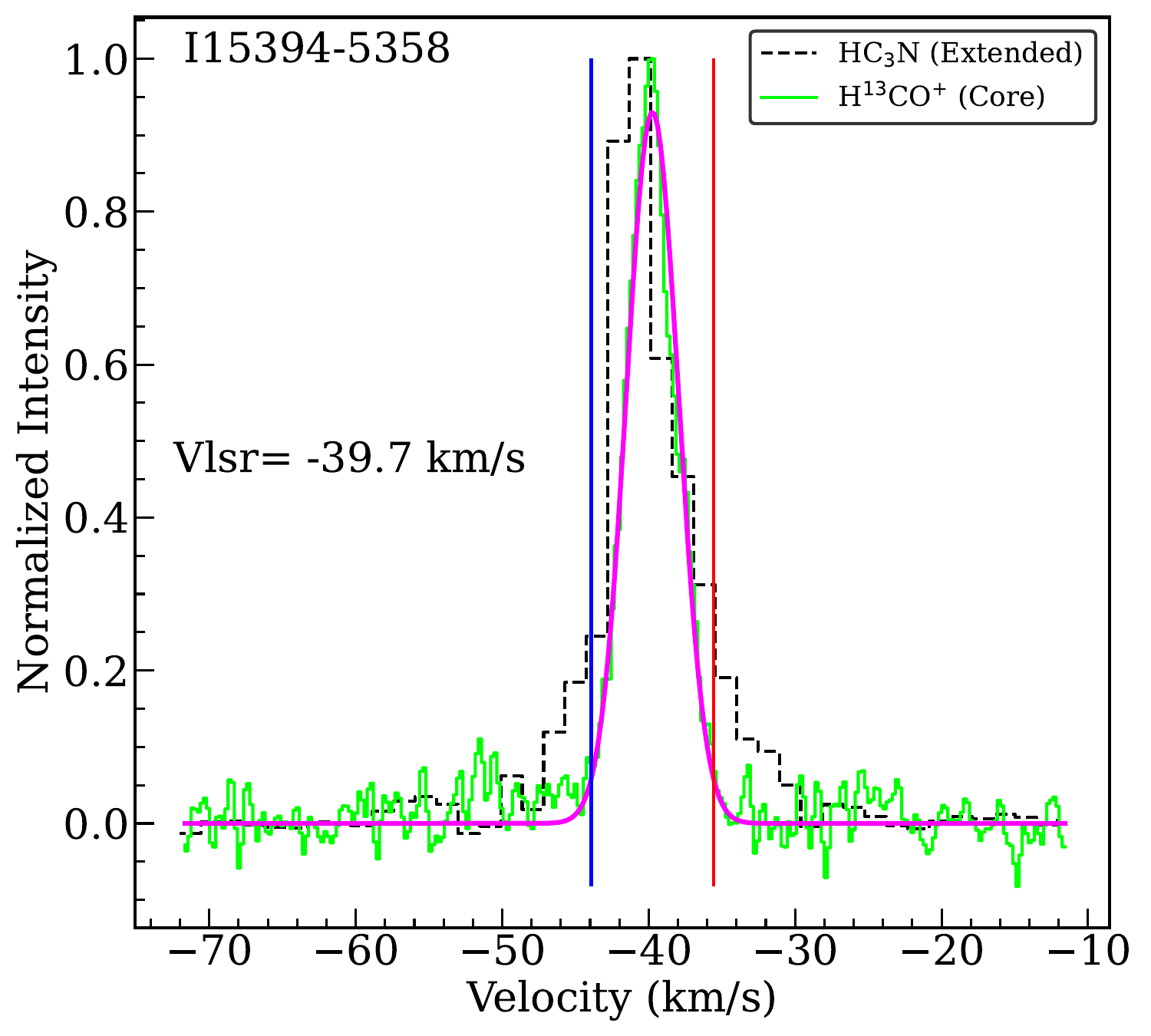}
    \includegraphics[width=0.32\textwidth]{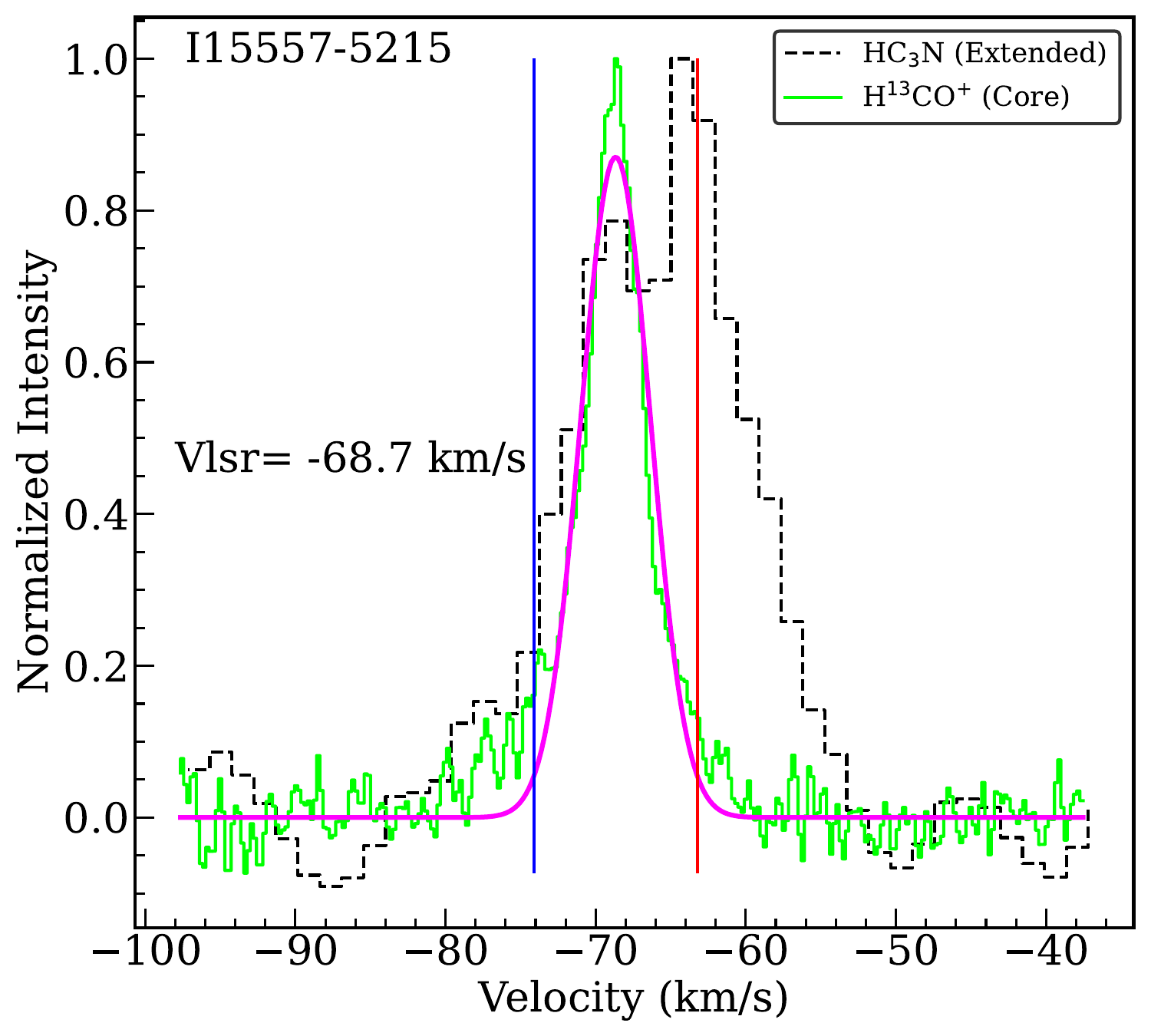}
    \includegraphics[width=0.32\textwidth]{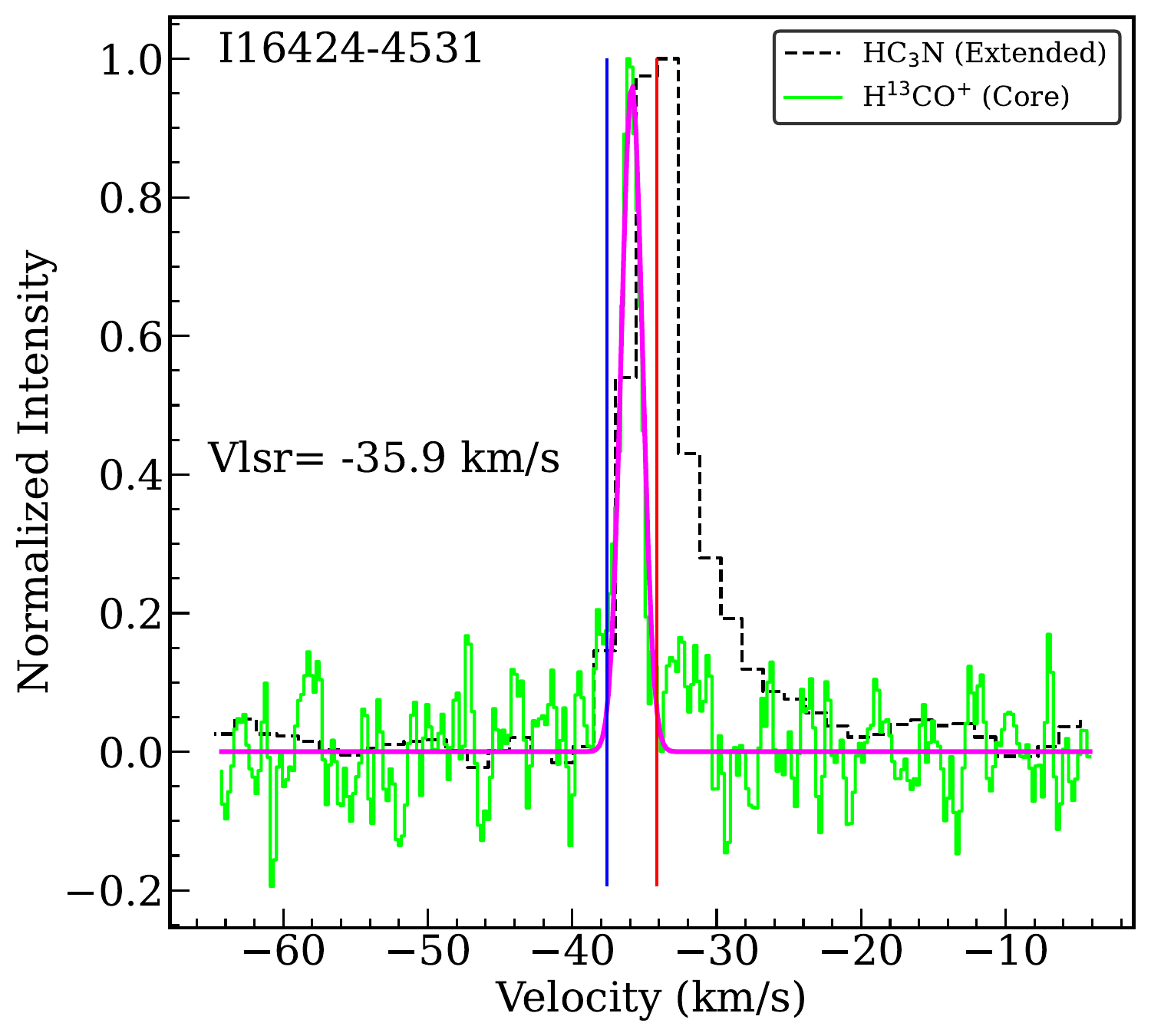}
\caption{Normalized HC$_3$N spectra integrated over the ALMA field of view (in black dashed line) for three target regions. The normalized H$^{13}$CO$^+$ spectra (in green) along the cores and the corresponding Gaussian fits  (in magenta curve) are also overlaid on the spectra. The red and blue lines represent V$_{lsr}\pm$FWHM of the Gaussian fit, respectively. The local standard of rest velocity (V$_{\text{lsr}}$) of the host core is also quoted in each panel. The target name is given at the top left corner of each panel where the initial I stands for IRAS.}
\label{fig-identification}
\end{figure}

Our primary aim in this work is to identify outflows using the HC$_3$N (J=11--10) molecular transition (hereafter HC$_3$N) line within a large sample of 146 star-forming clumps in the ATOMS survey \citep{liu2020}. Among the 146 sources, we found that 141 sources have significant HC$_3$N emission (i.e. emission $>10\sigma$ of the background rms) originating either from outflows or filamentary structures or dense cores. In 4 sources, we found weak HC$_3$N detections (i.e. emission is $\sim 5-8\sigma$ of the background rms) and no HC$_3$N was detected in one source. In our initial search for protostellar outflows, we generated HC$_3$N spectra integrated over ALMA field of view. However, we determined the velocity range of the central cloud component using optically thin H$^{13}$CO$^+$ (J=1--0) line that typically does not have contribution from outflowing material. The spectral profile of the H$^{13}$CO$^+$ line was generated for each of the dust cores (details of the dust cores were adopted from \citet[][]{HLiu2021}) in the field of view and overlaid on the HC$_3$N spectra. We fit Gaussian profiles to the H$^{13}$CO$^+$ (J=1--0) spectra, and considered $V_{\text{LSR}}\pm$FWHM as the velocity of the central cloud. Finally, we generated integrated emission maps of the wings (blue-shifted and red-shifted) for the spectral channels outside $V_{\text{LSR}}\pm$FWHM. Spectral profiles having wing emissions were selected as outflow candidates. Figure~\ref{fig-identification} represents the spectral profiles of HC$_3$N and H$^{13}$CO$^{+}$ with HC$_3$N showing high-velocity wing emissions in a few of the ATOMS sources.  However, identification of outflows using this method lacks the ability to detect those outflows away from 3 mm dust continuum cores or outflows without any detected dust continuum cores. Additionally, low-velocity outflow wings might get blended with the emission from the cloud components away from the host cores. Therefore, we also utilized the \texttt{spectral-cube} package from the \texttt{astropy} project \citep{astropy2013} to generate the integrated intensity (i.e., moment-0), the intensity-weighted velocity (i.e., moment-1), and the intensity-weighted variance (i.e., moment-2) maps for each of the ATOMS target regions. We searched for outflow candidates using these moment maps. As mentioned before, molecular outflows are high-velocity gas components that arise because of strong jets passing through the surrounding environments. Thus, in the moment-1 map (i.e., velocity map) they appear as elongated features with a substantially different velocity compared to their driving cores.
Thus, candidates having an elongated morphology in the moment maps with a blue-shifted and red-shifted component (relative to the systemic velocity of the host cloud)  
are primarily selected as outflow candidates.

An elongated high velocity feature is not always indicative of an outflow. %refer to outflows.
Thus, for further confirmation, we inspected the HC$_3$N data cubes and determined the outflow directions. We generated position-velocity (PV) diagrams along the probable outflow directions. Outflows generally show ``Hubble law'' velocity structures (i.e the maximum velocity of the outflowing gas increases linearly with distance from the host core) or ``Hubble law wedges'' in the PV diagram \citep[][]{Arce2001, li2020, mori2021}{}{}. We identified a total of 45 outflows within the 146 ATOMS targets. Among them, 44 are bipolar in nature, and one is explosive outflows \citep[IRAS 15520-5432; typically identified with multiple blue- and red-shifted outflow lobes diverging out from a single core; see][]{bally2016, guzman2022}. Based on their brightness in wing emission maps, morphology in moment maps and structure in PV diagrams, we marked the outflows as `Confirmed' and `Probable' outflow candidates. The `Confirmed' candidates have a definite signature of an outflow in all the moment maps and also in PV diagram. On the other hand, `Probable' candidates are those outflows that do either have a well-defined outflow structure in moment maps or a characteristics PV diagram of an outflow. A total of 31 outflows are marked as `Confirmed' and 14 outflows are marked as `Probable' outflows. Among the 45 outflows, 38 are also detected in SiO (J=2--1) emission of the ATOMS survey (Baug et al.(Private communication)). A catalog of all the identified outflows along with their detection in SiO tracer are listed in Table~\ref{tab:parameters}. An example of a bipolar-outflow identified using moment maps and PV diagram is shown in Figure~\ref{outflow}. Similar figures for all of the identified HC$_3$N outflows are shown in Appendix~\ref{sec:appendix-MomentMaps}.

\begin{figure}[ht!]
\centering
    \includegraphics[width=0.95\textwidth]{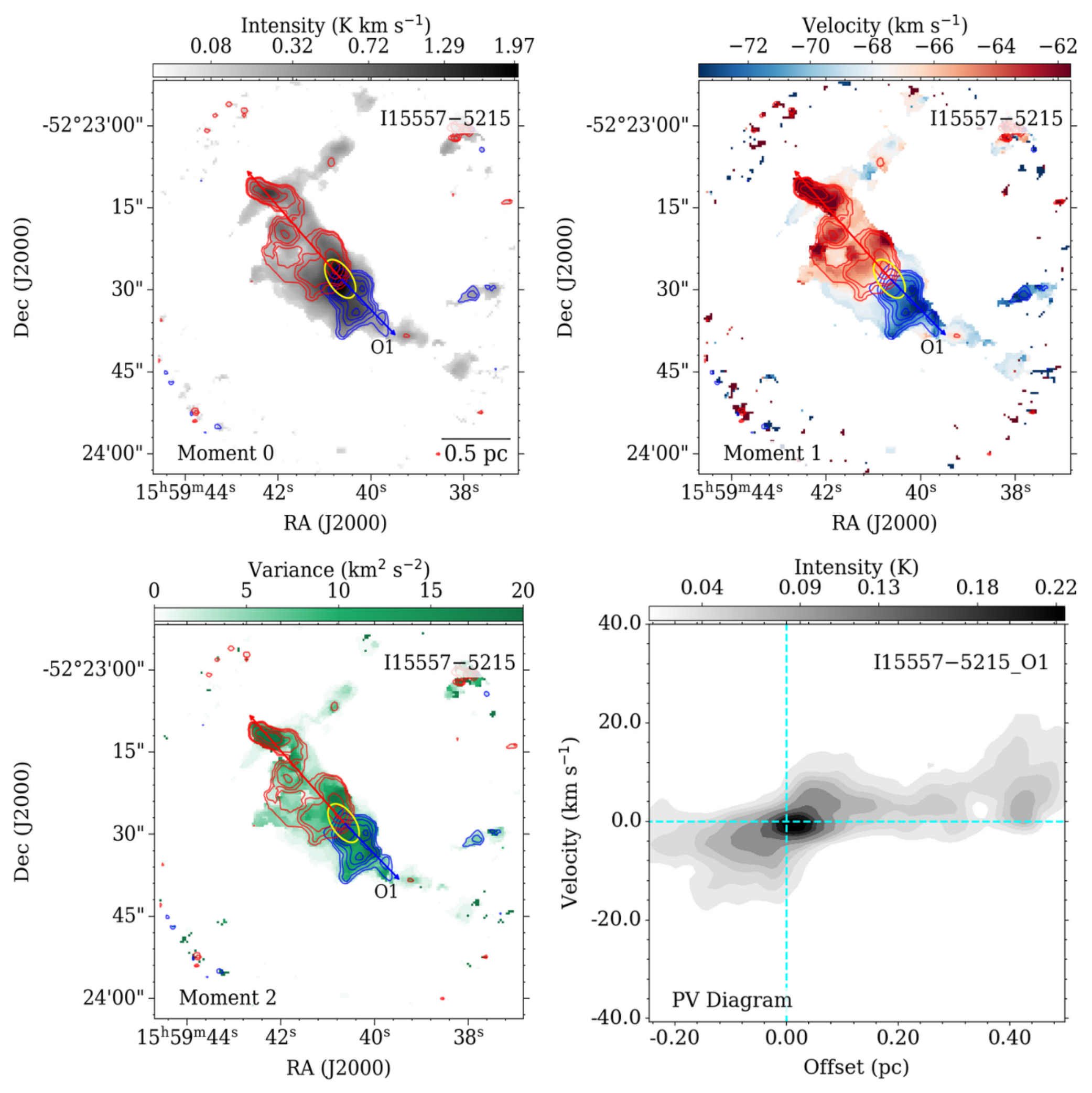}
\caption{Example of a bipolar outflow in IRAS 15557-5215 region. The top panels show the moment-0 map (left) and moment-1 map (right) of HC$_3$N line. The outflow axis is represented by thin blue (for blue-shifted lobe) and red arrows (for red-shifted lobe) in the moment maps, and the position of 3 mm dust continuum core (host core) is marked with a yellow ellipse. The red-shifted and blue-shifted components of the emission are also shown with red and blue contours, respectively. The bottom panels show the moment-2 map (left, with a color bar above it) and the PV diagram along the outflow axis (right). The vertical and horizontal cyan dashed lines represent the position and the V$_{\text{lsr}}$ of the host core, respectively. } 
\label{outflow}
\end{figure}

\subsection{Outflow host cores} \label{subsec:host cores}
We identified outflow host cores using the catalogue of 3 mm ALMA dust cores of \citet[][]{HLiu2021}{}{}. A total of 36 dust cores were assigned with outflows as their host cores, where the outflow-driving protostar is deeply embedded. Out of these 36 dust cores, one core is found to host an explosive outflow and one core is found to be associated with two bipolar outflows. No host core was identified for 8 outflows. We also investigated the possibility of these dense cores for hot molecular cores (hereafter, HMC) and ultra-compact H{\scriptsize II} region (UCH{\scriptsize II}). For this, we used the catalogue of HMC and UCH{\scriptsize II} from \citet[][]{HLiu2021}{}{} to characterize the dust cores. Accordingly, we found that 13 of these dust cores are associated only with HMC catalogues, 5 dust cores are associated only with UCH{\scriptsize II} catalogues, 6 dust cores are associated with both HMC and UCH{\scriptsize II} catalogues, and 12 dust cores do not have any association with any of HMC or UCH{\scriptsize II} regions. The details of all the identified driving cores are listed in Table \ref{tab:dustcore}.

\subsection{Column Density of HC$_3$N} \label{subsec:columndensity}
The HC$_3$N transitions are typically considered to be optically thin in dense molecular clouds \citep{morris1976}. For estimation of the column density of HC$_3$N, we assumed the gas is in local thermodynamic equilibrium (LTE) and also considered the HC$_3$N emission to be optically thin. We calculated the HC$_3$N column density in each pixel of the data cube using the following equation \citep[adpoted from][]{Garden1991,Sanhueza2012}.
\begin{equation}
N= \frac{8\pi\nu^3}{c^3}\frac{Q_{\text{rot}}}{g_{\text{u}} A_{\text{ul}}} 
\frac{\exp(E_{\text{l}}/kT_{\text{ex}})}{1-\exp(-h\nu/kT_{\text{ex}} )}  \frac{\int T_{\text{B}} dv} {J(T_{\text{B}})-J(T_{\text{bg}})}{\color{red} ~,}
\end{equation}
%and 
with $J(T)$ is defined as,
\begin{equation}
    J(T)= \frac{h\nu}{k} \frac{1}{e^{h\nu/kT}-1}{\color{red} ~,}
\end{equation}
where $\nu$ is the frequency of the transition, $c$ is the speed of light, $Q_{\text{rot}}$ is the partition function, $g_{\text{u}}$ is the statistical weight of the upper level, $A_{\text{ul}}$ is the Einstein coefficient for spontaneous emission, $E_{\text{l}}$ is the energy of the lower level, $T_{\text{ex}}$ is the excitation temperature, $T_{\text{B}}$ is the brightness temperature, $T_{\text{bg}}$=2.7 K is the cosmic microwave background temperature. In our analysis, we measured the integrated intensity over the velocity range of the outflows which is approximately equal to $\int T_{\text{B}} dv$. We assumed a $T_{\text{ex}}$ of 50 K  for the outflow materials (The choice of $T_{\text{ex}}$ and related uncertainties are discussed in Sec \ref{subsec:parameters} and the variation in HC$_3$N column density with $T_{\text{ex}}$ is shown in Figure~\ref{Fig-CdenTex}). The values of $Q_{\text{rot}}$($=kT_{\text{ex}}/hB\,+\,1/3$=686.7), $g_{\text{u}}(=21)$, $A_{\text{ul}}(=2.567\times10^{-5})$, $E_{\text{l}}(=24.01$ K) are obtained from the Cologne Database for Molecular Spectroscopy \citep[CDMS;][]{muller2001}.

\subsection{Estimation of Outflow Parameters} \label{subsec:parameters}
For each outflow, we defined an outflow mask for both blue-shifted and red-shifted lobes separately. We estimated the rms of the background emission ($\sigma$) using a few line-free channels in the data cube. Emission above a signal-to-noise ratio of 5$\sigma$ is considered to define the outflow mask. In general, outflow components at velocities close to the systemic velocity of the host cloud are hard to disentangle. Therefore, we excluded spectral channels corresponding to the velocity of the central cloud ($V_{\text{LSR}}\pm$FWHM) (FWHM is estimated using  H$^{13}$CO$^+$ line as discussed in Section~\ref{subsec:identification}) when estimating outflow parameters. In the case of non-detection of host cores, we took the average FWHM velocity of all the dust cores within the corresponding field as the cloud component. We determined the terminal velocity and the extent of each outflow lobe using the 5$\sigma$ masked data cubes. Outflow parameters were derived using the following equations given in \citet[][]{lopez2009, Wang2011}.
\begin{equation}
    M_{\text{out}}=\sum_i{M_{\text{i}}}=d^2 \left[\frac{\text{H}_2}{\text{HC}_3\text{N}}\right] \mu_{\text{H}} \, \text{m}_{\text{H}} \, A \, \sum_i {N_{\text{i}}}
\end{equation}
\begin{equation}
    P_{\text{out}}=\sum_i{M_{\text{i}} v_{\text{i}}} 
\end{equation}
\begin{equation}
    E_{\text{out}}=\frac{1}{2} \sum_i{M_{\text{i}} v_{\text{i}}^2}
\end{equation}
\begin{equation}
    t_{\text{dyn}}= \frac{L_{\text{lobe}}}{v_{\text{lobe}}}
\end{equation}
\begin{equation}
    \dot{M}_{\text{out}}=\frac{M_{\text{out}}}{t_{\text{dyn}}}
\end{equation}
\begin{equation}
    F_{\text{out}}=\frac{P_{\text{out}}}{t_{\text{dyn}}}
\end{equation}
\begin{equation}
    L_{\text{mech}}=\frac{E_{\text{out}}}{t_{\text{dyn}}}
\end{equation}
where $M_{\text{i}}$ and $v_{\text{i}}$ are the mass and velocity of each channel within the velocity range of the outflow lobes, $[\text{HC}_3\text{N}/\text{H}_2]$ is the relative abundance of HC$_3$N in comparison to H$_2$, $d$ is the distance to the source, $\mu_{\text{H}}$ is the mean molecular weight (adopted as 2.8), m$_{\text{H}}$ is the mass of hydrogen, $A$ is the angular sky area subtended by a single pixel, $N_{\text{i}}$ is the column density for each pixel within the outflow mask, $L_{\text{lobe}}$ is the extent of the outflow lobe from the host core and $v_{\text{lobe}}$ is the terminal velocity of the outflow lobe. The value of $[\text{HC}_3\text{N}/\text{H}_2]$ is adopted as $5\times10^{-9}$ i.e., the mean of the values $1.4\times10^{-8}$\citep[][]{mendoza2018}, $5.1\times10^{-11}$ \citep[][]{taniguchi2018b}, $5\times10^{-10}$ \citep[][]{Yu2019} reported in previous studies. Since the observed outflow velocity is the line-of-sight component of the actual velocity, and the length of the outflow lobe is in the plane of sky project of actual length, a correction for the inclination angle is required. For our study, we assumed a mean inclination angle, $\theta$, of 53.7$^{\circ}$ \citep[see][for detailed discussion]{dunham2014} of the outflow axis with respect to the line of sight direction and corrected the outflow parameters $L_{\text{lobe}}$, $P_{\text{out}}$, $E_{\text{out}}$, $t_{\text{dyn}}$, $\dot{M}_{\text{out}}$, $F_{\text{out}}$, $L_{\text{mech}}$ by multiplying 1/sin$\theta$, 1/cos$\theta$, 1/cos$^2\theta$, cos$\theta$/sin$\theta$, sin$\theta$/cos$\theta$, sin$\theta$/cos$^2\theta$, sin$\theta$/cos$^3\theta$.
Statistics of all the outflow parameters are listed in Table~\ref{tab:statistics}, while detailed information for individual outflows can be found in Table~\ref{tab:parameters}

\begin{figure}[ht!]
\centering
\includegraphics[width=0.40\textwidth]{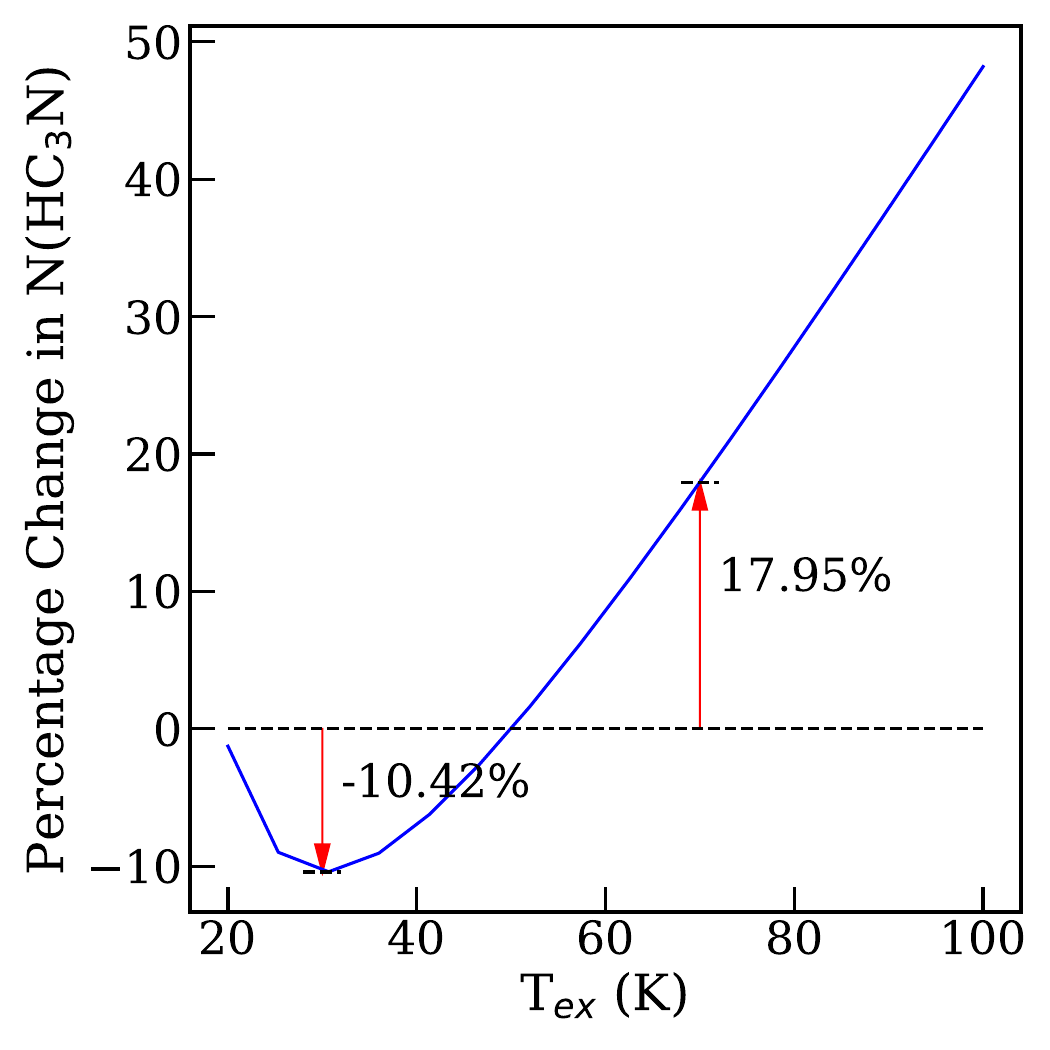}
\caption{The blue line shows the variation of column density (N(HC$_3$N)) with excitation temperature (T$_{\text{ex}}$). The dashed black line shows the adopted value of the column density obtained by adopting a temperature of 50 K. The red arrows indicate the percentage of increase or decrease in column density if the temperature is 30 K or 70 K, respectively.}
\label{Fig-CdenTex}
\end{figure}

A number of factors contribute to the uncertainty in the estimation of the outflow parameters, including 
(i) uncertainty in the distance and flux measurement, (ii) assumption of a single, constant excitation temperature for the outflowing gas, (iii) missing low-velocity outflow component %mixed 
 blended with the ambient cloud material,  (iv) the assumption that the HC$_3$N is optically thin, and (v) the assumption of a constant relative abundance of HC$_3$N with respect to H$_2$.
Even though a systematic procedure is followed to determine the cloud component by fitting a Gaussian to the spectral profile, the low-velocity outflowing gas always mixes with the cloud component and it is next to impossible to disentangle such emission.
If we consider a typical uncertainty of about 10\% in the distance to the source and also in the measured flux, the uncertainty in the estimation of outflow parameters would be $\sim$30\%. The temperature of the outflowing material generally lies within $\sim20-100$~K and it increases towards high velocity jets \citep[][]{hatchell1999, Shimajiri2009}. We assumed a constant excitation temperature ($T_{\text{ex}}$= 50~K) in the estimation of column density. Similar outflow temperatures ($\sim$ 50~K) were estimated earlier by \citet[][]{Kempen2009} using CO line ratios in several low-mass protostars. In high-mass star-forming clumps, \citet[][]{tang2018} also found the gas kinetic temperature in the range 30K to $>200$K with an average of $62\pm2$K. Since different parts of the ambient medium can have different excitation temperatures, our assumption of constant excitation temperature would lead to overestimation or underestimation of outflow parameters. Fig \ref{Fig-CdenTex} shows the percentage variation in column density plotted against $T_{\text{ex}}$, where we took the column density at $T_{\text{ex}}$=50 K as reference. Fig \ref{Fig-CdenTex} suggests that the maximum uncertainty due to the assumption of constant excitation temperature in the outflow parameters estimated with $T_{\text{ex}}$= 50~K would be $\sim50\%$.  

\begin{deluxetable*}{cccc}
\tabletypesize{\scriptsize}
\tablecaption{Statistics of the outflow parameters \label{tab:statistics}}
\tablehead{
\colhead{$^{\dagger}$Parameters} & \colhead{Minimum} & \colhead{Maximum}  & \colhead{Mean} %& \colhead{Std. deviation} 
}
\startdata
$N_{\text{tot,Blue}}$ $[10^{14}$cm$^{-2}]$ & 0.73 & 317.3 & 29.98 \\%& 56.18\\
$N_{\text{tot,Red}}$ $[10^{14}$cm$^{-2}]$ & 1.78 & 301.4 & 29.49 \\%& 47.60\\
$M_{\text{out,Blue}}$ $[\text{M}_{\odot}]$ & 8.5$\times10^{-3}$ & 7.5 & 0.46 \\%& 1.12\\
$M_{\text{out,Red}}$ $[\text{M}_{\odot}]$ & 25.1$\times10^{-3}$ & 7.1 & 0.59 \\%& 1.14\\
$P_{\text{out,Blue}}$ $[\text{M}_{\odot} \text{km s}^{-1}]$ & 0.13 & 203.8 & 17.22 \\%& 31.53\\
$P_{\text{out,Red}}$ $[\text{M}_{\odot} \text{km s}^{-1}]$ & 0.06 & 69.9 & 8.46 \\%& 16.43\\
$E_{\text{out,Blue}}$ $[\text{M}_{\odot}$ km$^2$ s$^{-2}$] & 1.0 & 2789.5 & 412.98 \\%& 598.39\\
$E_{\text{out,Red}}$ $[\text{M}_{\odot}$ km$^2$ s$^{-2}$] & 0.1 & 2316.3 & 129.05 \\%& 377.29\\
$t_{\text{dyn,Blue}}$ [10$^3$yr] & 1.9 & 93.4 & 26.65 \\%& 23.96\\
$t_{\text{dyn,Red}}$ [10$^3$yr] & 1.5 & 354.6 & 33.72 \\%& 55.58\\
$v_{\text{lobe,Blue}}$ [km s$^{-1}$] & 3.2 & 40.1 & 16.09 \\%& 9.79\\
$v_{\text{lobe,Red}}$ [km s$^{-1}$] & 3.4 & 55.4 & 16.23 \\%& 9.85\\
${\dot{M}}_{\text{out,Blue}}$ $[10^{-5}\text{M}_{\odot}$ yr$^{-1}]$ & 0.06 & 16.0 & 2.64 \\%& 3.30\\
$\dot{M}_{\text{out,Red}}$ $[10^{-5}\text{M}_{\odot}$ yr$^{-1}]$ & 0.09 & 21.3 & 2.71 \\%& 4.02\\
$F_{\text{out,Blue}}$ $[10^{-4}\text{M}_{\odot}$ km s$^{-1}$ yr$^{-1}]$ & 0.06 & 120.1 & 15.33 \\%& 24.17\\
$F_{\text{out,Red}}$ $[10^{-4}\text{M}_{\odot}$ km s$^{-1}$ yr$^{-1}]$ & 0.03 & 139.2 & 7.17 \\%& 21.23\\
$L_{\text{mech,Blue}}$ $[10^{-4}\text{M}_{\odot}$ km$^2$ s$^{-2}$ yr$^{-1}]$ & 0.29 & 4576.4 & 526.80 \\%& 1005.69\\
$L_{\text{mech,Red}}$ $[10^{-4}\text{M}_{\odot}$ km$^2$ s$^{-2}$ yr$^{-1}]$ & 0.60 & 5220.6 & 173.36 \\%& 778.73\\
\enddata
\tablenotetext{\dagger}{$N_{\text{tot,Blue}}$ and $N_{\text{tot,Red}}$ represent the total HC$_3$N column density within the blue-shifted and red-shifted outflow lobes, respectively.}
\end{deluxetable*}

\subsection{Outflow Opening Angle} \label{subsec:opening angle}

The collimation of outflowing gas (measured as outflow opening angle) in the early stages of protostellar evolution shows different morphology and kinematics depending on the evolutionary state and choice of outflow tracer used. At the very early stages, young protostars drive powerful, well collimated (i.e., with small opening angle) outflows. With time, as the protostars evolve, the outflows tend to be less collimated \citep[][]{Arce2006, Arce2007}. The increase in outflow opening angle with the source age is also verified in several observational studies \citep[][]{Velusamy1998, Richer2000, Arce2006}. The degree of collimation of outflow also depends on the tracer used for detection. For example, tracers like CO and HCO$^+$ generally trace less collimated outflows, whereas SiO typically detects highly collimated outflows, closer to the jet axis. We examined the outflow lobes (i.e the blue-shifted and red-shifted wing emission maps) using SAOds9 software \citep[][]{ds9} to estimate the opening angle ($\phi$) of our identified HC$_3$N outflows. We visually marked two `vectors' to denote the outer edges of 5$\sigma$ emission for each individual outflow lobes, and then measured the angle between those vectors for the estimation of the opening angles, $\phi$. The typical uncertainty in estimated opening angle is $\sim$10$^{\circ}$.

\begin{figure}[ht!]
    \centering
    \includegraphics[width=0.45\textwidth]{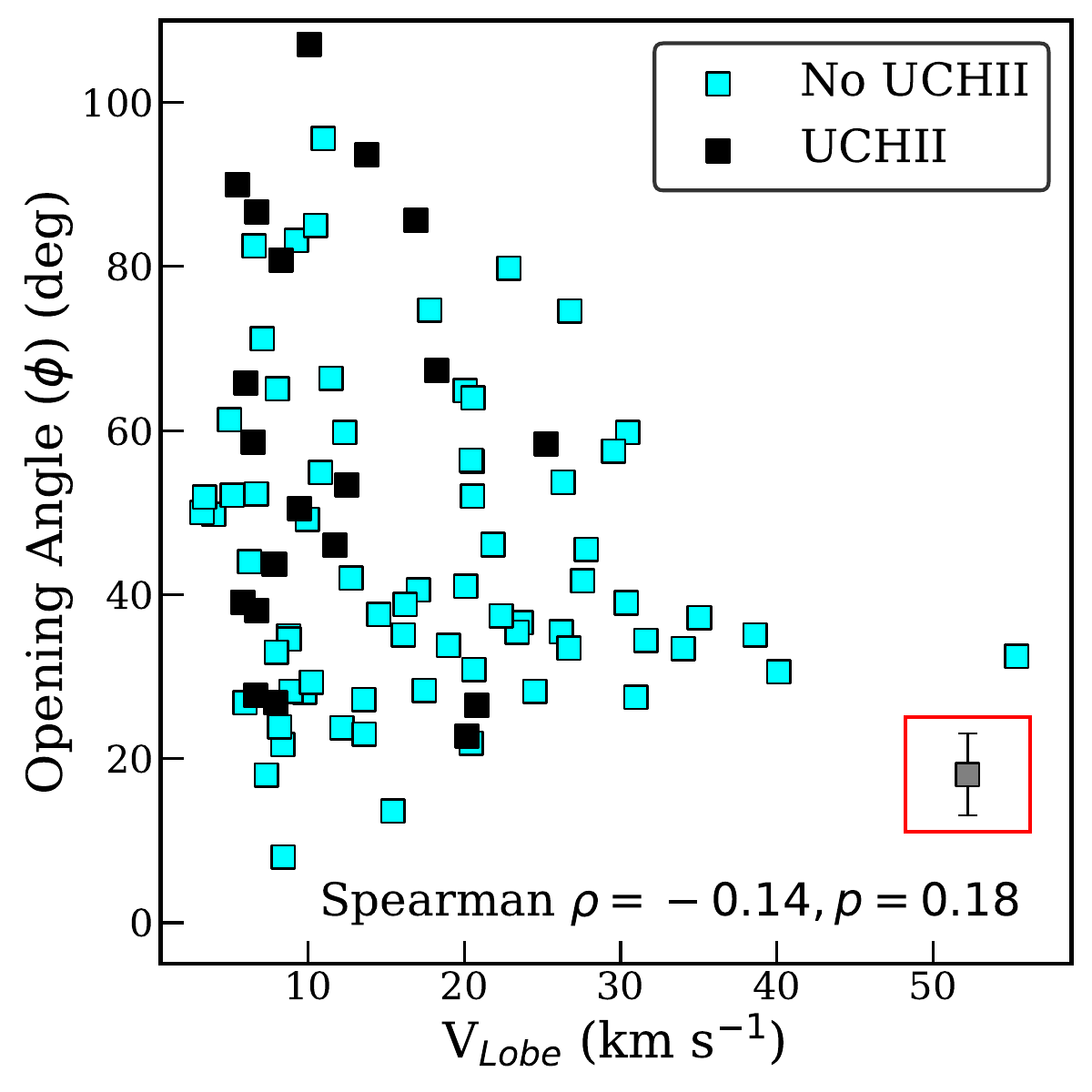}
    \includegraphics[width=0.5\textwidth]{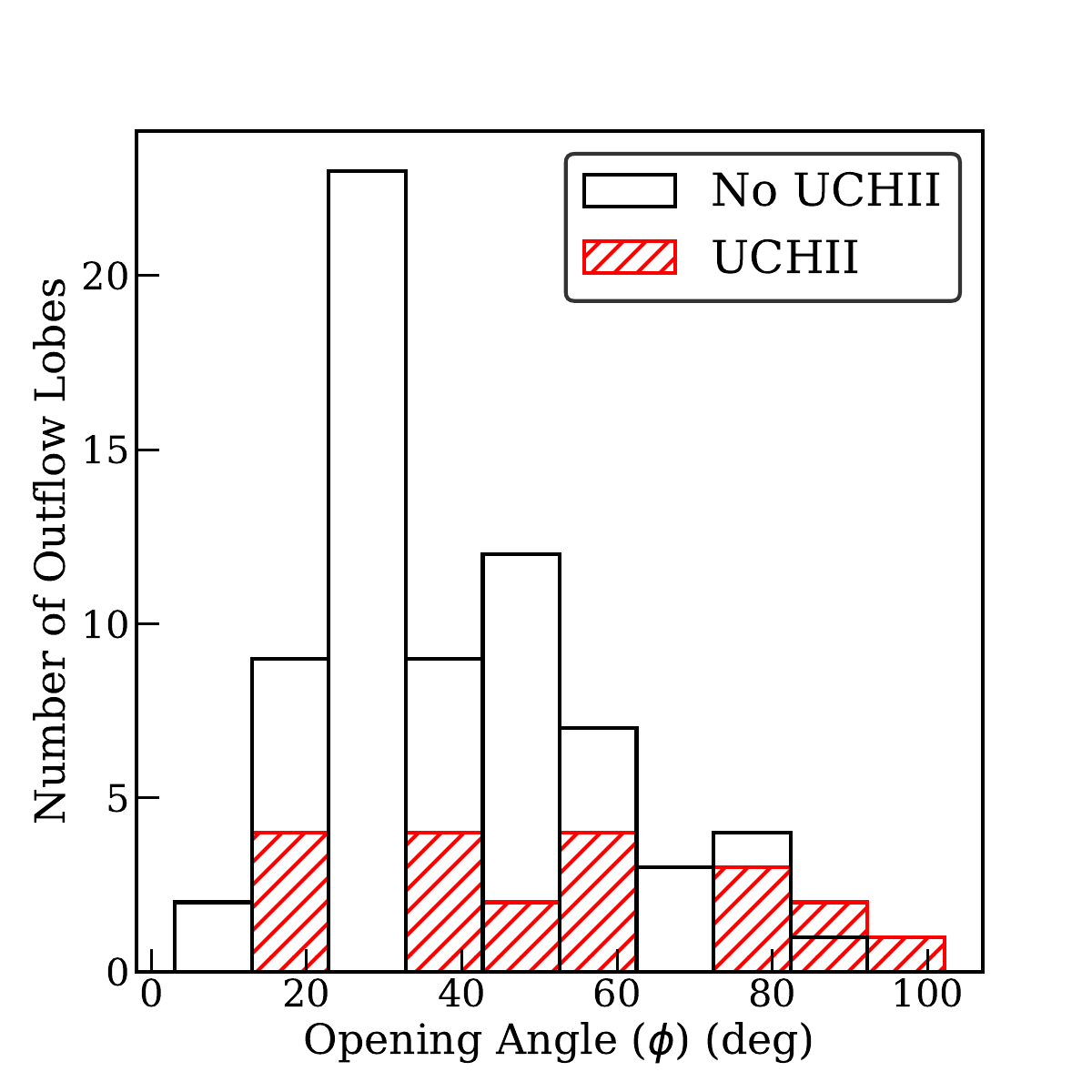}
    \caption{(Left) Scatter plot between outflow opening angle ($\phi$) and terminal velocity ($v_{\text{lobe}}$). The black and cyan squares represent the outflows associated and not associated with UCH{\scriptsize II} emissions, respectively. A representative error bar is added at the bottom right corner. (Right) Histogram of outflow opening angle ($\phi$). Outflows associated with UCH{\scriptsize II} emissions are shown with red bars, while those not associated are shown with black bars.}

    \label{fig-OpeningAngle}
\end{figure}

In Figure~\ref{fig-OpeningAngle} (Left), we plot the opening angle ($\phi$) against the terminal velocity ($v_{\text{lobe}}$) of the outflowing lobes. Although opening angles show an increasing trend of collimation with increasing $v_{\text{lobe}}$, we found no correlation (Spearman correlation coefficient, $\rho\sim-0.14$, p $<0.2$) between $\phi$ and $v_{\text{lobe}}$. This non-correlation of the opening angle with the outflow terminal velocity indicates that HC$_3$N may be tracing less well-collimated outflows. However, we note that our sample includes sources at different evolutionary stages (discussed in section~\ref{subsec:host cores}) which could also lead to such non-correlation. In Figure~\ref{fig-OpeningAngle} (Left), we found that a few outflows have large opening angles ($\phi >90^\circ$). Such large values of opening angle arise possibly because the low-velocity outflows are primarily less collimated, as previously reported in \citet{Arce2006}. The distribution of outflows associated with UCH{\scriptsize II} regions in the $\phi$ vs $v_{\text{lobe}}$ plot are mostly concentrated toward lower velocities ($<$20 km s$^{-1}$) compared to those not associated with UCH{\scriptsize II} regions. 
In Figure~\ref{fig-OpeningAngle} (Right), we presented the histograms of $\phi$ for outflows associated with UCH{\scriptsize II} emission and outflows not associated with UCH{\scriptsize II} emission. The outflows associated with UCH{\scriptsize II} regions show no trend in the distribution of $\phi$, while outflows not associated with UCH{\scriptsize II} regions show a peak around $\phi\sim$30$^{\circ}$. 
 
\subsection{Outflow Position Angles}\label{subsec:position angles}
We estimated the position angle (PA) of the blue- and red-shifted lobes separately, measured from the celestial north to the east. The values of PAs range from $-90^{\circ}$ to $90^{\circ}$, where PAs in the first quadrant have negative signs and PAs in the second quadrant have positive signs, similar to the convention used earlier by \citet[][]{baug2020}. Since we determined the axis of the outflow lobes by visual inspection, the uncertainty in the measurement of position angles is $\sim$10$^\circ$. The PAs of all the outflow lobes are listed in Table~\ref{tab:parameters}.

\section{Discussion} \label{sec:discussion}
\subsection{Outflow parameters with clump properties} \label{dis:parameters}
\begin{figure}[ht!]
    \centering
    \includegraphics[width=0.95\textwidth]{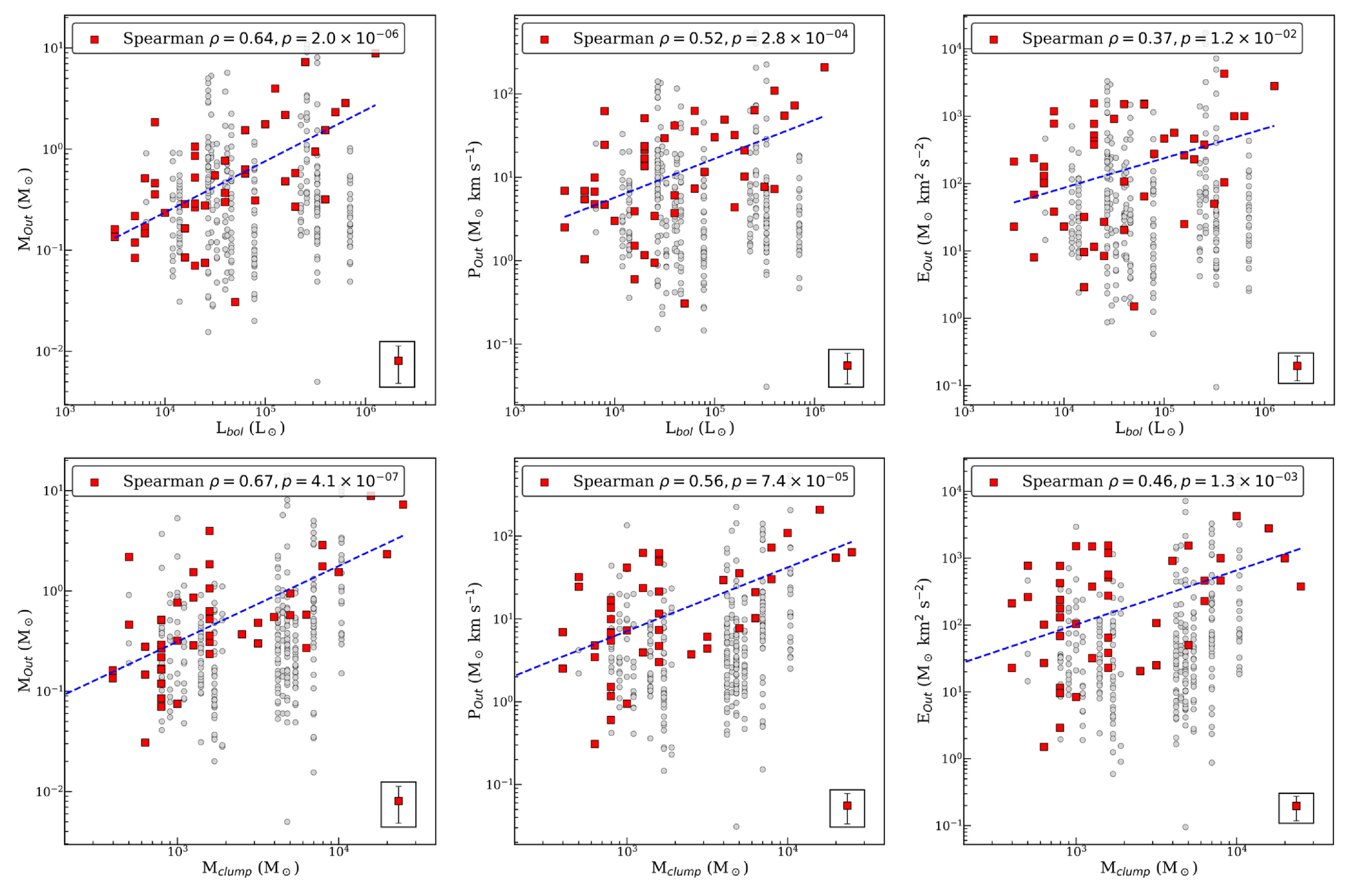}
    \caption{(Top) Variation of Outflow mass (M$_{\text{out}}$), momentum (P$_{\text{out}}$) and energy (E$_{\text{out}}$) with clump luminosity (L$_{\text{bol}}$). (Bottom) Variation of M$_{\text{out}}$, P$_{\text{out}}$ and E$_{\text{out}}$ with the mass of the clumps (M$_{\text{clump}}$). The red squares in each panel indicate the outflow parameters of our targets while the gray circles are obtained from \citet{towner2024}. The blue dashed line in each panel shows the least square fit to the distribution (i.e., the red data points). The Spearman correlation coefficients are indicated in the legends of each panel. A representative error bar is also added to the bottom-right corner of each panel.}
    \label{fig:Mclump}
\end{figure}

In general, outflow parameters are compared with the corresponding host core mass. However, we did not estimate the core mass in this study because of the unavailability of the well-determined dust temperature values of the cores. Our identified outflow catalogue covers a wide range of clump masses ($10^{2.2}-10^{4.4}$ M$_\odot$) and clump luminosities ($10^{3.5}-10^{6.3}$ L$_\odot$) in the high-mass star-forming regions. This gave us an ideal opportunity to investigate the dependency of the derived outflow parameters on their host clumps. Figure~\ref{fig:Mclump} shows the derived outflow properties as a function of clump bolometric luminosity ($L_{\text{bol}}$) and clump mass ($M_{\text{clump}}$). The values of $L_{\text{bol}}$ and $M_{\text{clump}}$ are adapted from \citet{liu2020}. The basic properties of the clumps associated with HC$_3$N outflows are listed in Appendix~\ref{Appendix:Clumps}. We derived the Spearman correlation coefficients ($\rho$) for each outflow parameter (i.e., outflow mass, momentum, energy) with clump parameters (i.e., clump bolometric luminosity, clump mass). We found a moderate correlation (Spearman $\rho$ $\sim$0.5--0.6, p $<10^{-4}$ ) between outflow parameters and clump properties, except for $E_{\text{out}}$-$L_{\text{bol}}$ where we found a comparatively weak correlation (Spearman $\rho$ $\sim$0.4, p $\sim$0.01). Similar correlations were also reported previously in \citet[][]{lopez2009, maud2015, towner2024}{}{}. These correlations suggest that more luminous and massive sources tend to drive more powerful outflows. This is also consistent with the assumption that outflows are entrained material within molecular clouds \citep[][]{bally2016}. We also fit a linear function to find the linear dependency of outflow parameters on clump luminosity and mass in log--log scale. The best-fit relationships between outflow parameters and clump properties are,
\begin{equation}
    \log(M_{\text{out}})= (0.5\pm0.1)\, \log(L_{\text{bol}}) -2.4\pm0.4
\end{equation}
\begin{equation}
     \log(P_{\text{out}})= (0.4\pm0.1)\, \log(L_{\text{bol}}) -1.1\pm0.5
\end{equation}
\begin{equation}
    \log(E_{\text{out}})= (0.4\pm0.2)\, \log(L_{\text{bol}}) +0.3\pm0.7
\end{equation}
\begin{equation}
    \log(M_{\text{out}})= (0.7\pm0.1)\, \log(M_{\text{clump}}) -2.6\pm0.4
\end{equation}
\begin{equation}
    \log(P_{\text{out}})= (0.7\pm0.2)\, \log(M_{\text{clump}}) -1.3\pm0.5
\end{equation}
%\centering and,$\,$
\begin{equation}
     \log(E_{\text{out}})= (0.8\pm0.2)\, \log(M_{\text{clump}}) -0.3\pm0.7
\end{equation}

\begin{figure}[ht!]
    \centering
    \includegraphics[width=0.45\textwidth]{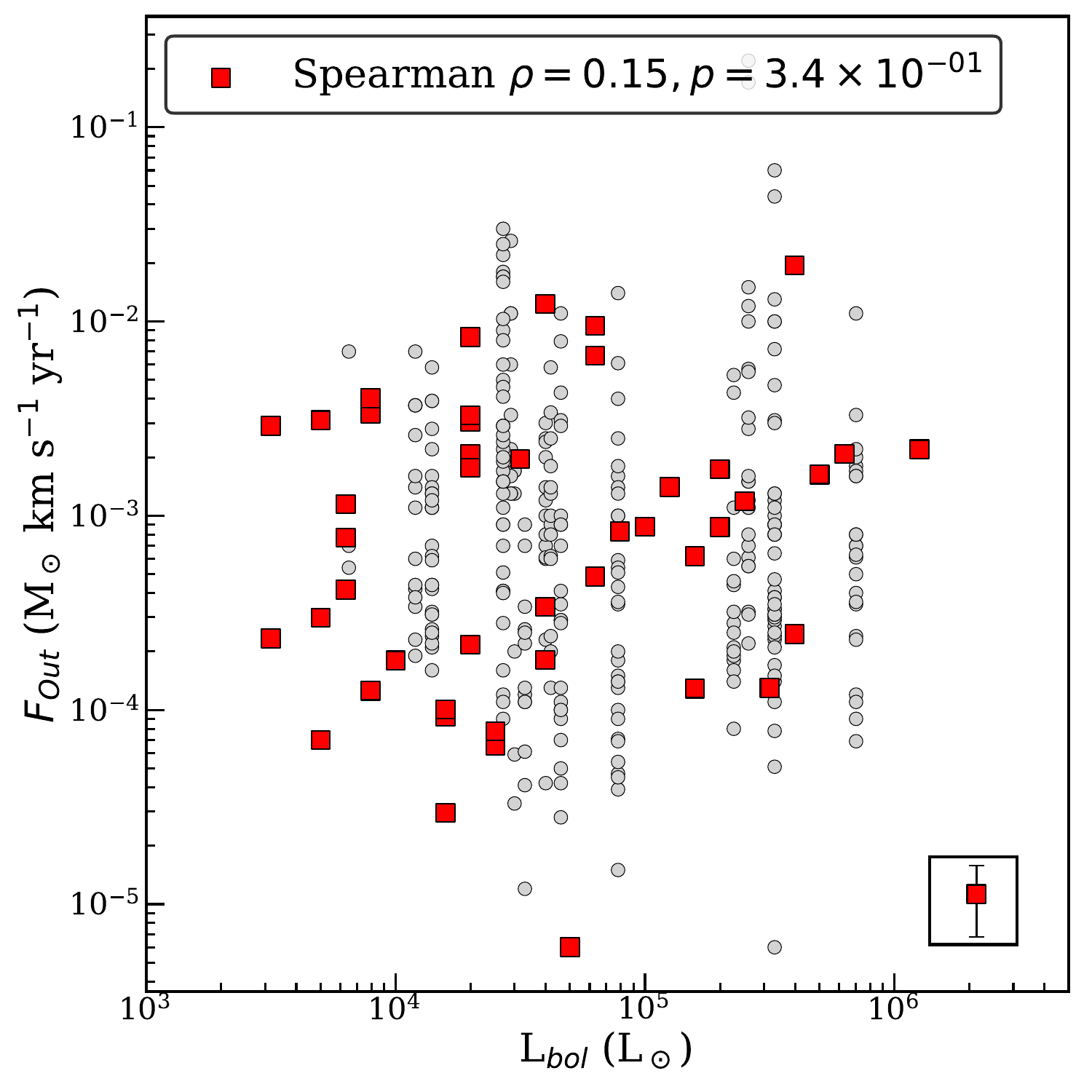}
    \includegraphics[width=0.45\textwidth]{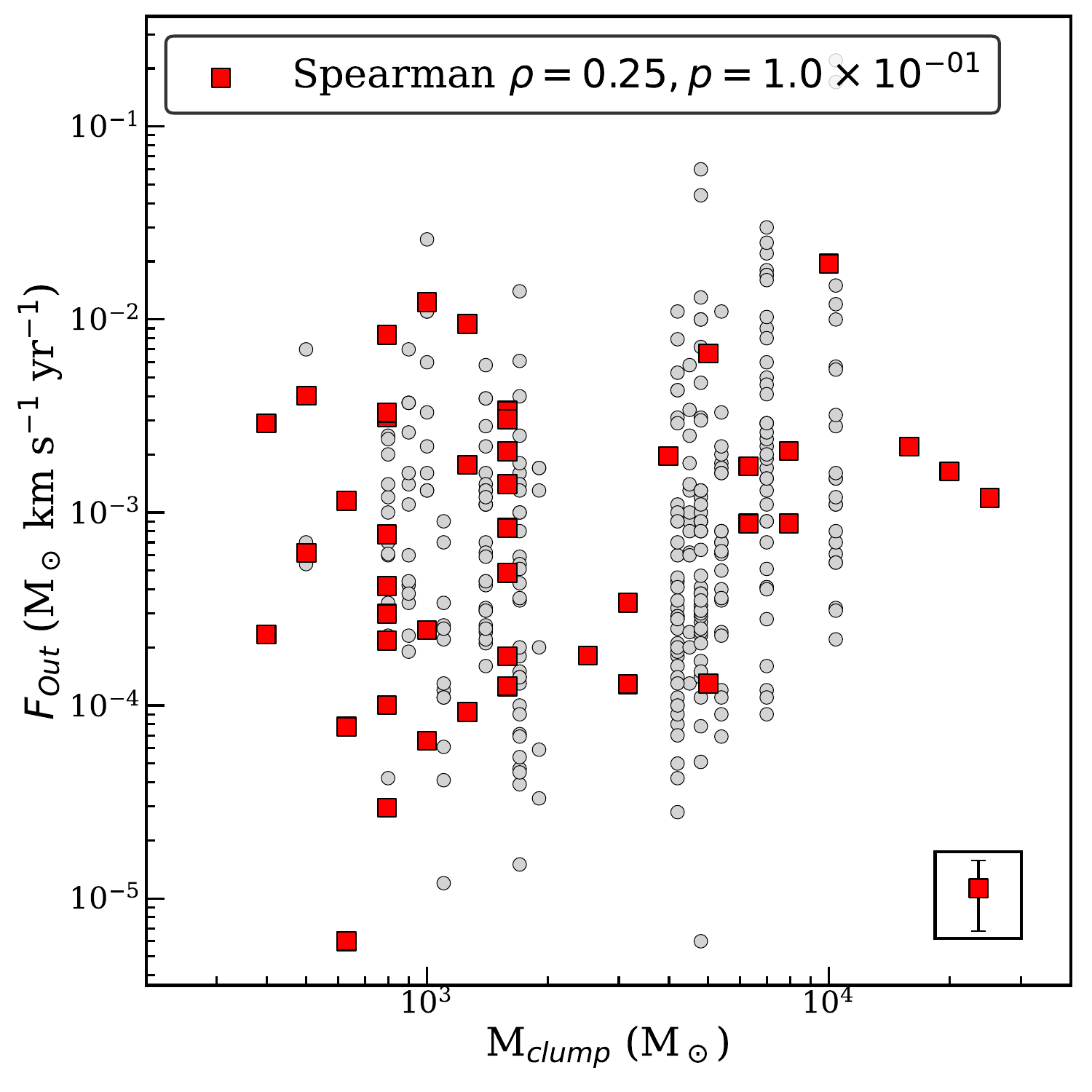}
    \caption{(Left) Scatter plot between outflow mechanical force (F$_{\text{out}}$) vs clump luminosity (L$_{\text{bol}}$). (Right) Scatter plot between outflow mechanical force (F$_{\text{out}}$) vs clump mass (M$_{\text{clump}}$). The red squares and gray circles in each panel represent outflow mechanical force from our targets and \citet[][]{towner2024} respectively. A representative error bar is also added to the bottom-right corner of each panel.}
    \label{fig:Force}
\end{figure}

We compared our outflow parameters to the SiO outflows identified by \citet[][]{towner2024} in 15 massive protoclusters. The outflow parameters from \citet[][]{towner2024} are marked with gray circles in Figure~\ref{fig:Mclump}. Their derived the outflow mass, momentum and energy ranges from $5\times10^{-2}$ to $ 1.4\times10^{1}$ M$_\odot$, $5.7\times10^{-2}$ to $10^{3}$ M$_\odot$ km s$^{-1}$, $10^{-1}$ to $ 2\times10^{4}$ M$_\odot$ km$^{2}$ s$^{-2}$, respectively. Our derived outflow parameters agree well with the range reported in \citet[][]{towner2024}. We also plot outflow mechanical force (F$_{\text{out}}$) against L$_{\text{bol}}$ and M$_{\text{clump}}$ for our identified outflows along with the values reported in \citet[][]{towner2024} (see Figure~\ref{fig:Force}). Although, our derived F$_{\text{out}}$ values are consistent with \citet[][]{towner2024}, we did not find any correlation for F$_{\text{out}}$ with L$_{\text{bol}}$ and M$_{\text{clump}}$ \citep[as also the case for values of][]{towner2024}. One primary reason for such non-correlation is the large scatter in the estimated outflow parameters resulted from the assumption of a constant relative abundance of HC$_3$N with respect to H$_2$. Note that the relative abundance of HC$_3$N varies based on the physical conditions, such as shock velocity, density, and temperature. 

\begin{figure}[ht!]
\centering
\includegraphics[width=0.95\textwidth]{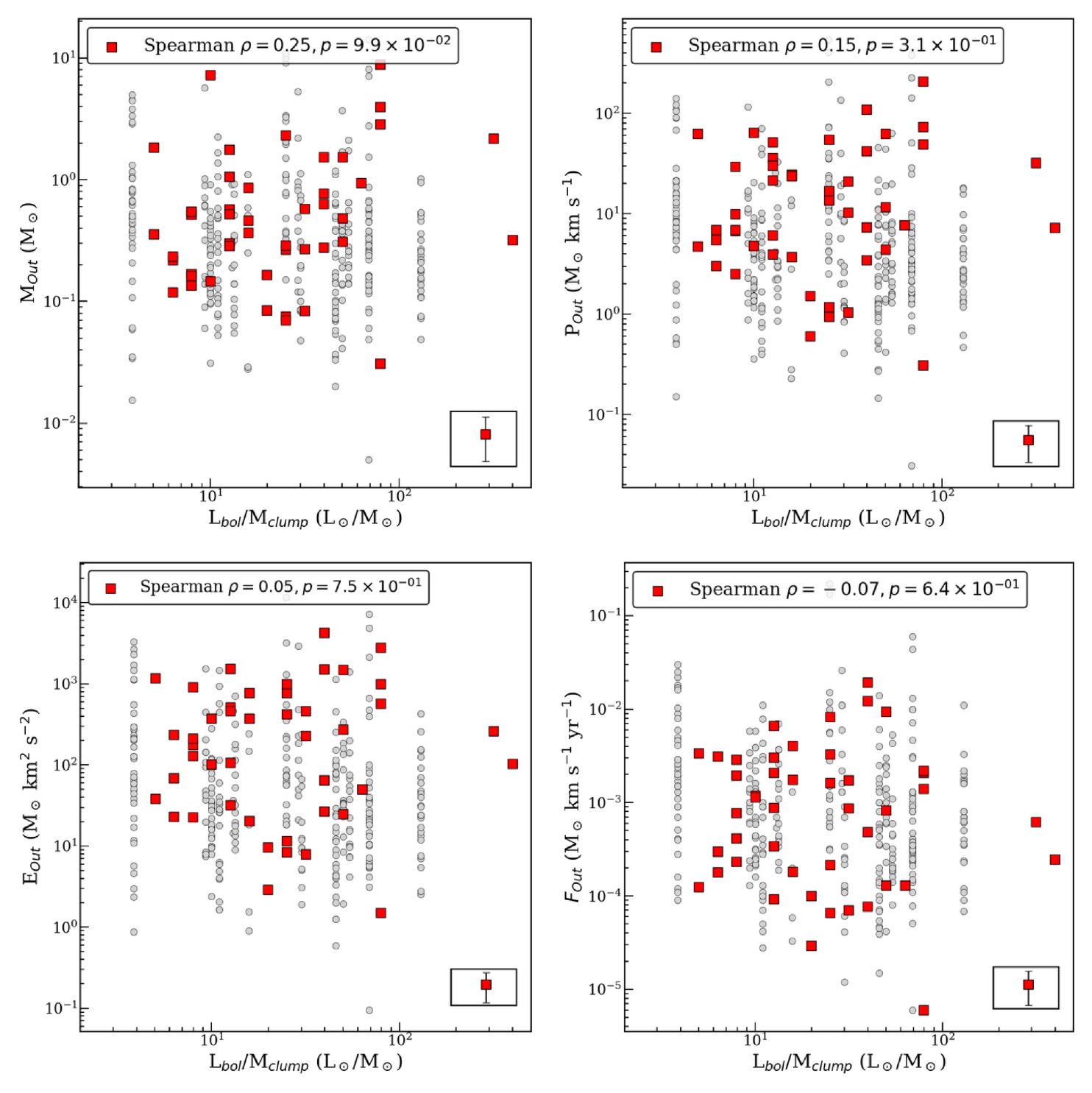}
\caption{Variation of Outflow mass (M$_{\text{out}}$), momentum (P$_{\text{out}}$), energy (E$_{\text{out}}$) and mechanical force (F$_{\text{out}}$) with clump L$_{\text{bol}}$/M$_{\text{clump}}$ ratio.
The symbols are the same as they are in Figure~\ref{fig:Mclump}.}
\label{fig:clump_L_M}
\end{figure}

It is now important to examine at what evolutionary stages of the host clumps HC$_3$N outflows can be detected. For that reason, we explored the correlation of outflow properties with the clump evolutionary stage. We used the  luminosity-to-mass ratio (L$_{\text{bol}}$/M$_{\text{clump}}$) of the clumps as a proxy to their evolutionary stage \citep[][]{molinari2008, molinari2019}. We found no significant correlation (Spearman $\rho \lesssim0.2$, p$\sim$0.5) between outflow parameters and clump evolutionary stage (see Figure~\ref{fig:clump_L_M}). Such non-correlations between outflow parameters and clump L$_{\text{bol}}$/M$_{\text{clump}}$ ratio was also reported in \citet[][]{maud2015, guerra2023, towner2024}. It implies that the detection of HC$_3$N outflows does not represent a specific evolutionary stage of the host clumps, and can arise in all the evolutionary stages.


\subsection{HC$_3$N as outflow tracer} \label{dis:outflow_tracer}
\subsubsection{Line-width and Spatial Distribution}
\label{sec:linewidth}
Theoretical models of \citet[][]{benedettini2013, mendoza2018}{}{} and observations of broad spectral lines with wings \citep[][]{beltran2004,taniguchi2018b,Yu2019}{}{} suggest that HC$_3$N also originate in outflow shocks. \citet[][]{Zinchenko2021}{}{} detected broad HC$_3$N (J=24--23) lines in one massive young stellar object, G18.88MME, and in their study, they compared the molecular gas traced by $^{13}$CO (J=2--1), SiO (J=5--4) and HC$_3$N (J=24--23). Among these tracers, the spectra of $^{13}$CO show wider line-widths compared to HC$_3$N. However, the spatial distributions of molecular gas traced by all three molecules are similar. 

\begin{figure}[ht!]
    \centering
    \includegraphics[width=0.44\textwidth]{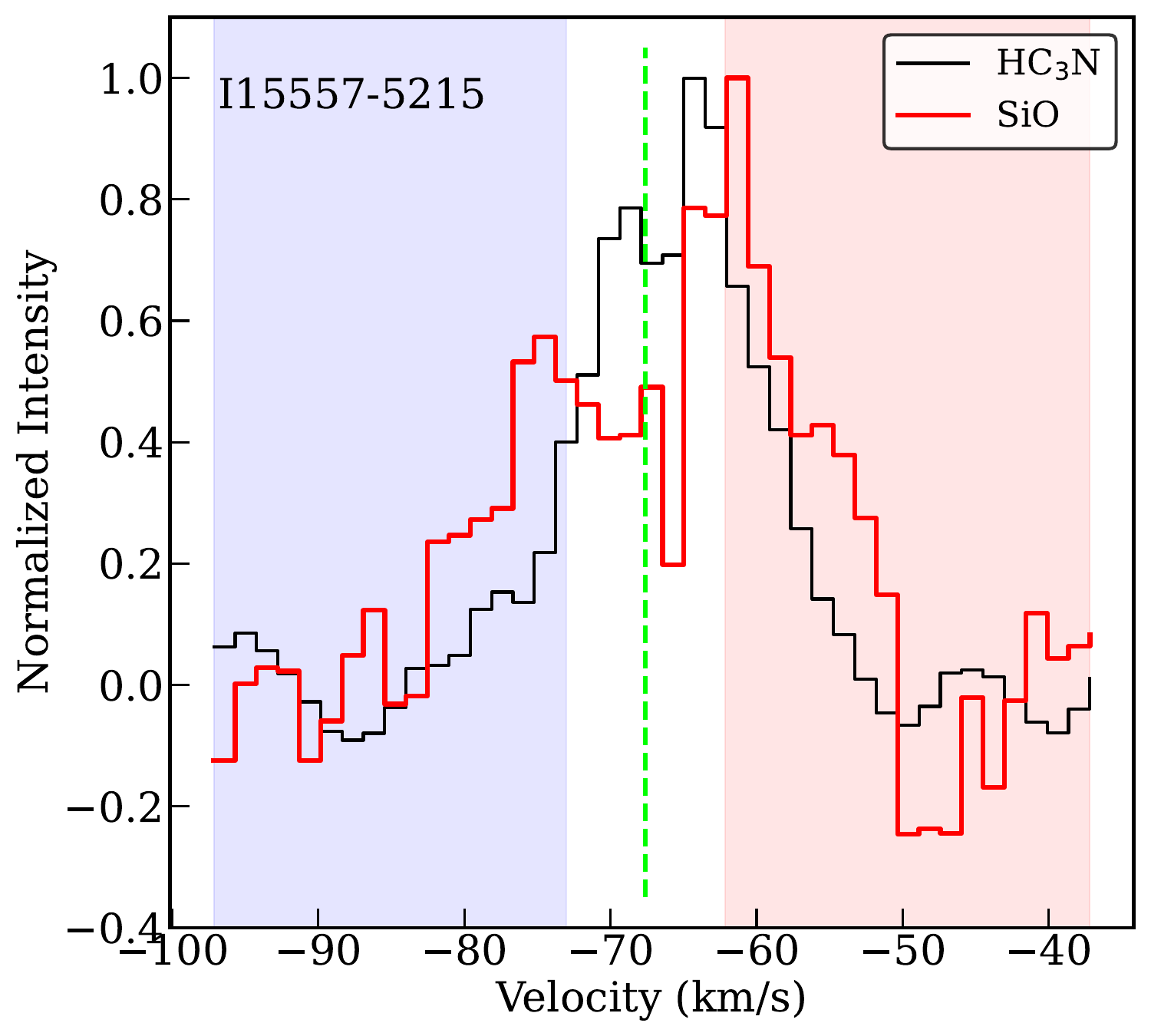}
\includegraphics[width=0.46\textwidth]{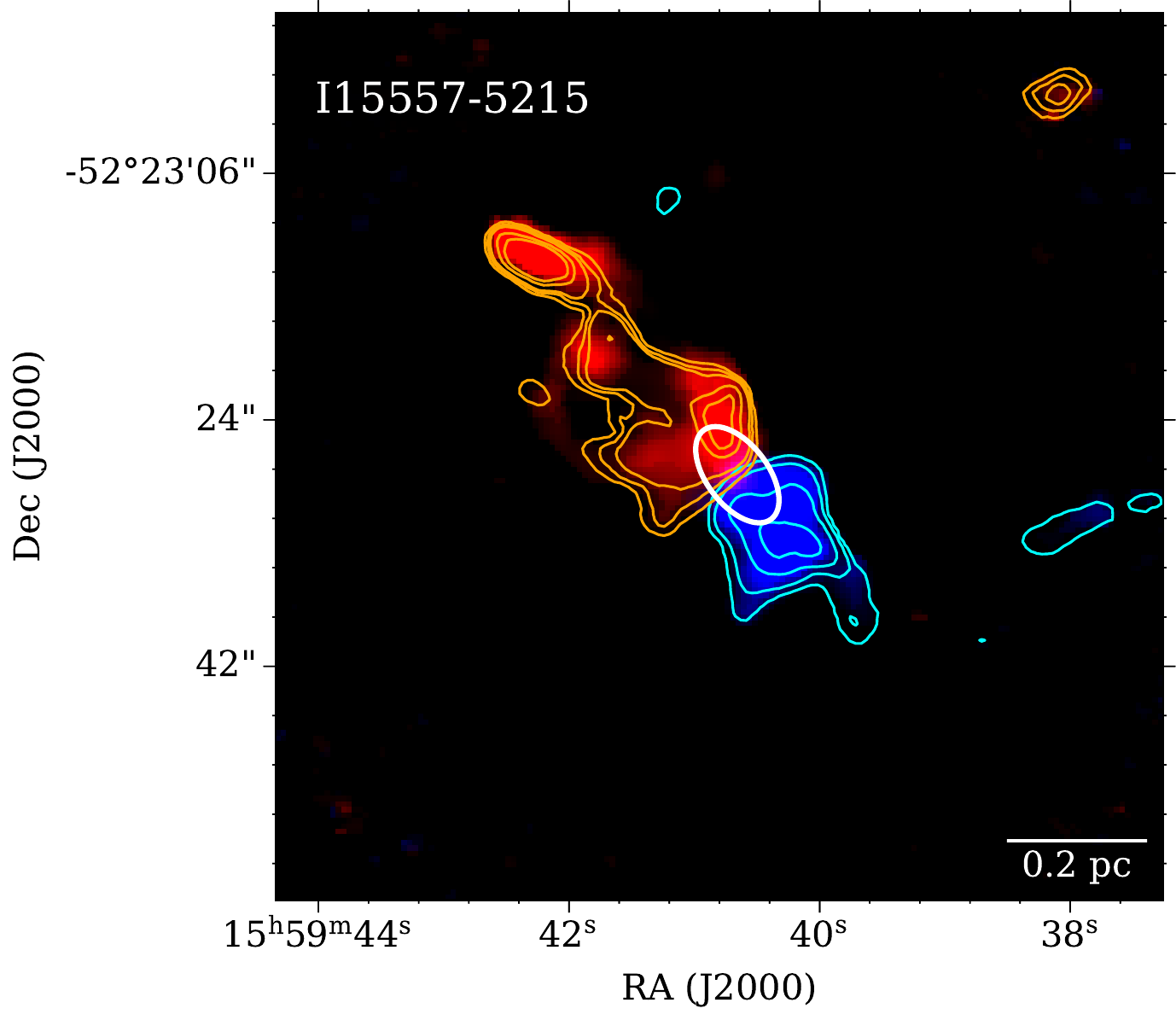}\\

    \includegraphics[width=0.44\textwidth]{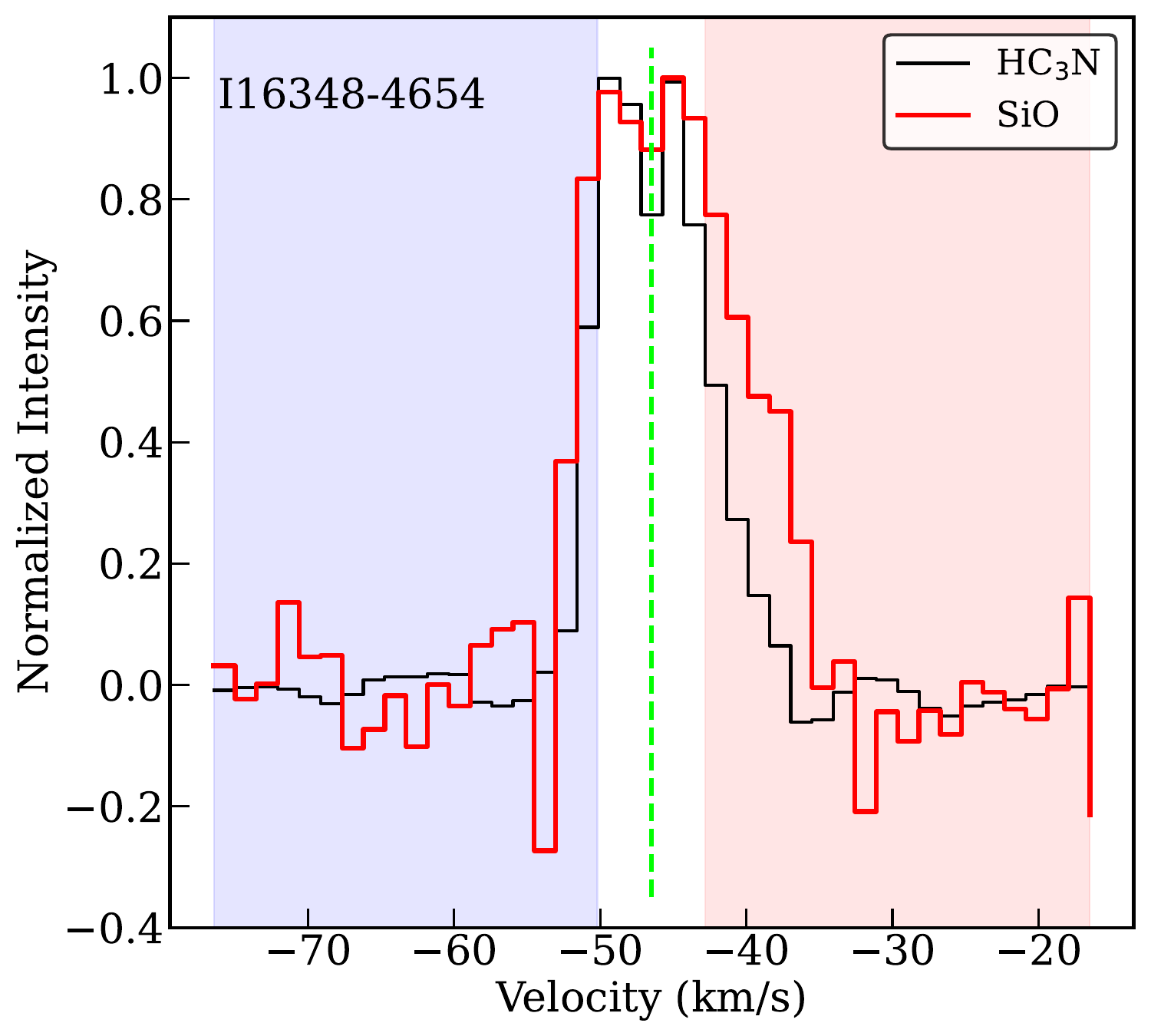}
\includegraphics[width=0.46\textwidth]{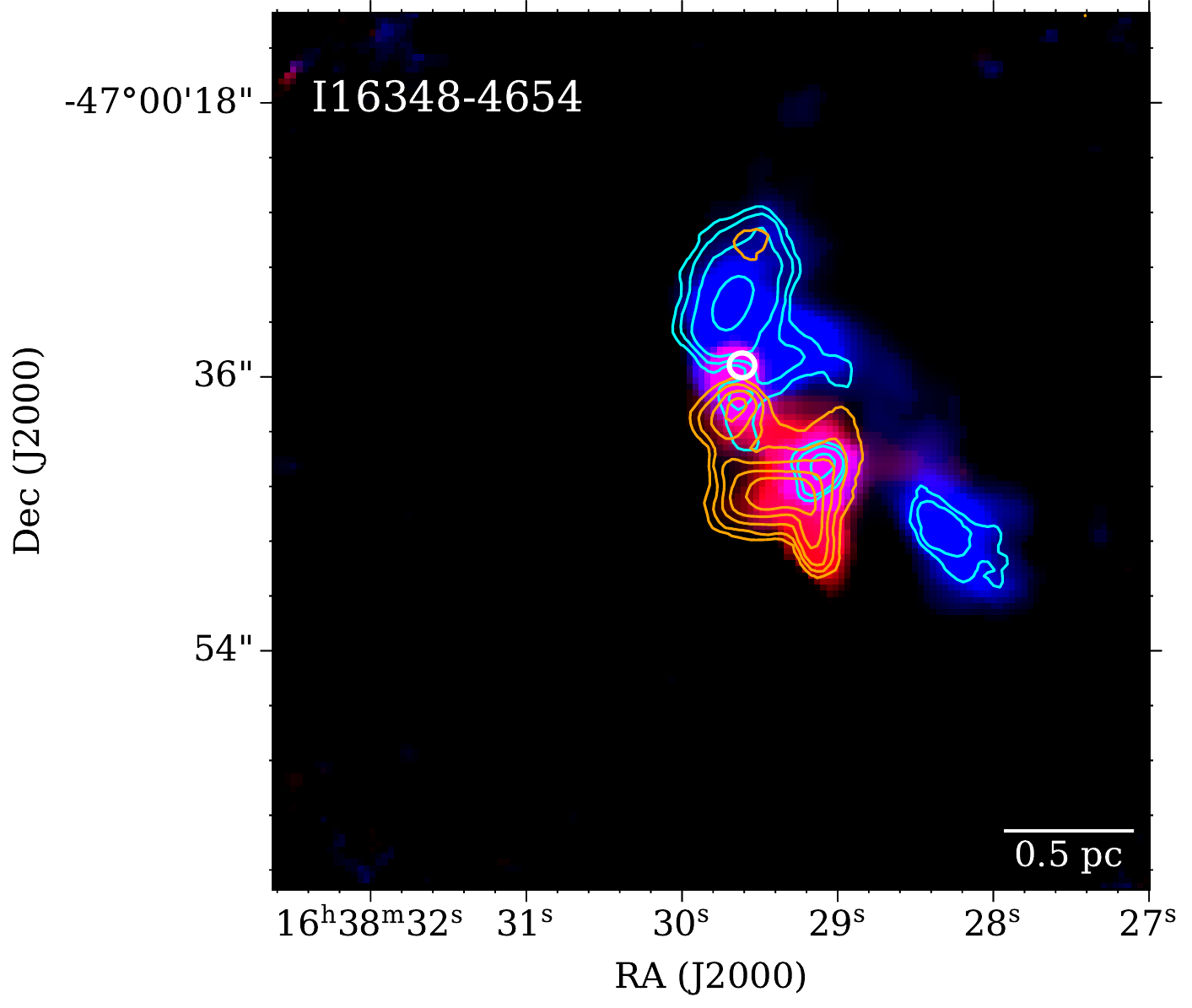}\\

    \includegraphics[width=0.44\textwidth]{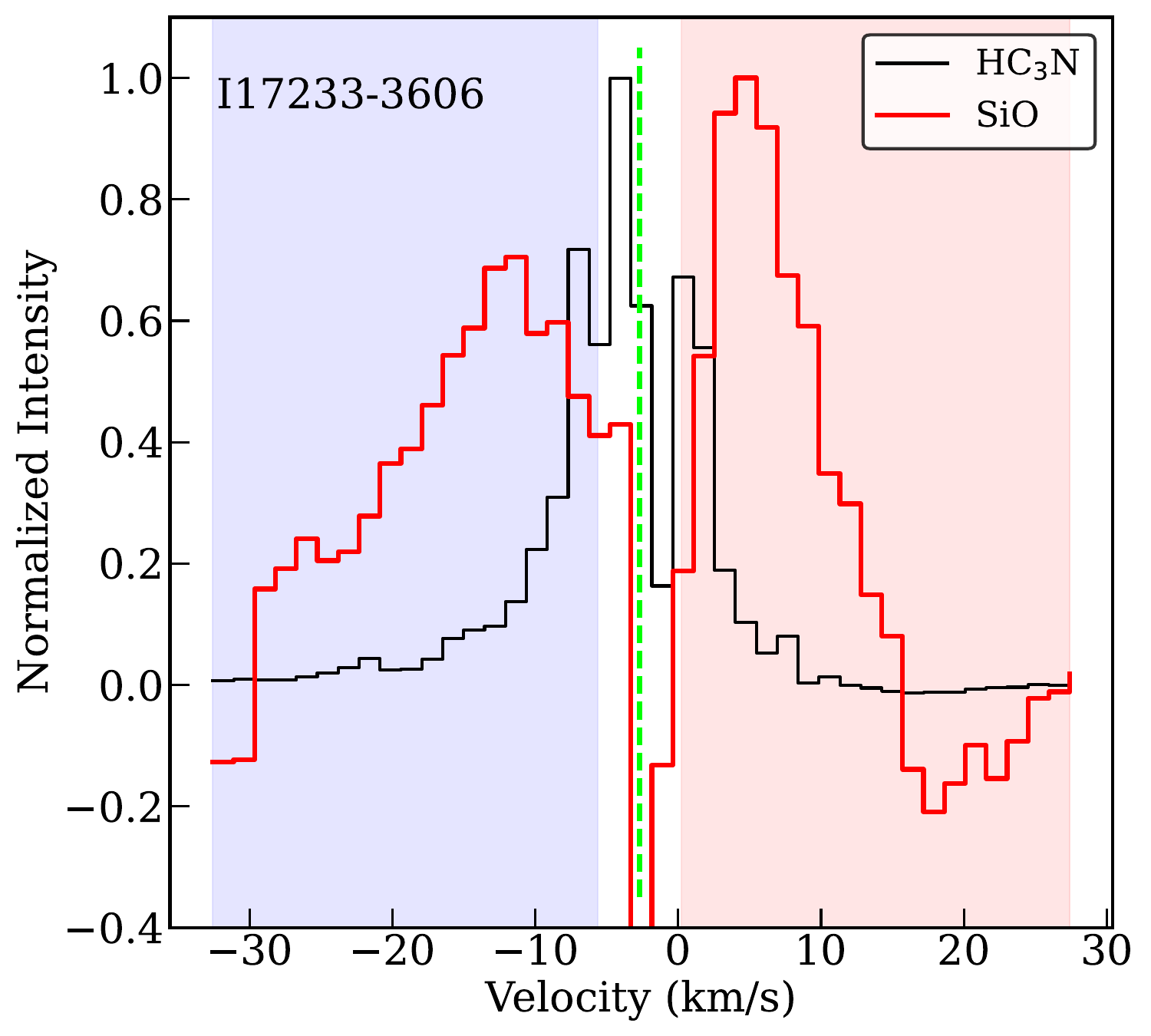}
\includegraphics[width=0.46\textwidth]{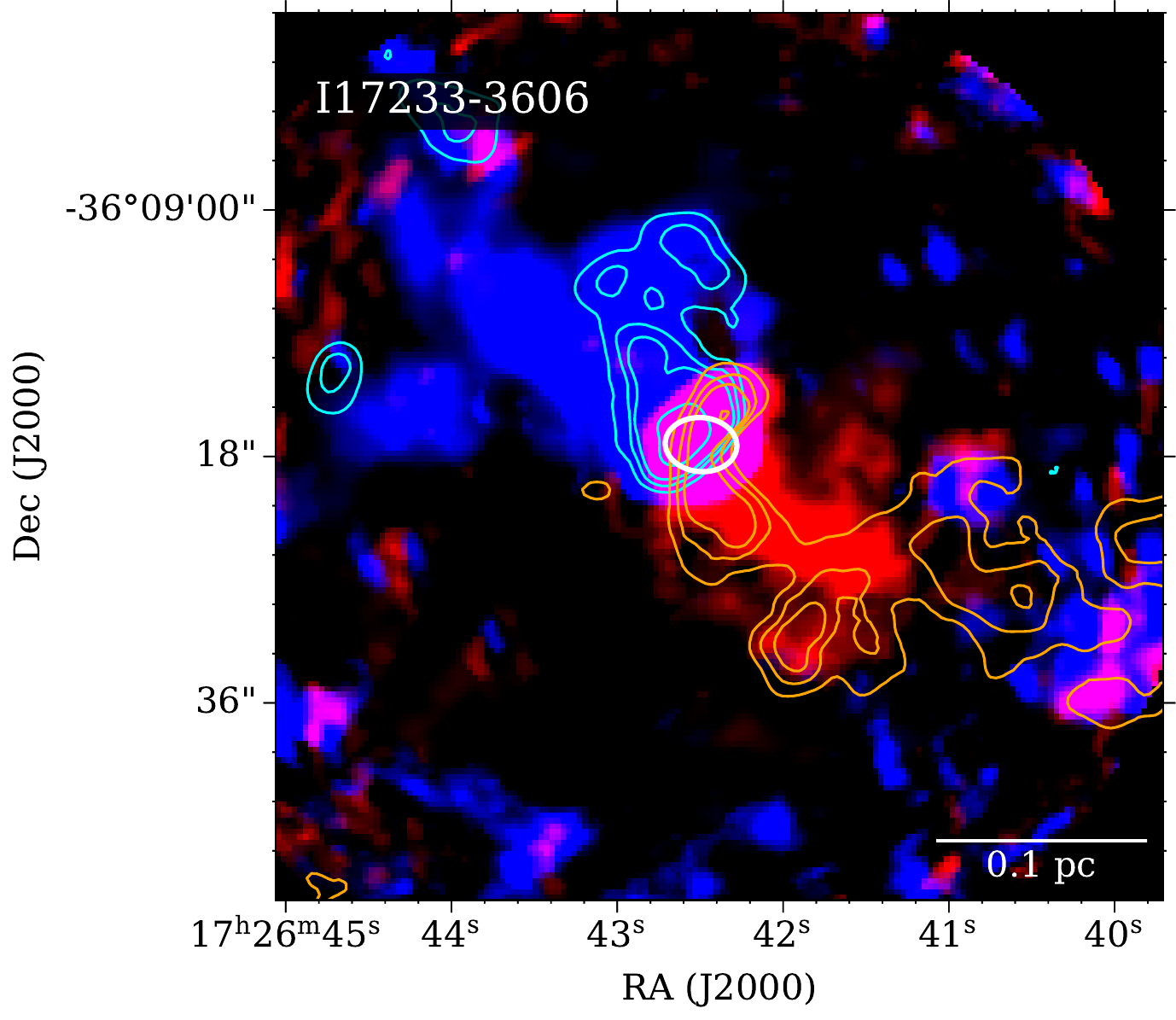}

    \caption{(Left) Normalized HC$_3$N (in black) and SiO (in red) spectra integrated over the ALMA field-of-view for the ATOMS sources. The green dashed line represents local standard of rest velocity (V$_{\text{lsr}}$) of the ATOMS sources. The blue and red-shaded regions indicate the velocity range of the blue and red-shifted lobes, respectively. (Right) Color composite image made using blue-shifted (blue) and red-shifted (red) emissions of HC$_3$N outflow. Cyan and orange contours represent the blue-shifted and red-shifted lobes detected in SiO tracer. The cyan and orange contours are drawn at $[5, 10, 20, 30, 50]\times\sigma$, where $\sigma$ is the background rms of each of the wing (red-shifted/blue-shifted) emission maps. The white ellipse in each panel represents the location of the continuum sources associated with the outflow lobes. A scale bar is also added at the bottom right corner of each panel. The field names are displayed at the top left corner of each panel.}
    \label{fig-spectra}
\end{figure}

In our study, we examined the spectra and the spatial distributions of outflowing gas traced in HC$_3$N and SiO. We found a comparable spectral profile for both HC$_3$N and SiO, with HC$_3$N having a narrower line-width in some cases. Spatial distributions of outflowing gas for both the tracers also appear similar. Figure~\ref{fig-spectra} illustrates line-widths and the spatial distribution of outflows detected in HC$_3$N and SiO towards a few ATOMS targets. For further confirmation of the observed similarity between the line-widths in HC$_3$N and SiO, we performed the Kolmogorov-Smirnov (KS) test on the line-widths of both the tracers. The estimated KS-test statistics is 0.31, with an associated p-value of 0.087. This suggests that a moderate difference between the empirical cumulative distribution functions is present, but this difference is not statistically significant (as p$>$0.05). Therefore, the KS-test do not provide enough evidence to claim the similarity between the line-widths of HC$_3$N and SiO.

\subsubsection{Comparison with a traditional shock tracer} \label{dis:comparison}
We compared our outflow sample identified in HC$_3$N with one of the commonly used shocked-gas tracers, SiO (J=2--1) (hereafter, SiO). The critical density of SiO is $\sim10^5-10^6$ cm$^{-3}$ \citep[][]{langer1990} while HC$_3$N has a critical density of $\sim 10^5$ cm$^{-3}$ \citep[][]{Shirley2015} for typical temperatures (10--100 K) in molecular clouds. A strong correlation between SiO and HC$_3$N column densities has been previously found by \citet[][]{wang2022, He2021}. \citet[][]{He2021} found similar line-widths and integrated intensities between SiO and HC$_3$N. 

\begin{figure}[ht!]
    \centering
    \includegraphics[width=0.46\textwidth]{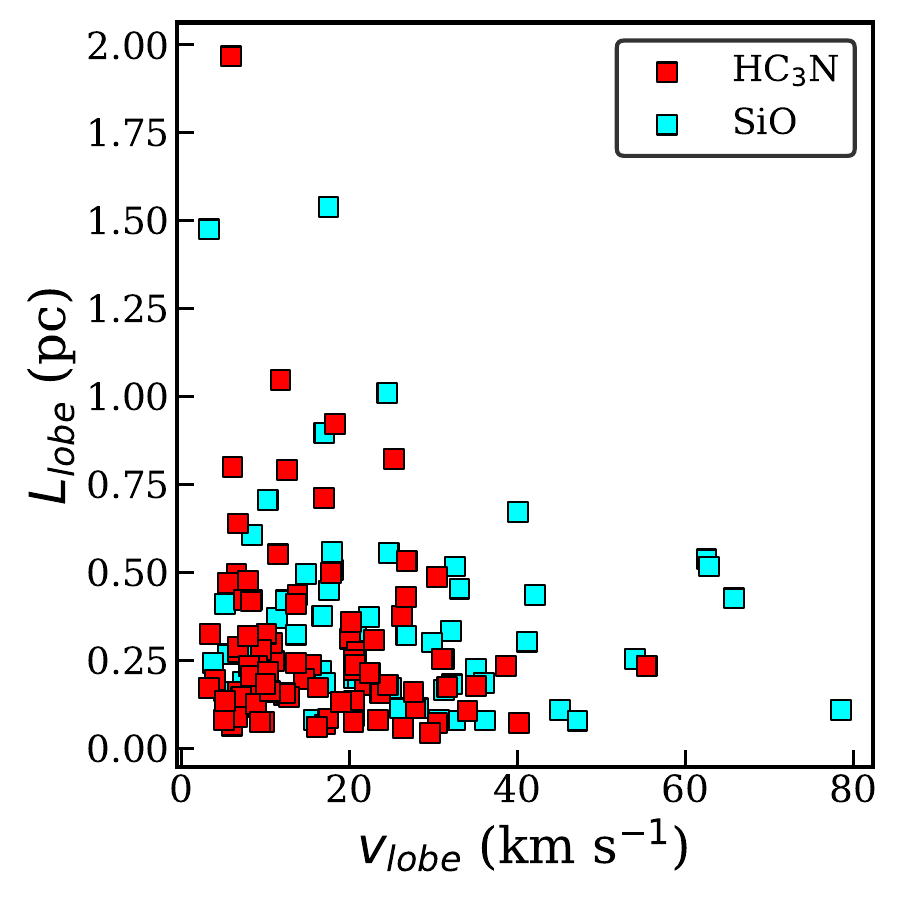}
\includegraphics[width=0.46\textwidth]{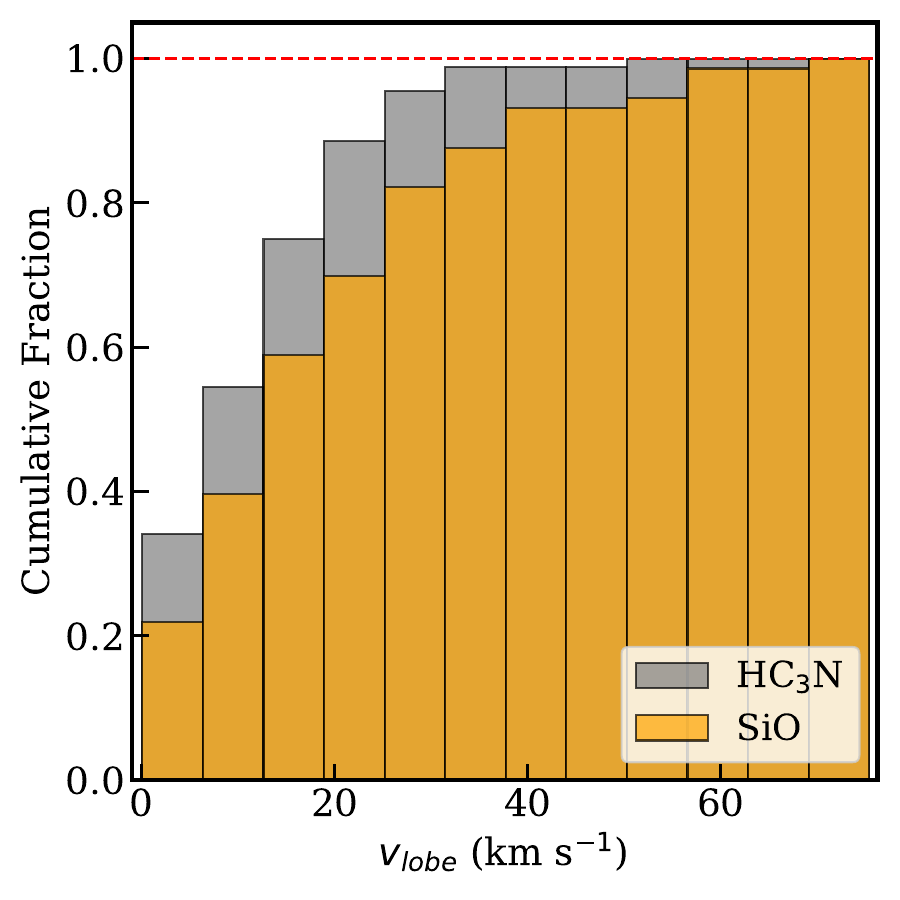}
    \caption{(Left) Comparison of $L_{\text{lobe}}$ with $v_{\text{lobe}}$ for outflows identified using HC$_3$N (in red) and SiO (in cyan). (Right) Distribution of cumulative fraction of $v_{\text{lobe}}$ of outflows detected in HC$_3$N (in grey) and SiO (in orange). The red line indicates a cumulative fraction of 100\%.}
    \label{fig-sio}
\end{figure}

We compared HC$_3$N outflow properties with the SiO outflows of the same targets in ATOMS survey identified by Baug et al.(private communication). Baug et al.(private communication) identified a total of 153 outflows in 146 sources of ATOMS survey. For both the tracers, we found a nearly similar distribution of the extent ($L_{\text{lobe}}$) and terminal velocity of outflow lobes ($v_{\text{lobe}}$; see Figure~\ref{fig-sio} (left)). For the confirmation of the similarity between both tracers, we performed the KS test on the outflow properties (i.e., PA, $L_{\text{lobe}}$ and $v_{\text{lobe}}$) of HC$_3$N with those of SiO. The test on PAs returns a KS test statistics of 0.05, with an associated p-value of $\sim0.997$, while the KS test statistics and p-values for $L_{\text{lobe}}$ are $\sim0.19$ and $\sim0.527$ and for $v_{\text{lobe}}$ the values are $\sim0.24$ and $\sim0.225$. This suggests that the outflow lobes traced by both HC$_3$N and SiO have almost the same orientation (when detected) but they trace outflows at different velocities and extents.  In fact, in case of HC$_3$N, we found a slightly larger fraction of outflows towards low velocities (see Figure~\ref{fig-sio} (right)). The same characteristics, i.e. significant HC$_3$N emission at low velocities without any high velocity component, can be seen in the moment maps and position-velocity diagrams in $\sim$15 targets (see Appendix~\ref{sec:appendix-MomentMaps}). 
%A detailed comparative study of multiple outflow tracers (e.g., HC$_3$N, SiO, HCO$^+$ etc.) will be presented in a follow-up paper.

\subsubsection{Detection of High-velocity Collimated flows}\label{sec:narrow jets}
We further examined the possibility of whether HC$_3$N also traces the fast narrow jets and slow wide outflow components separately as reported in \citet{Zinchenko2021}. For this purpose, we constructed the PV diagrams across the outflow lobes (see figures in Appendix~\ref{sec:transverse}). Among the 45 sources, such components have been identified in 10 sources. Earlier, \citet{torrelles2011} reported the presence of such two-wind outflows in a massive protostar Cep A HW2. Similar structures were also seen in a handful of other regions by \citet{Zinchenko2020, Zinchenko2021, dutta2024}. Note that protostellar outflows are generally explained by X-wind (i.e., jet-driven) and disk-wind (i.e., wind-driven) models \citep[see e.g.][]{shu1994, Arce2007}. As detection of such two components of outflows in a single source is rare, observations of such two components' outflows in HC$_3$N besides other outflow tracers will provide important clues in detailed understanding of the physical condition of jet-driven and wind-driven protostellar outflows.
%\subsubsection{Comparison of outflow mass}
\subsubsection{Limitations in detecting HC$_3$N outflows}
While HC$_3$N can be a potential tracer of outflows, its emission is generally weaker compared to more commonly used outflow tracers, primarily due to the lower relative abundance of HC$_3$N in molecular clouds. In our study, HC$_3$N showed comparatively lower detection of outflows (identifying 45 bipolar outflows among 146 sources) than SiO, which traced 153 outflows in the same 146 targets.
Studies have been performed earlier on protostellar outflows by \citet[][]{lu2021} using HC$_3$N along with multiple other outflow tracers in four massive clouds toward the Central Molecular Zone (CMZ). In that study, the authors identified a total of 43 outflows, of which 38 were detected in HC$_3$N (33 bipolar and 5 unipolar). They also noted that the emission from the in HC$_3$N outflows is fainter compared to the other outflow tracers like SiO, SO etc. A similar weaker detection of HC$_3$N compared to SiO emissions was also reported earlier in \citet[][]{Zinchenko2021}. Despite the fact that HC$_3$N emissions were weak in outflows, the outflow detection rate in CMZ reported by \citet[][]{lu2021}, is much higher compared to the detection rate in our study. One possible explanation for such higher detection is because of the sensitivity of the CMZ data (rms of $\sim1.6-2$ mJy beam$^{-1}$, with beam size $\sim0.28''\times0.19''$) were better than that of the ATOMS survey (rms of $\sim5-10$ mJy beam$^{-1}$, with beam size $\sim1.5''\times1.1''$).

The weaker detection of outflows in HC$_3$N could be related to the physical condition of the surrounding gas. Thus, we examined the relationship between the detection rate of HC$_3$N outflows with various clump parameters like clump mass, luminosity, average dust temperature (as adopted from \citet[][]{liu2020}) and velocity dispersion (using velocity FWHM of the H$^{13}$CO$^+$ line as a proxy). The distribution of the HC$_3$N outflows with clump parameters is presented in Appendix~\ref{sec:detection}. We however did not find any significant correlation in the distribution of HC$_3$N outflows with clump parameters. Therefore, it appears that HC$_3$N emission in outflowing gas is unbiased to the clump properties, and its emission is typically weaker in outflows as compared to the traditionally known outflow tracers. Thus, detection of outflows in HC$_3$N require much improved sensitivity of the observed data in comparison to typical tracer like SiO.

Overall, our study reveals that the typically known dense gas tracer HC$_3$N is capable of tracing the outflows in massive star-forming regions. 
Also, HC$_3$N tends to trace the low-velocity components of outflows in comparison to well-known shocked gas tracers like SiO.

\section{Summary} \label{sec:summary}
In this study, we have presented the first systematic study of protostellar outflows detected in HC$_3$N (J=11--10). About 30\% of the massive protoclusters in our sample (45 out of 146) show an outflow signature in HC$_3$N. Among the 45 identified outflows, 44 are bipolar while one is explosive in nature. We also identified the outflow host cores using the catalogue of ATOMS 3-mm dust continuum cores. We derived the HC$_3$N column density and several other outflow parameters, assuming the gas is optically thin and in local thermodynamic equilibrium. We examined the correlation of our derived outflow parameters with the clump luminosity and mass to find the dependence of an outflow's parameters on the properties of its host cloud. We found a moderate correlation of outflow parameters with both clump luminosity and clump mass. Detection of outflows in clumps with a wide range of masses (10$^{2.2}$-10$^{4.4}$ M$_\odot$) and luminosities (10$^{3.5}$-10$^{6.3}$ L$_\odot$) indicates that HC$_3$N can be used as a potential tracer of outflows. We determined the evolutionary stages of the host clumps using L$_{\text{bol}}$/M$_{\text{clump}}$ as a proxy. No particular correlation or trend is found for the detection of HC$_3$N outflows with the evolutionary stages of the host clumps, implying HC$_3$N could appear in shocked outflowing gas at any evolutionary stages. The non-correlation of HC$_3$N outflows with clump evolutionary stages also implies that HC$_3$N is an unbiased tracer of outflows.

We cross-matched and compared our catalog of HC$_3$N outflows with the outflows detected in a well--known outflow tracer, SiO (J=2--1). We found 
%a similar distribution for lobe extent and outflow velocity in both HC$_3$N and SiO. However, 
that HC$_3$N tends to detect slightly lower velocity components of outflows compared to SiO. We also found that HC$_3$N, while tracing slow wide-angle outflows can also detect fast narrow jets. Overall, HC$_3$N can be a valuable complement to other traditionally known outflow tracers, especially for detecting the low-velocity components of outflows.
\begin{longrotatetable}
\begin{deluxetable*}{cccccccccccccccccccc}
\tabletypesize{\scriptsize}
\tablecaption{Parameters of the identified outflows \label{tab:parameters}}
\tablehead{
\colhead{Outflow Name} & \multicolumn{2}{c}{$v$} & \multicolumn{2}{c}{PA} & \multicolumn{2}{c}{$M_{\text{out}}$} & \multicolumn{2}{c}{$P_{\text{out}}$ (10$^{-1}$} & \multicolumn{2}{c}{$E_{\text{out}}$ (M$_\odot$} & \multicolumn{2}{c}{$t_{\text{dyn}}$} & \multicolumn{2}{c}{$\dot{M}_{\text{out}}$ (10$^{-5}$} & \multicolumn{2}{c}{$F_{\text{out}}$ (10$^{-4}$M$_\odot$ } & \multicolumn{2}{c}{$L_{\text{mech}}$ (10$^{-4}$M$_\odot$}  & \colhead{Detection}\\
\colhead{} & \multicolumn{2}{c}{(km s$^{-1}$)} & \multicolumn{2}{c}{($^\circ$)} & \multicolumn{2}{c}{(10$^{-2}$M$_\odot$)} & \multicolumn{2}{c}{M$_\odot$ km s$^{-1}$)} & \multicolumn{2}{c}{ km$^2$ s$^{-2}$)} & \multicolumn{2}{c}{(10$^3$ yr)} & \multicolumn{2}{c}{M$_\odot$ yr$^{-1}$)} & \multicolumn{2}{c}{km s$^{-1}$ yr$^{-1}$)} & \multicolumn{2}{c}{ km$^2$ s$^{-2}$ yr$^{-1}$)}  & {in SiO}\\
\colhead{} & \colhead{Blue} & \colhead{Red} & \colhead{Blue} & \colhead{Red} & \colhead{Blue} & \colhead{Red} & \colhead{Blue} & \colhead{Red} & \colhead{Blue} & \colhead{Red} & \colhead{Blue} & \colhead{Red} & \colhead{Blue} & \colhead{Red} & \colhead{Blue} & \colhead{Red} & \colhead{Blue} & \colhead{Red} & {}
}
\startdata
\multicolumn{20}{c}{Confirmed Outflows}\\
\hline
$^a$I12320-6122\_O1 &   [-53.3, -43.7] &   [-38.5, -31.3] &-80.1 & 85.8 &       23.0 &        8.9 &       67.9 &        4.7 &  102.5 &    1.6 &       29.3 &       34.0 &        0.8 &        0.3 &        2.3 &        0.1 &        35.0 &         0.5 & $\checkmark$\\
I13291-6249\_O1 &   [-40.0, -36.1] &   [-30.9, -16.6] & 37.1 & 10.4    &  70.8 &      147.2 &      108.8 &      212.8 &   84.0 &  178.8 &       80.1 &       44.1 &        0.9 &        3.3 &        1.4 &        4.8 &        10.5 &        40.5   & $\checkmark$\\
I15394-5358\_O1 &   [-51.6, -45.0] &   [-36.6, -17.9] &  -56.7 & -49.3    &  9.7 &       12.1 &       24.5 &       30.1 &   31.6 &   36.9 &       29.1 &       14.1 &        0.3 &        0.9 &        0.8 &        2.1 &        10.9 &        26.2 & $\checkmark$\\
I15394-5358\_O2 &   [-72.0, -43.4] &   [-39.8, -12.1] & -11.0 & 2.1      & 7.6 &        4.3 &       56.8 &       12.3 &  216.8 &   20.9 &        2.5 &        1.5 &        3.1 &        2.8 &       22.9 &        8.1 &       874.0 &       138.3 & $\checkmark$\\
$^*$I15520-5234\_O1 &   -- &   [-36.3, -34.2] &   --      &  -- & -68.7    & 28.4 &  -- &        7.9 &  -- &    1.4 &    -- &       37.5 &      -- &        0.8 &     -- &        0.2 &    -- &         0.4  & $\checkmark$\\
$^*$I15520-5234\_O2 &   [-59.0, -47.5] &  -- &   66.7 & --  &  95.5 &        -- &      384.7 &   -- &  795.2 &  -- &       15.2 &       -- &        6.3 &   -- &       25.4 &   -- &       524.3 &        --  & $\checkmark$\\
$^*$I15520-5234\_O3 &   -- &   [-36.3, -22.5] &   -- &  26.0 & --   & 67.9 &    -- &      216.2 &  -- &  360.6 &   -- &       13.0 &       -- &        5.2 &   -- &       16.6 &   -- &       276.8  & $\checkmark$\\
$^*$I15520-5234\_O4 &   [-62.0, -47.5] &   -- &  -22.0 & --    & 31.3 &        -- &      145.1 &   -- &  345.6 &  -- &       14.6 &       -- &        2.1 &    -- &        9.9 &   -- &       237.1 &     --  & $\checkmark$\\
$^*$I15520-5234\_O5 &   [-51.7, -47.5] &  -- &  -49.8 & --    & 82.4 &   -- &      198.4 &  -- &  244.6 & -- &       52.4 &   -- &        1.6 &   -- &        3.8 &   -- &        46.7 &   -- & $\checkmark$\\
$^*$I15520-5234\_O6 &   -- &   [-36.3, -34.2] & -- & -69.8 & -- &      105.8 &   -- &      111.8 &  -- &   63.8 &       -- &       35.8 &    -- &        3.0 &    -- &        3.1 &   -- &        17.8 & $\checkmark$\\
$^*$I15520-5234\_O7 &   [-51.7, -47.5] &  -- &  30.4 & --    & 68.9 &   -- &      166.7 &    -- &  206.4 &  -- &       19.4 &  -- &        3.6 &   -- &        8.6 &   -- &       106.4 &    -- & $\checkmark$\\
I15557-5215\_O1 &   [-88.4, -72.5] &   [-64.1, -41.6] & 46.1 & 43.2     & 59.8 &      125.0 &      278.5 &      344.6 &  661.4 &  523.4 &       16.3 &       20.8 &        3.7 &        6.0 &       17.1 &       16.5 &       406.1 &       251.3 & $\checkmark$\\
I16071-5142\_O1 &  [-119.5, -89.7] &   [-79.2, -63.9] & 2.8 & 37.7     & 38.9 &       18.5 &      339.6 &       18.3 & 1532.8 &   11.8 &        5.3 &        6.9 &        7.3 &        2.7 &       64.0 &        2.6 &      2889.3 &        17.0 & $\checkmark$\\
I16158-5055\_O1 &   [-56.8, -52.1] &   [-48.3, -40.7] &  -85.3 & 85.6     & 8.2 &       39.9 &       14.1 &       29.8 &   12.4 &   12.6 &       42.9 &       30.9 &        0.2 &        1.3 &        0.3 &        1.0 &         2.9 &         4.1 & $\checkmark$\\
I16164-5046\_O1 &   [-65.6, -62.5] &   [-53.0, -49.5] & 2.7 & -37.5     & 29.0 &       65.6 &       50.6 &       26.6 &   44.2 &    5.9 &       63.7 &       53.0 &        0.5 &        1.2 &        0.8 &        0.5 &         6.9 &         1.1 & $\checkmark$\\
I16272-4837\_O1 &   [-76.5, -49.5] &   [-42.7, -19.4] &  26.3 & 17.0    & 55.3 &       50.7 &      384.2 &      128.0 & 1363.2 &  188.3 &       16.8 &       16.9 &        3.3 &        3.0 &       22.8 &        7.6 &       810.8 &       111.3 & $\checkmark$\\
I16348-4654\_O1 &   [-61.8, -54.7] &   [-43.9, -31.1] & -9.1 & 22.6     & 14.9 &      710.5 &       51.0 &      587.7 &   88.4 &  289.6 &       66.3 &       52.9 &        0.2 &       13.4 &        0.8 &       11.1 &        13.3 &        54.7 & $\checkmark$\\
I16424-4531\_O1 &   [-67.7, -39.1] &     [-33.2, 2.5] & -54.9 &  -74.8    & 13.4 &       32.7 &      106.5 &      138.6 &  436.0 &  339.1 &        5.8 &        6.4 &        2.3 &        5.1 &       18.5 &       21.8 &       755.7 &       532.9 & $\checkmark$\\
$^a$I16458-4512\_O1 &   [-76.6, -53.7] &   [-47.2, -41.5] & -46.2 & -28.9     & 48.0 &        7.0 &      293.0 &        2.1 &  912.6 &    0.4 &       15.0 &       27.5 &        3.2 &        0.3 &       19.5 &        0.1 &       607.0 &         0.2 & $\times$\\
$^a$I16487-4423\_O1 &   [-53.2, -46.3] &   [-42.5, -35.6] &  69.5 & 66.3     & 2.8 &        4.7 &        6.3 &        3.2 &    7.1 &    1.3 &       14.2 &       14.9 &        0.2 &        0.3 &        0.4 &        0.2 &         5.0 &         0.9 & $\checkmark$\\
I16489-4431\_O1 &   [-63.5, -44.4] &   [-38.9, -25.4] &  23.1 & 31.7    & 11.5 &        5.3 &       62.1 &        5.4 &  173.4 &    4.0 &        8.6 &       11.2 &        1.3 &        0.5 &        7.2 &        0.5 &       201.7 &         3.6 & $\checkmark$\\
I16571-4029\_O1 &   [-43.3, -17.3] &     [-13.6, 7.9] &  -47.3 & -58.5    & 16.1 &       10.5 &      110.2 &       26.0 &  386.6 &   38.1 &        4.3 &        3.6 &        3.8 &        2.9 &       25.7 &        7.3 &       900.5 &       106.4 & $\checkmark$\\
I16571-4029\_O2 &   [-55.0, -17.7] &     [-12.1, 5.6] & -37.0 & -37.9     & 14.6 &       14.4 &      145.4 &       23.1 &  745.7 &   24.3 &        1.9 &        3.8 &        7.8 &        3.7 &       77.1 &        6.0 &      3954.1 &        63.0 & $\checkmark$\\
I16571-4029\_O3 &   [-22.8, -18.7] &     [-15.0, 0.6] &  -49.5 & -46.3     & 0.9 &        6.2 &        1.3 &       10.4 &    1.0 &   10.5 &       11.3 &        5.1 &        0.1 &        1.2 &        0.1 &        2.1 &         0.8 &        20.7 & $\checkmark$\\
I17006-4215\_O1 &   [-29.4, -24.1] &   [-21.2, -14.8] & -75.7 & 77.3     & 16.5 &       11.2 &       28.5 &        6.0 &   24.9 &    2.1 &       45.0 &       42.2 &        0.4 &        0.3 &        0.6 &        0.1 &         5.5 &         0.5 & $\checkmark$\\
I17016-4124\_O1 &   [-48.2, -31.6] &    [-23.9, -7.3] & 18.2 & 51.9     & 33.8 &       24.2 &      166.3 &       44.1 &  418.2 &   45.5 &       12.4 &       11.4 &        2.7 &        2.1 &       13.5 &        3.9 &       338.5 &        39.9 & $\checkmark$\\
I17016-4124\_O2 &   [-48.2, -27.7] &    [-27.7, -8.7] & -33.8 & -13.0     & 37.8 &       14.7 &      187.2 &       40.8 &  476.0 &   60.6 &       12.9 &        7.3 &        2.9 &        2.0 &       14.5 &        5.6 &       369.6 &        83.1 & $\checkmark$\\
$^a$I17220-3609\_O1 &  [-111.5, -96.2] &   [-91.1, -82.2] & -50.1 & -63.8     & 83.3 &      149.2 &      381.5 &      167.5 &  898.1 &  105.7 &       29.4 &       50.3 &        2.8 &        3.0 &       13.0 &        3.3 &       305.6 &        21.0 & $\times$\\
I17233-3606\_O1 &   [-38.4, -13.0] &      [4.2, 20.1] & 33.1 & 47.8     & 52.4 &       24.3 &      393.3 &       24.3 & 1499.2 &   14.2 &        3.3 &        7.7 &       16.0 &        3.1 &      120.1 &        3.2 &      4576.4 &        18.4 & $\checkmark$\\
$^a$I17271-3439\_O1 &   [-48.5, -20.0] &    [-15.2, 37.8] & 72.7 & 74.1    &  59.6 &       94.7 &      473.6 &      617.9 & 1946.6 & 2316.3 &        8.6 &        4.4 &        6.9 &       21.3 &       55.2 &      139.3 &      2269.3 &      5220.6 & $\checkmark$\\
I18117-1753\_O1 &     [21.4, 35.0] &     [38.6, 44.8] & -4.6 & -2.4     & 11.7 &       18.2 &       48.0 &       13.0 &  101.6 &    5.6 &       16.1 &       30.7 &        0.7 &        0.6 &        3.0 &        0.4 &        63.1 &         1.8 & $\checkmark$\\
I18159-1648\_O1 &      [3.8, 17.2] &     [24.5, 30.1] & -22.5 & -44.7     & 11.2 &        3.5 &       47.2 &        0.6 &  101.2 &    0.1 &        4.1 &        8.5 &        2.7 &        0.4 &       11.4 &        0.1 &       244.7 &         0.1 & $\checkmark$\\
I18182-1433\_O1 &     [45.2, 56.5] &     [63.0, 80.4] & -51.1 & -21.0     & 46.1 &       39.7 &      163.0 &       73.6 &  296.1 &   81.1 &       14.2 &       12.0 &        3.3 &        3.3 &       11.5 &        6.1 &       209.0 &        67.6 & $\checkmark$\\
I18290-0924\_O1 &     [74.6, 79.1] &     [83.3, 93.6] & 33.0 & 30.4      & 7.2 &       16.3 &       13.3 &       16.8 &   12.7 &   10.3 &       25.2 &       13.2 &        0.3 &        1.2 &        0.5 &        1.3 &         5.0 &         7.8 & $\checkmark$\\
I18479-0005\_O1 &      [5.0, 13.4] &     [20.0, 22.5] & 38.5 & 11.9    & 753.3 &      134.5 &     2038.2 &       46.6 & 2789.5 &    8.1 &       93.4 &      354.6 &        8.1 &        0.4 &       21.8 &        0.1 &       298.6 &         0.2 & $\times$\\
I18507+0121\_O1 &     [49.1, 52.1] &     [62.9, 79.8] & 52.1 & 46.1      & 1.8 &       14.3 &        3.5 &       21.7 &    3.4 &   19.5 &       19.2 &       10.1 &        0.1 &        1.4 &        0.2 &        2.2 &         1.8 &        19.4 & $\checkmark$\\
$^a$I18507+0121\_O2 &     [31.6, 55.8] &     [60.0, 74.0] & -13.7 & 14.9     & 10.5 &        3.0 &       64.9 &        4.3 &  208.1 &    3.6 &        2.3 &        4.0 &        4.5 &        0.8 &       27.9 &        1.1 &       893.7 &         9.0 & $\checkmark$\\
\hline
\multicolumn{20}{c}{Probable Outflows}\\
\hline
I13140-6226\_O1 &   [-45.4, -37.4] &   [-31.5, -26.4] & 25.8 & 48.6     & 36.5 &  15.0 &   96.5 &        2.9 &  129.6 &    0.4 &       23.8 &       30.4 &        1.5 &        0.5 &        4.1 &        0.1 &        54.6 &         0.1   & $\checkmark$\\
I14164-6028\_O1 &   [-53.1, -49.9] &   [-43.1, -41.4] & -40.4 & -62.8      & 5.9 &        2.5 &        9.5 &        0.9 &    7.8 &    0.2 &       14.3 &       27.2 &        0.4 &        0.1 &        0.7 &        0.0 &         5.5 &         0.1 & $\times$\\
$^a$I14453-5912\_O1 &   [-45.0, -42.0] &   [-38.1, -31.9] &  -70.4 & -64.2     & 2.5 &        5.9 &        3.2 &        2.8 &    2.1 &    0.8 &       17.1 &       26.2 &        0.1 &        0.2 &        0.2 &        0.1 &         1.2 &         0.3 & $\checkmark$\\
$^a$I15584-5247\_O1 &   [-84.3, -78.5] &   [-75.4, -66.7] & 54.9 & 57.1     & 16.3 &       12.4 &       29.7 &        9.6 &   27.4 &    4.5 &       60.2 &       22.3 &        0.3 &        0.6 &        0.5 &        0.4 &         4.6 &         2.0  & $\times$\\
I16060-5146\_O1 &  [-101.9, -99.3] &   [-93.5, -71.2] & -83.8 & 83.8     & 21.4 &      265.0 &       29.4 &      698.8 &   20.4 &  984.2 &       89.8 &       34.1 &        0.2 &        7.8 &        0.3 &       20.5 &         2.3 &       288.3 & $\checkmark$\\
I16119-5048\_O1 &   [-68.3, -50.8] &   [-45.6, -42.0] & 80.1 & -72.0     & 45.1 &        7.2 &      212.7 &        1.9 &  514.4 &    0.3 &       10.3 &       29.0 &        4.4 &        0.2 &       20.7 &        0.1 &       500.5 &         0.1 & $\times$\\
I16351-4722\_O1 &   [-60.5, -47.0] &   [-32.4, -19.6] & -28.5 & -59.3     & 22.8 &        8.3 &      110.7 &        5.4 &  273.4 &    2.3 &       13.8 &       18.6 &        1.7 &        0.4 &        8.0 &        0.3 &       198.0 &         1.2 & $\checkmark$\\
I16385-4619\_O1 & [-127.7, -120.2] & [-115.1, -111.6] &  -54.5 & -74.8   & 199.1 &      197.4 &      470.7 &       21.7 &  567.3 &    1.5 &       33.9 &      139.1 &        5.9 &        1.4 &       13.9 &        0.2 &       167.3 &         0.1 & $\checkmark$\\
I16445-4459\_O1 & [-136.5, -125.8] & [-119.6, -116.0] & 88.1 & -73.0     & 90.1 &       86.1 &      285.5 &       17.1 &  462.9 &    2.1 &       33.2 &       99.6 &        2.7 &        0.9 &        8.6 &        0.2 &       139.6 &         0.2 & $\checkmark$\\
I16547-4247\_O1 &   [-43.8, -35.6] &   [-26.5, -17.4] & -66.4 & -56.4     & 10.5 &       52.4 &       30.3 &       43.1 &   44.6 &   20.0 &       11.9 &       18.6 &        0.9 &        2.8 &        2.5 &        2.3 &        37.4 &        10.7 & $\checkmark$\\
I17175-3544\_O1 &   [-36.5, -15.7] &    [-10.0, 14.8] & 80.4 & 53.6     & 79.0 &       75.0 &      429.6 &      197.7 & 1189.3 &  309.3 &        6.9 &        6.1 &       11.5 &       12.4 &       62.3 &       32.6 &      1724.5 &       509.6 & $\checkmark$\\
I17204-3636\_O1 &   [-25.2, -21.1] &    [-15.1, -7.6] & 22.7 & 56.3      & 5.2 &       11.3 &        8.6 &        6.5 &    7.3 &    2.3 &       25.5 &        9.8 &        0.2 &        1.2 &        0.3 &        0.7 &         2.9 &         2.4 & $\checkmark$\\
I18264-1152\_O1 &     [35.2, 40.9] &     [46.3, 57.1] & 88.7 & 84.7      & 9.0 &       26.8 &       18.4 &       28.7 &   19.1 &   19.2 &       52.4 &       31.7 &        0.2 &        0.8 &        0.4 &        0.9 &         3.6 &         6.0 & $\checkmark$\\
I18445-0222\_O1 &     [83.4, 84.4] &     [88.9, 96.6] &  -41.2 & -79.0     & 4.3 &       32.7 &        3.7 &       33.6 &    1.6 &   18.9 &       55.9 &       19.2 &        0.1 &        1.7 &        0.1 &        1.7 &         0.3 & 9.8 & $\times$
\enddata
\tablenotetext{a}{No Outflow driving continuum source detected}
\tablenotetext{*}{Explosive outflows}
\end{deluxetable*}
\end{longrotatetable}

\begin{deluxetable*}{ccccccccc}
\tabletypesize{\scriptsize}
\tablecaption{Details of the driving cores \label{tab:dustcore}}
\tablehead{
\colhead{Outflow Name} & \colhead{$^b$RA (J2000)} & \colhead{$^b$DEC (J2000)}  & \colhead{Vlsr} & \colhead{$^b$F$_{int}$} &  \multicolumn{3}{c}{$^b$Associated with} \\
{} & (h:m:s) & (d:m:s) & (km s$^{-1}$) & (mJy) &Young Core & HMC & UCHII
}
\startdata
I13140-6226\_O1 & 13:17:15.48 & -062:42:24.4 & -34.7 & 59.6 & N & Y & N \\
I13291-6249\_O1 & 13:32:31.02 & -063:05:20.2  & -34.0 & 562.4 & N & N & Y \\
I14164-6028\_O1 & 14:20:8.59  & -060:42:1.4 & -46.8 & 24.1 & Y & N & N \\
I15394-5358\_O1 & 15:43:16.55 & -054:07:13.7 & -39.7 & 226.7 & N & Y & N \\
I15394-5358\_O2 & 15:43:17.99 & -054:07:32.2 & -42.6 & 33.3 & Y & N & N \\
I15520-5234\_O1 & 15:55:48.47 &  -052:43:6.7 & -43.5 & 1703.0 & N & Y & Y \\
I15520-5234\_O2 & 15:55:48.47 &  -052:43:6.7 & -43.5 & 1703.0 & N & Y & Y \\
I15520-5234\_O3 & 15:55:48.47 &  -052:43:6.7 & -43.5 & 1703.0 & N & Y & Y \\
I15520-5234\_O4 & 15:55:48.47 &  -052:43:6.7 & -43.5 & 1703.0 & N & Y & Y \\
I15520-5234\_O5 & 15:55:48.47 &  -052:43:6.7 & -43.5 & 1703.0 & N & Y & Y \\
I15520-5234\_O6 & 15:55:48.47 &  -052:43:6.7 & -43.5 & 1703.0 & N & Y & Y \\
I15520-5234\_O7 & 15:55:48.47 &  -052:43:6.7 & -43.5 & 1703.0 & N & Y & Y \\
I15557-5215\_O1 & 15:59:40.72 & -052:23:28.1 & -68.7 & 33.1 & N & Y & N \\
I16060-5146\_O1 & 16:09:52.64 & -051:54:54.5 & -95.5 & 6366.0 & N & Y & Y \\
I16071-5142\_O1 & 16:10:59.26 & -051:50:10.5 & -85.0 & 15.9 & Y & N & N \\
I16119-5048\_O1 & 16:15:45.01 & -050:55:56.6 &  -48.2 & 59.2 & Y & N & N \\
I16158-5055\_O1 & 16:19:36.71 & -051:03:23.7 & -49.8 & 289.0 & N & N & Y \\
I16164-5046\_O1 & 16:20:11.08 & -050:53:14.8 & -57.3 & 4904.0 & N & Y & Y \\
I16272-4837\_O1 & 16:30:58.77 & -048:43:53.6 & -47.5 & 113.7 & N & Y & N \\
I16348-4654\_O1 & 16:38:29.65 & -047:00:35.7 & -44.1 & 414.4 & N & Y & Y \\
I16351-4722\_O1 & 16:38:50.48 & -047:28:2.8 & -40.7 & 184.8 & N & Y & Y \\
I16385-4619\_O1 & 16:42:13.86 & -046:25:29.5 & -117.7 & 189.9 & N & N & Y \\
I16424-4531\_O1 & 16:46:7.28  & -045:36:40.7 & -35.9 &  16.4 & Y & N & N \\
I16445-4459\_O1 & 16:48:5.16 & -045:05:10.0 & -122.2 & 123.5 & N & N & Y \\
I16489-4431\_O1 & 16:52:33.98 & -044:36:28.3 & -42.1 & 3.8 & Y & N & N \\
I16547-4247\_O1 & 16:58:17.18  & -042:52:7.6 & -31.1 & 144.1 & N & Y & N \\
I16571-4029\_O1 & 17:00:32.28 & -040:34:10.2 & -14.7 & 45.4 & Y & N & N \\
I16571-4029\_O2 & 17:00:31.94 & -040:34:12.9  & -14.8 & 27.9 & N & Y & N \\
I16571-4029\_O3 & 17:00:33.00 & -040:34:13.8 & -16.4 & 17.4 & Y & N & N \\
I17006-4215\_O1 &  17:04:12.86 & -042:19:53.1 & -21.9 & 695.1 & N & N & Y \\
I17016-4124\_O1 &  17:05:10.90 & -041:29:06.80 & -26.7 & 100.3 & N & Y & N \\
I17016-4124\_O2 &  17:05:10.90 & -041:29:06.80 & -26.7 & 100.3 & N & Y & N \\
I17175-3544\_O1 & 17:20:53.43 & -035:46:57.9 & -7.2 & 611.0 & N & Y & N \\
I17204-3636\_O1 & 17:23:50.31 & -036:38:54.4 & -16.7 & 3.5 & Y & N & N \\
I17233-3606\_O1 & 17:26:42.46 & -036:09:17.8 & -4.3 & 529.2 & N & Y & N \\
I18117-1753\_O1 & 18:14:38.46 & -017:51:57.1 & 36.5 & 3.2 & Y & N & N \\
I18159-1648\_O1 & 18:18:54.67 & -016:47:50.3 & 21.7 & 51.4 & N & Y & N \\
I18182-1433\_O1 &  18:21:9.05 & -014:31:47.9 & 59.5 & 53.2 & N & Y & N \\
I18264-1152\_O1 & 18:29:14.40 & -011:50:23.2 & 43.6  & 71.8 & N & Y & N \\
I18290-0924\_O1 & 18:31:43.29 & -009:22:25.8 & 83.6 & 4.4 & Y & N & N \\
I18445-0222\_O1 & 18:47:9.99  & -002:18:45.6 & 86.3 & 6.9 & Y & N & N \\
I18479-0005\_O1 & 18:50:30.80 & -000:01:53.7 & 16.3 & 1184.0 & N & Y & Y \\
I18507+0121\_O1 &  18:53:18.01 & +001:25:25.6 & 57.6 & 89.5 & N & Y & N
\enddata
\tablenotetext{b}{Reference:\citet[][]{HLiu2021}{}{}}
\end{deluxetable*}
\begin{acknowledgments}
We thank the anonymous referee for the valuable comments and suggestions, which improved the quality of the manuscript. AH and TB thank the support of the S. N. Bose National Centre for Basic Sciences under the Department of Science and Technology, Govt. of India. AH also thanks the CSIR-HRDG, Govt. of India for funding the fellowship. This work was performed in part at the Jet Propulsion Laboratory, which is operated by the California Institute of Technology under contract with the National Aeronautics and Space Administration (NASA). MJ acknowledges support from the Research Council of Finland, grant No. 348342. G.G. and LB gratefully acknowledge support by the ANID BASAL project FB210003. SRD acknowledges support from the Fondecyt Postdoctoral fellowship (project code 3220162) and ANID BASAL project FB210003.

This paper utilizes the following ALMA data: ADS/JAO.ALMA 2019.1.00685.S and
2017.1.00545.S. ALMA is a partnership of ESO (representing its member states), NSF (USA),
and NINS (Japan), together with NRC (Canada), MOST and ASIAA (Taiwan), and KASI (Republic of Korea), in cooperation with the Republic of Chile. The Joint ALMA Observatory is operated by ESO, AUI/NRAO, and NAOJ.
\end{acknowledgments}

\facilities{ALMA}

\software{astropy \citep{astropy2013}, spectral-cube \citep{spectralcube2019}, APLpy \citep{aplpy2012}, SAOds9 \citep{ds9}}

\bibliography{references}{}
\bibliographystyle{aasjournal}

\appendix
\section{Moment Maps and Position-Velocity Diagrams of All Identified Outflows}\label{sec:appendix-MomentMaps}
As presented in Section~\ref{subsec:identification} we identified a total of 45 outflows within 146 massive star-forming clumps. An example figure was presented in Figure~\ref{fig-identification}. In this section, we present the moment maps and position-velocity diagrams of all the identified outflows (See Fig. Set 1).

\figsetstart
\figsetnum{1}
\figsettitle{Moment Maps and Position-Velocity Diagrams of Identified Outflows}

\figsetgrpstart
\figsetgrpnum{1.1}
\figsetgrptitle{Outflows in I12320-6122}
\figsetplot{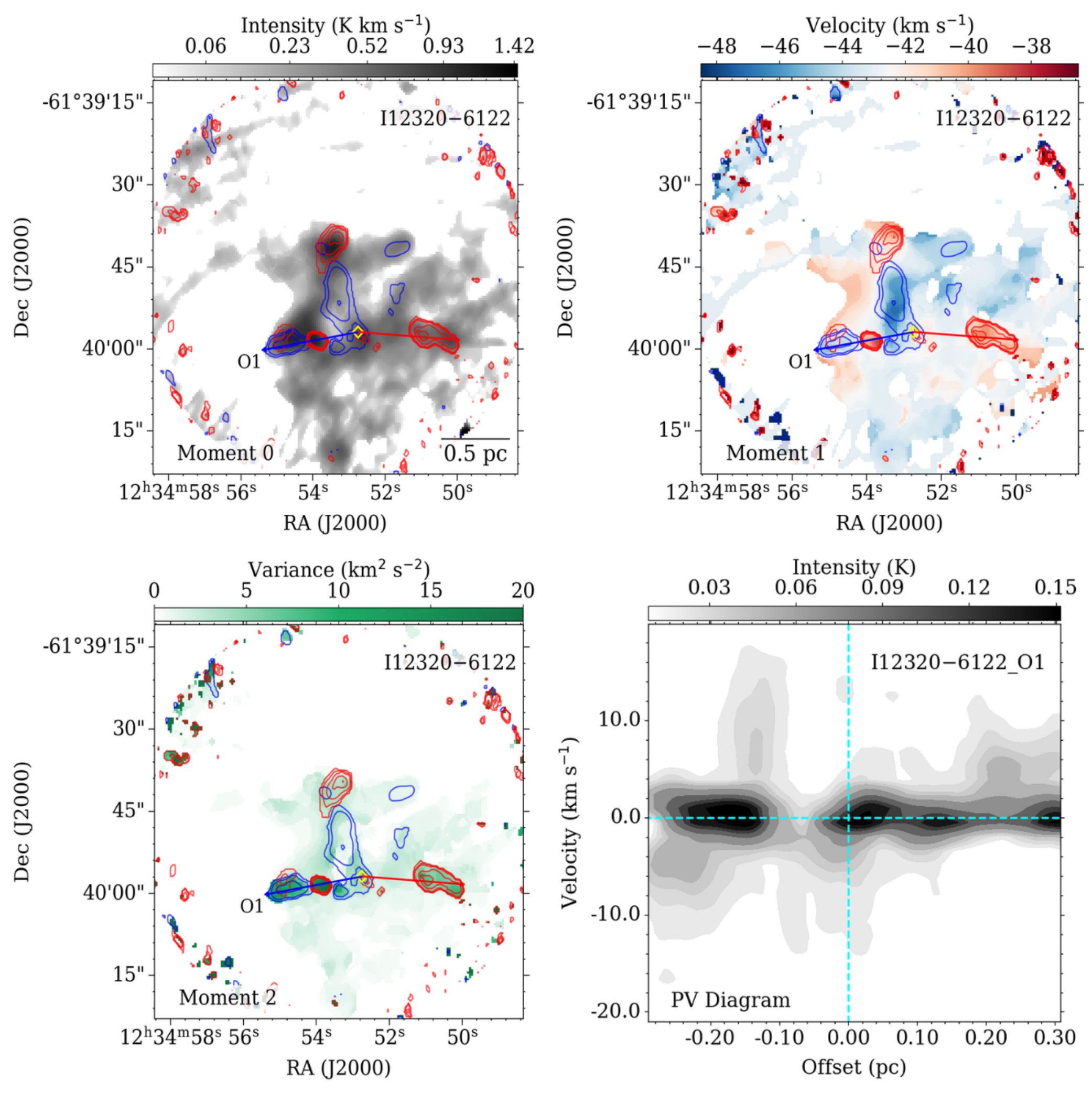}
\figsetgrpnote{ The top panels show the moment-0 map (left) and moment-1 map (right) of HC$_3$N line. The bottom panels show the moment-2 map (left, with a color bar above it) and the PV diagram along the outflow axis (right).}
\figsetgrpend

\figsetgrpstart
\figsetgrpnum{1.2}
\figsetgrptitle{Outflows in I13140-6226}
\figsetplot{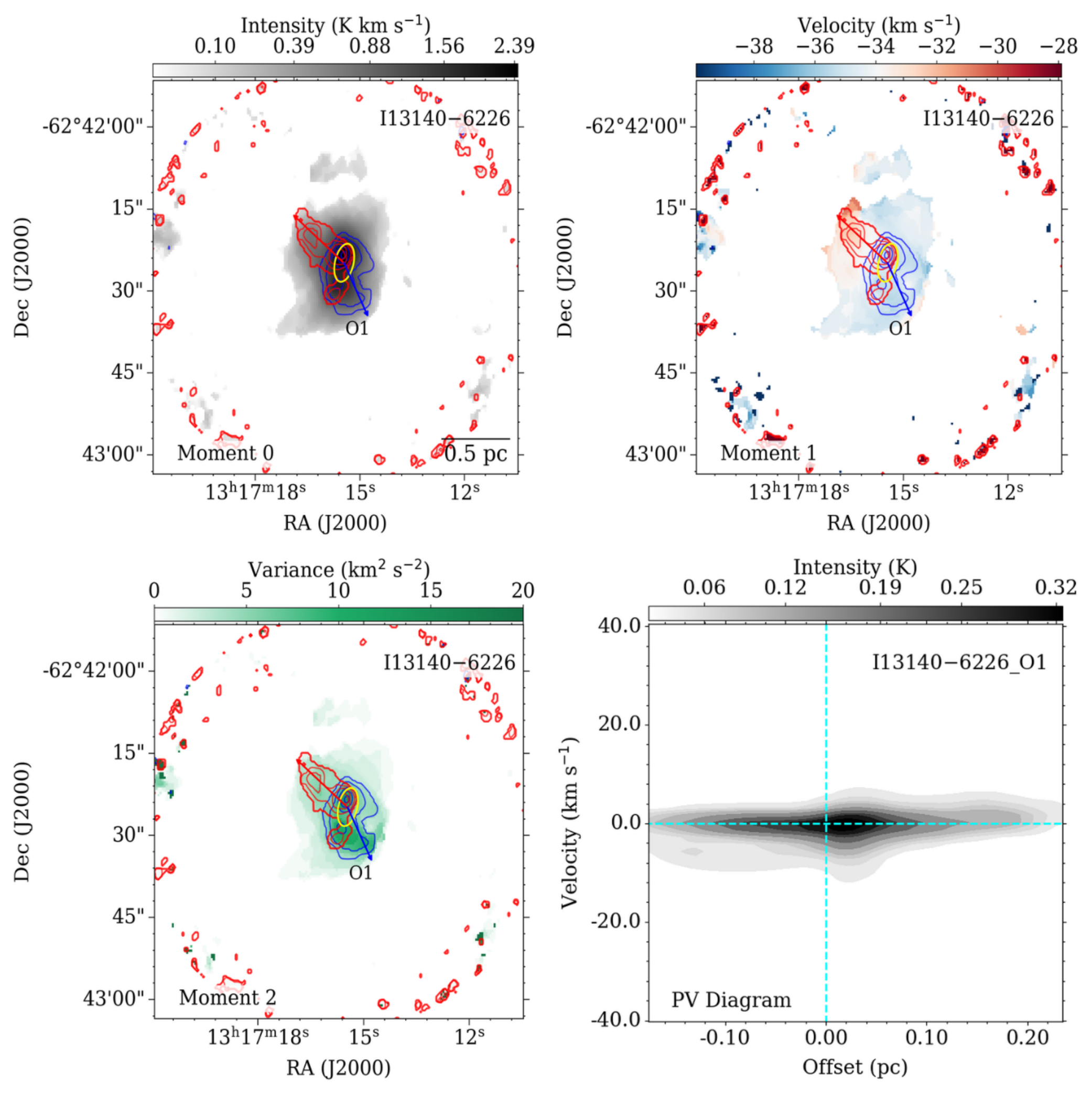}
\figsetgrpnote{ The top panels show the moment-0 map (left) and moment-1 map (right) of HC$_3$N line. The bottom panels show the moment-2 map (left, with a color bar above it) and the PV diagram along the outflow axis (right).}
\figsetgrpend

\figsetgrpstart
\figsetgrpnum{1.3}
\figsetgrptitle{Outflows in I13291-6249}
\figsetplot{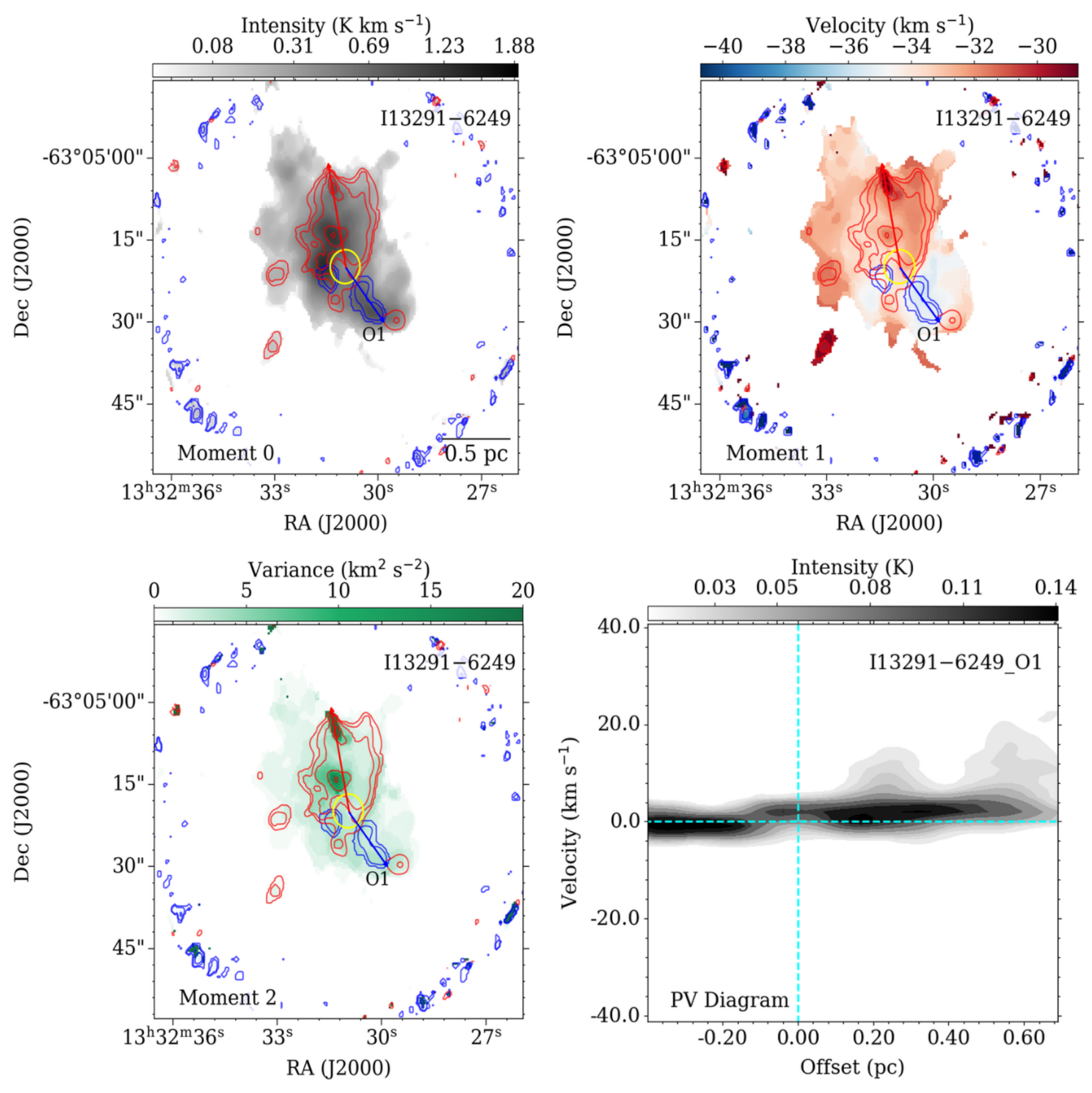}
\figsetgrpnote{ The top panels show the moment-0 map (left) and moment-1 map (right) of HC$_3$N line. The bottom panels show the moment-2 map (left, with a color bar above it) and the PV diagram along the outflow axis (right).}
\figsetgrpend

\figsetgrpstart
\figsetgrpnum{1.4}
\figsetgrptitle{Outflows in I14164-6028}
\figsetplot{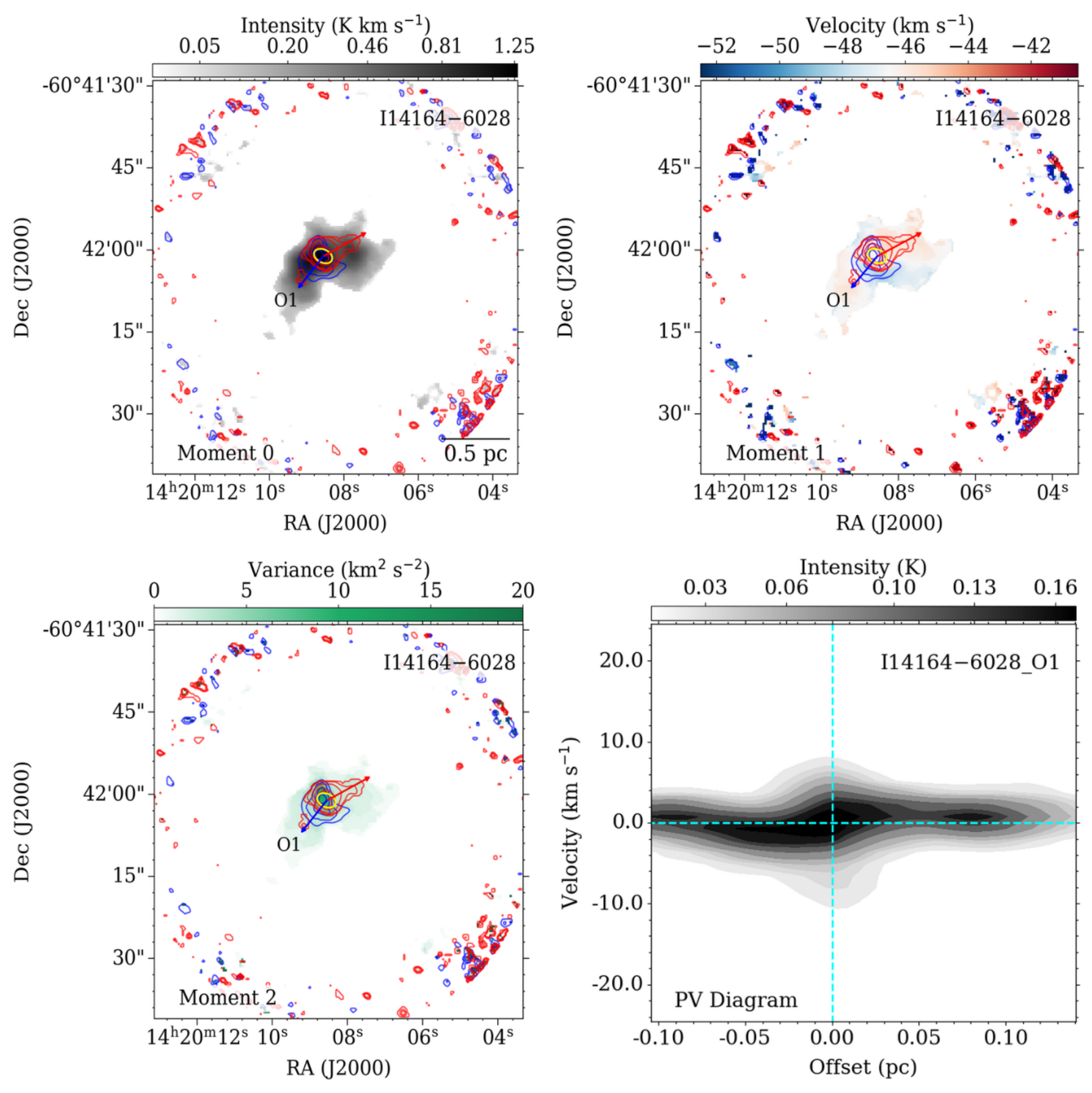}
\figsetgrpnote{ The top panels show the moment-0 map (left) and moment-1 map (right) of HC$_3$N line. The bottom panels show the moment-2 map (left, with a color bar above it) and the PV diagram along the outflow axis (right).}
\figsetgrpend

\figsetgrpstart
\figsetgrpnum{1.5}
\figsetgrptitle{Outflows in I14453-5912}
\figsetplot{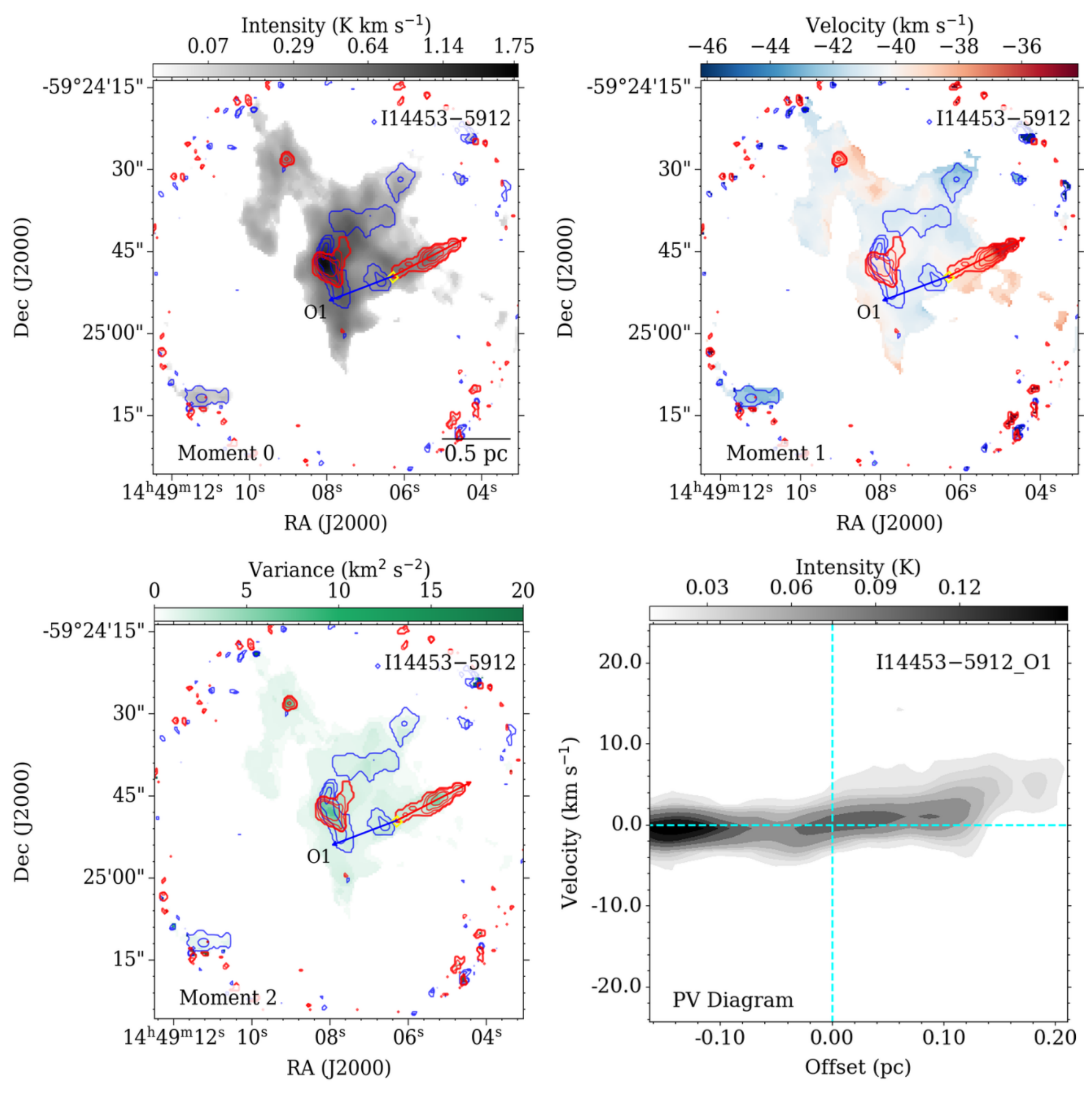}
\figsetgrpnote{ The top panels show the moment-0 map (left) and moment-1 map (right) of HC$_3$N line. The bottom panels show the moment-2 map (left, with a color bar above it) and the PV diagram along the outflow axis (right).}
\figsetgrpend

\figsetgrpstart
\figsetgrpnum{1.6}
\figsetgrptitle{Outflows in I15394-5358}
\figsetplot{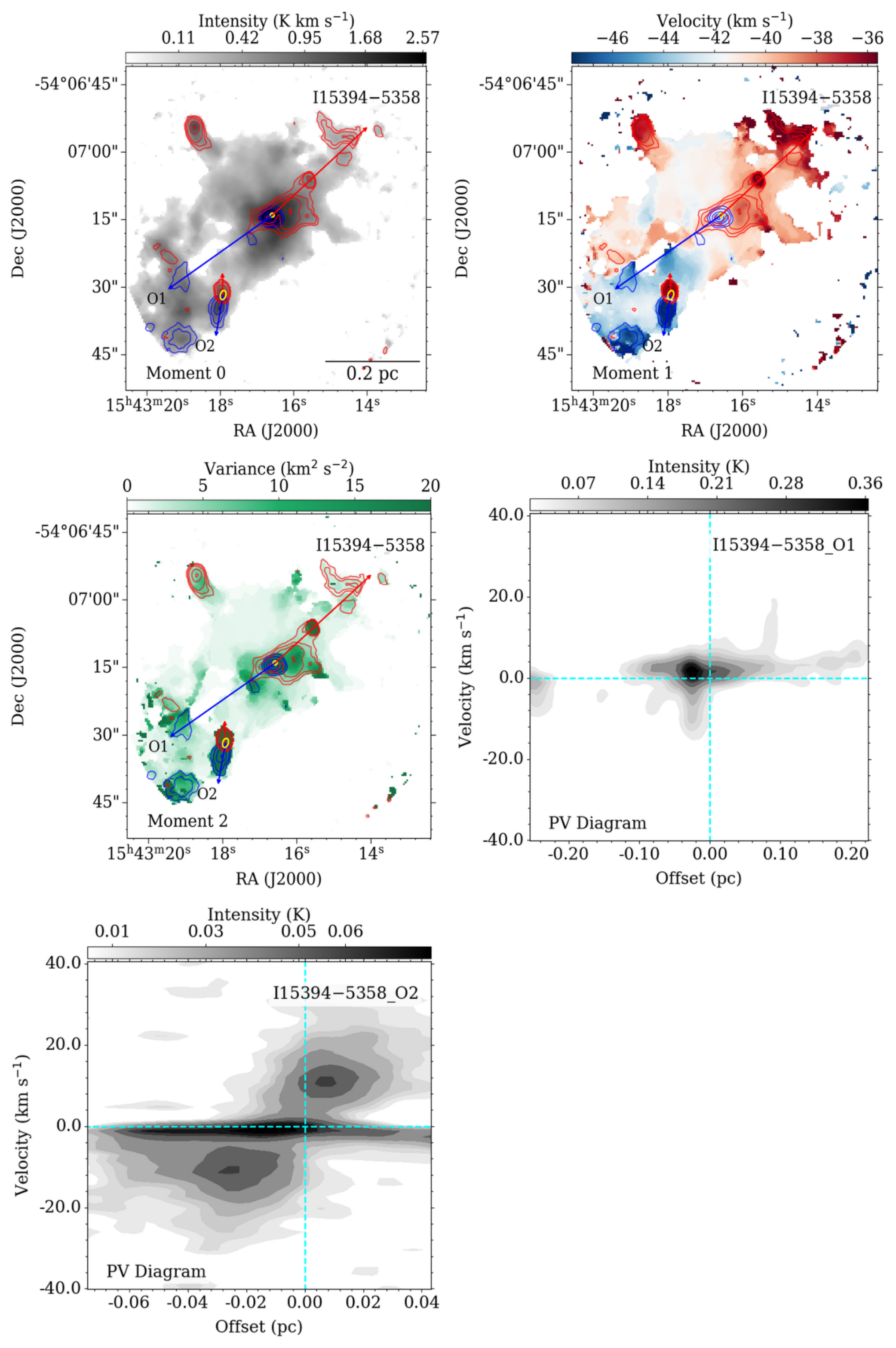}
\figsetgrpnote{ The top panels show the moment-0 map (left) and moment-1 map (right) of HC$_3$N line. The bottom panels show the moment-2 map (left, with a color bar above it) and the PV diagram along the outflow axis (right).}
\figsetgrpend

\figsetgrpstart
\figsetgrpnum{1.7}
\figsetgrptitle{Outflows in I15520-5234}
\figsetplot{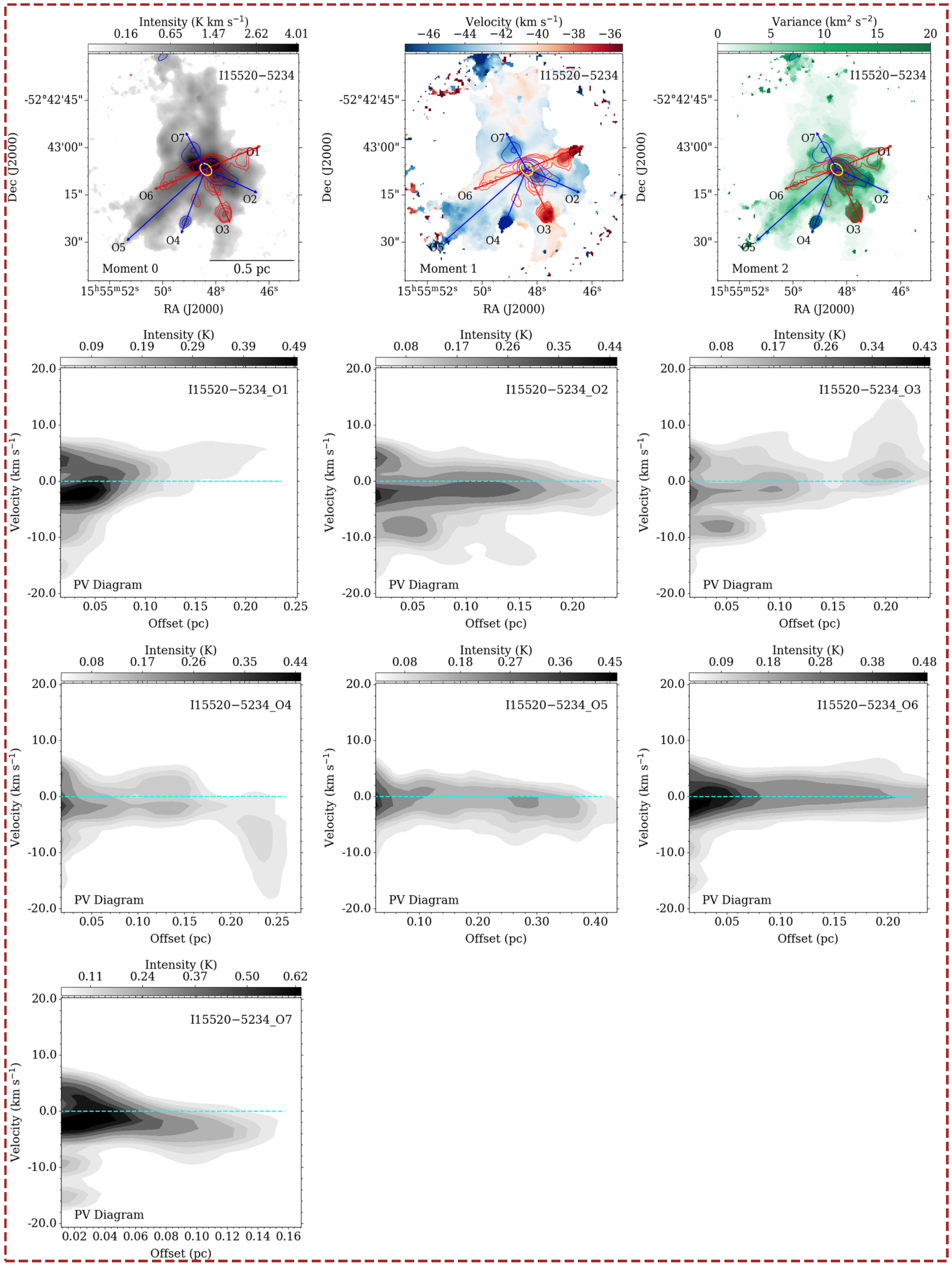}
\figsetgrpnote{ The top panels show the moment-0 map (left) and moment-1 map (right) of HC$_3$N line. The bottom panels show the moment-2 map (left, with a color bar above it) and the PV diagram along the outflow axis (right).}
\figsetgrpend

\figsetgrpstart
\figsetgrpnum{1.8}
\figsetgrptitle{Outflows in I15557-5215}
\figsetplot{figures/MomentMapsTogether/I15557-5215_Together.pdf}
\figsetgrpnote{ The top panels show the moment-0 map (left) and moment-1 map (right) of HC$_3$N line. The bottom panels show the moment-2 map (left, with a color bar above it) and the PV diagram along the outflow axis (right).}
\figsetgrpend

\figsetgrpstart
\figsetgrpnum{1.9}
\figsetgrptitle{Outflows in I15584-5247}
\figsetplot{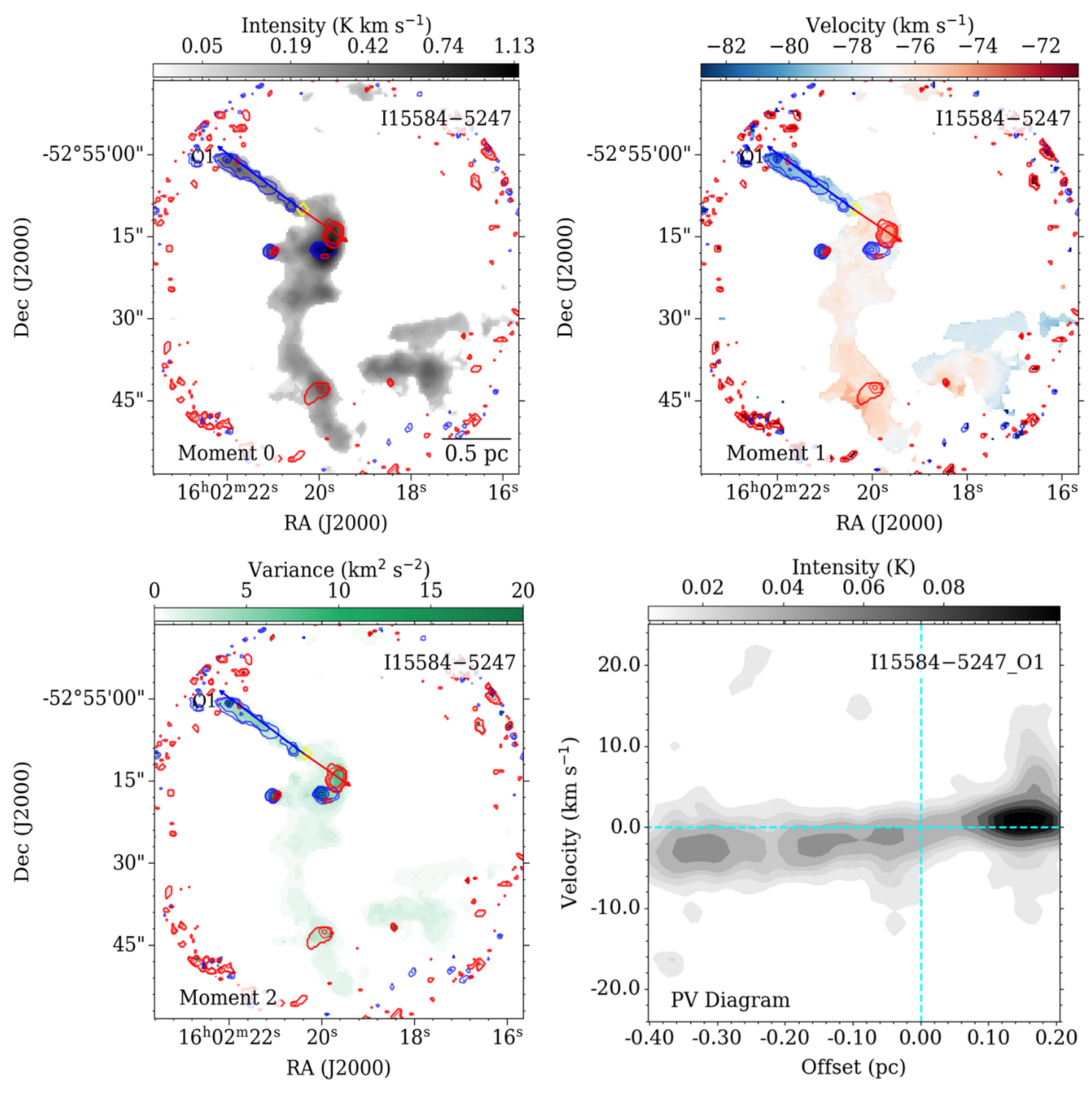}
\figsetgrpnote{ The top panels show the moment-0 map (left) and moment-1 map (right) of HC$_3$N line. The bottom panels show the moment-2 map (left, with a color bar above it) and the PV diagram along the outflow axis (right).}
\figsetgrpend

\figsetgrpstart
\figsetgrpnum{1.10}
\figsetgrptitle{Outflows in I16060-5146}
\figsetplot{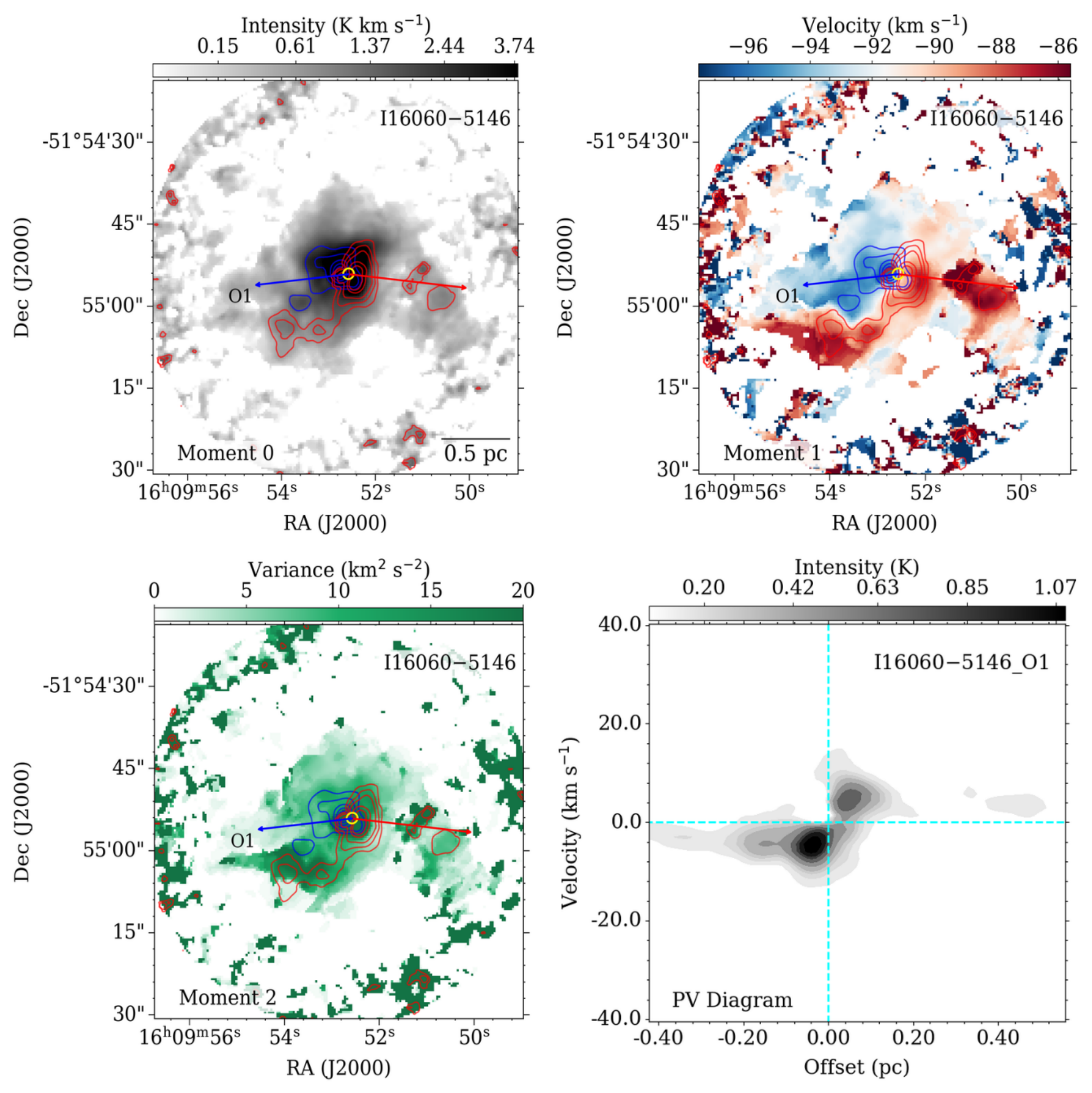}
\figsetgrpnote{ The top panels show the moment-0 map (left) and moment-1 map (right) of HC$_3$N line. The bottom panels show the moment-2 map (left, with a color bar above it) and the PV diagram along the outflow axis (right).}
\figsetgrpend

\figsetgrpstart
\figsetgrpnum{1.11}
\figsetgrptitle{Outflows in I16071-5142}
\figsetplot{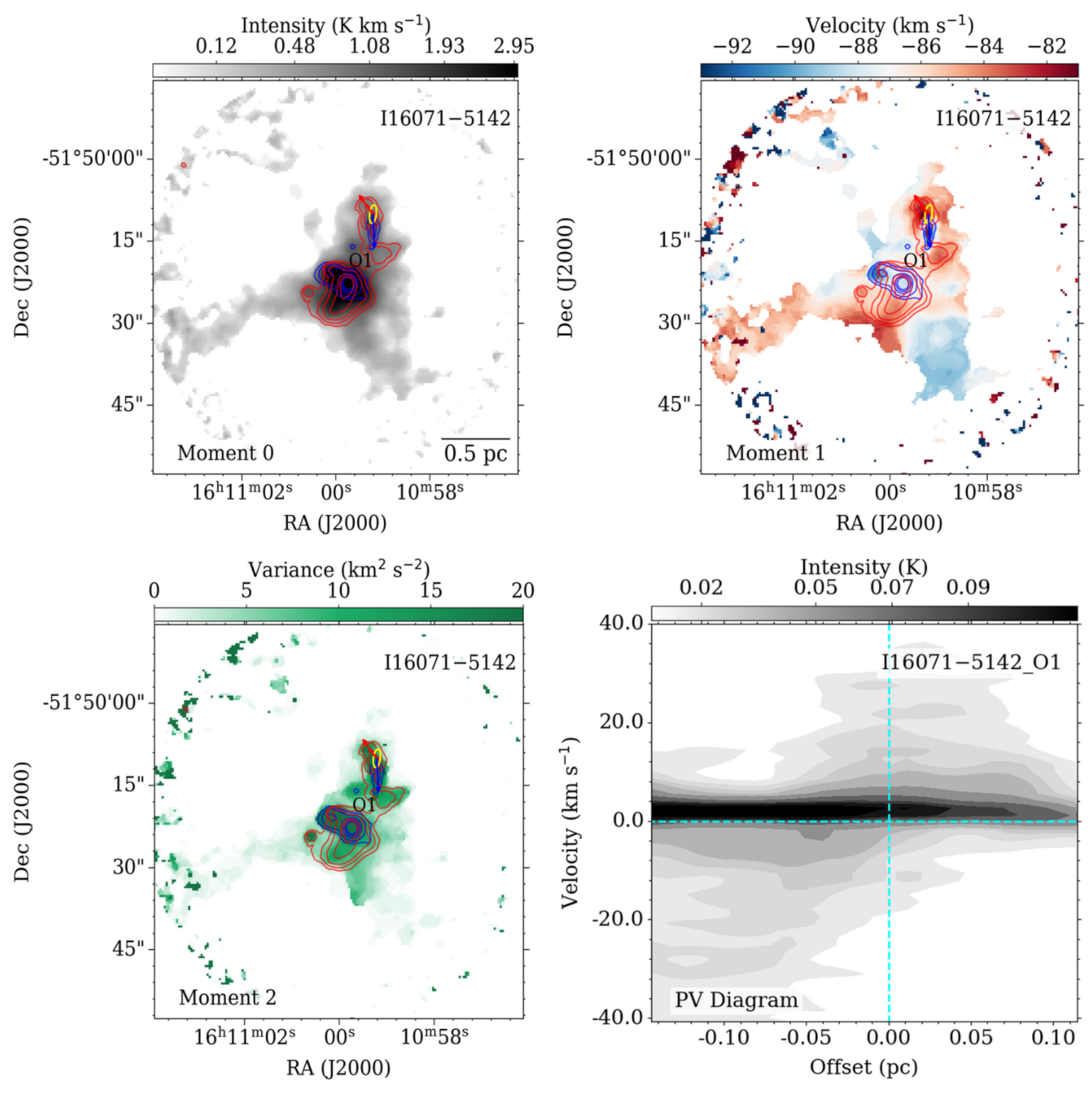}
\figsetgrpnote{ The top panels show the moment-0 map (left) and moment-1 map (right) of HC$_3$N line. The bottom panels show the moment-2 map (left, with a color bar above it) and the PV diagram along the outflow axis (right).}
\figsetgrpend

\figsetgrpstart
\figsetgrpnum{1.12}
\figsetgrptitle{Outflows in I16119-5048}
\figsetplot{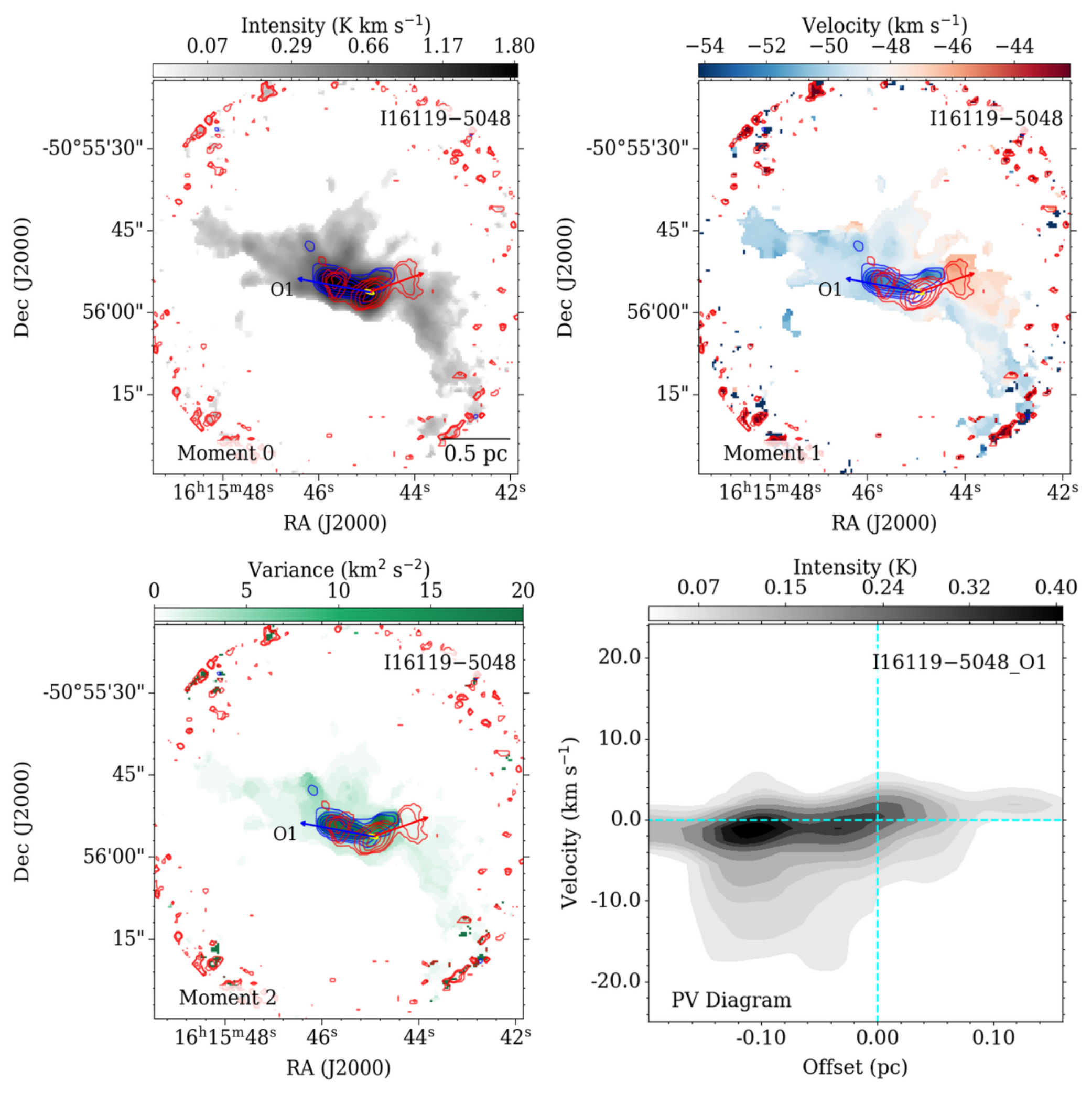}
\figsetgrpnote{ The top panels show the moment-0 map (left) and moment-1 map (right) of HC$_3$N line. The bottom panels show the moment-2 map (left, with a color bar above it) and the PV diagram along the outflow axis (right).}
\figsetgrpend

\figsetgrpstart
\figsetgrpnum{1.13}
\figsetgrptitle{Outflows in I16158-5055}
\figsetplot{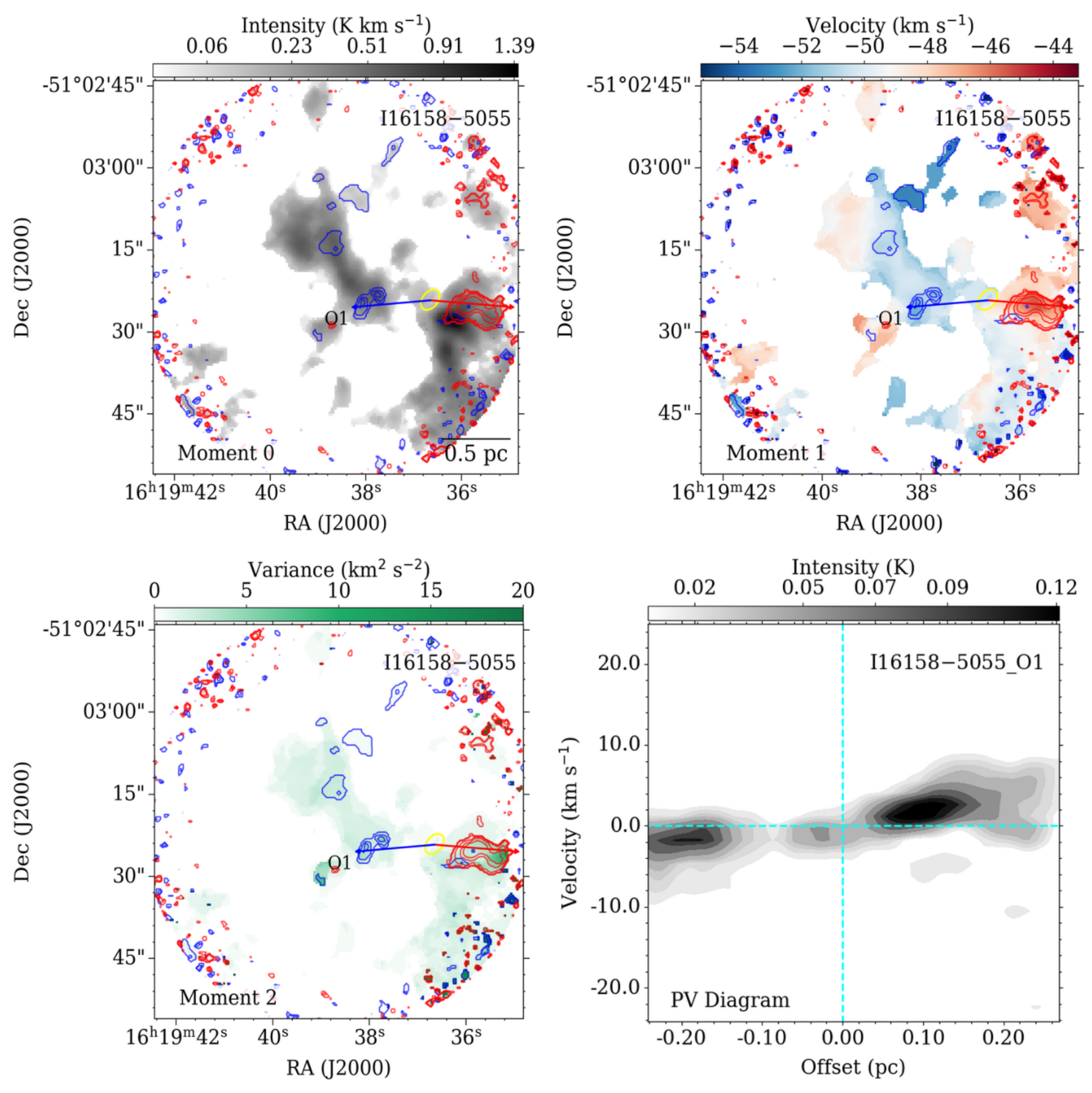}
\figsetgrpnote{ The top panels show the moment-0 map (left) and moment-1 map (right) of HC$_3$N line. The bottom panels show the moment-2 map (left, with a color bar above it) and the PV diagram along the outflow axis (right).}
\figsetgrpend

\figsetgrpstart
\figsetgrpnum{1.14}
\figsetgrptitle{Outflows in I16164-5046}
\figsetplot{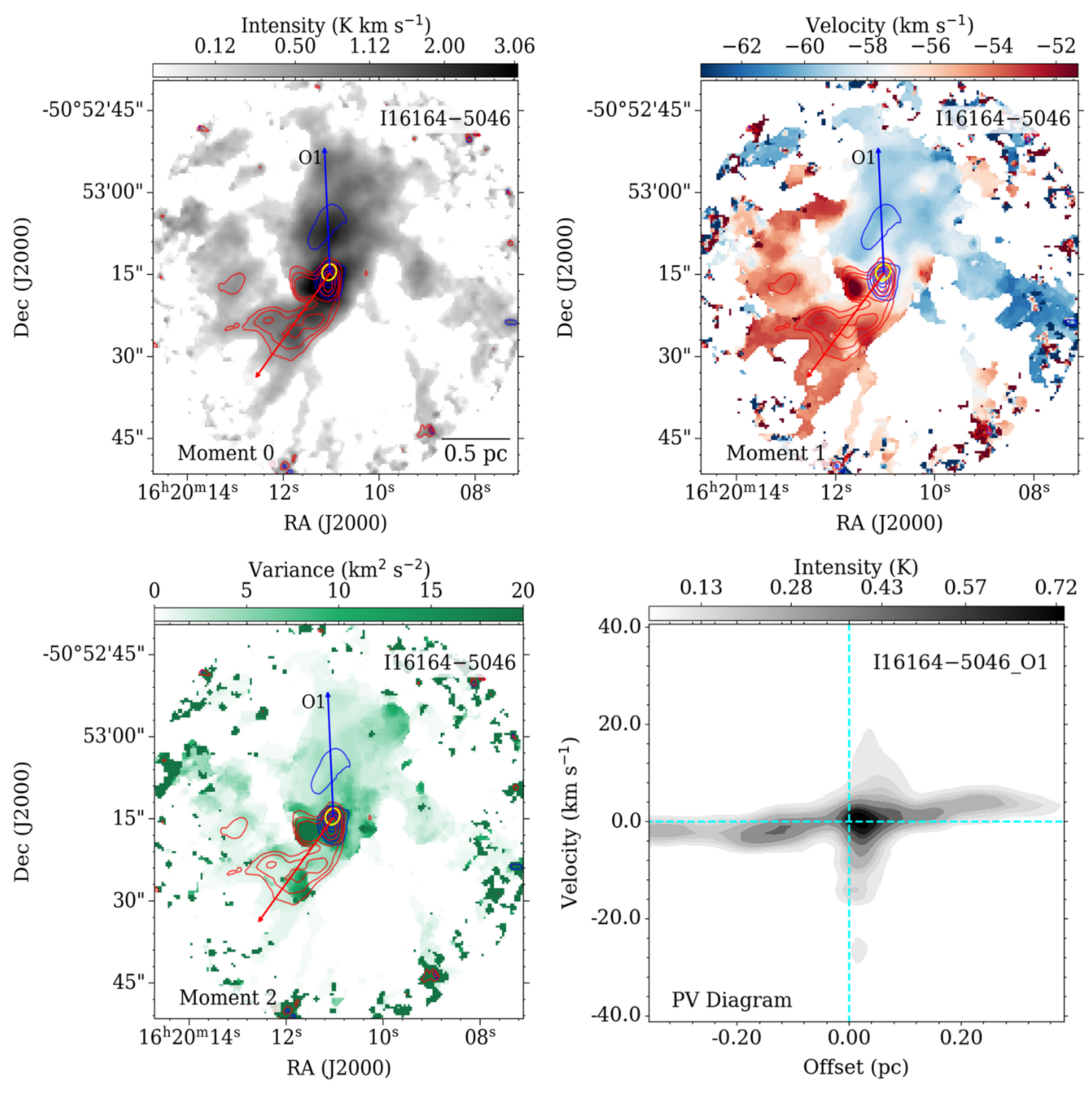}
\figsetgrpnote{ The top panels show the moment-0 map (left) and moment-1 map (right) of HC$_3$N line. The bottom panels show the moment-2 map (left, with a color bar above it) and the PV diagram along the outflow axis (right).}
\figsetgrpend

\figsetgrpstart
\figsetgrpnum{1.15}
\figsetgrptitle{Outflows in I16272-4837}
\figsetplot{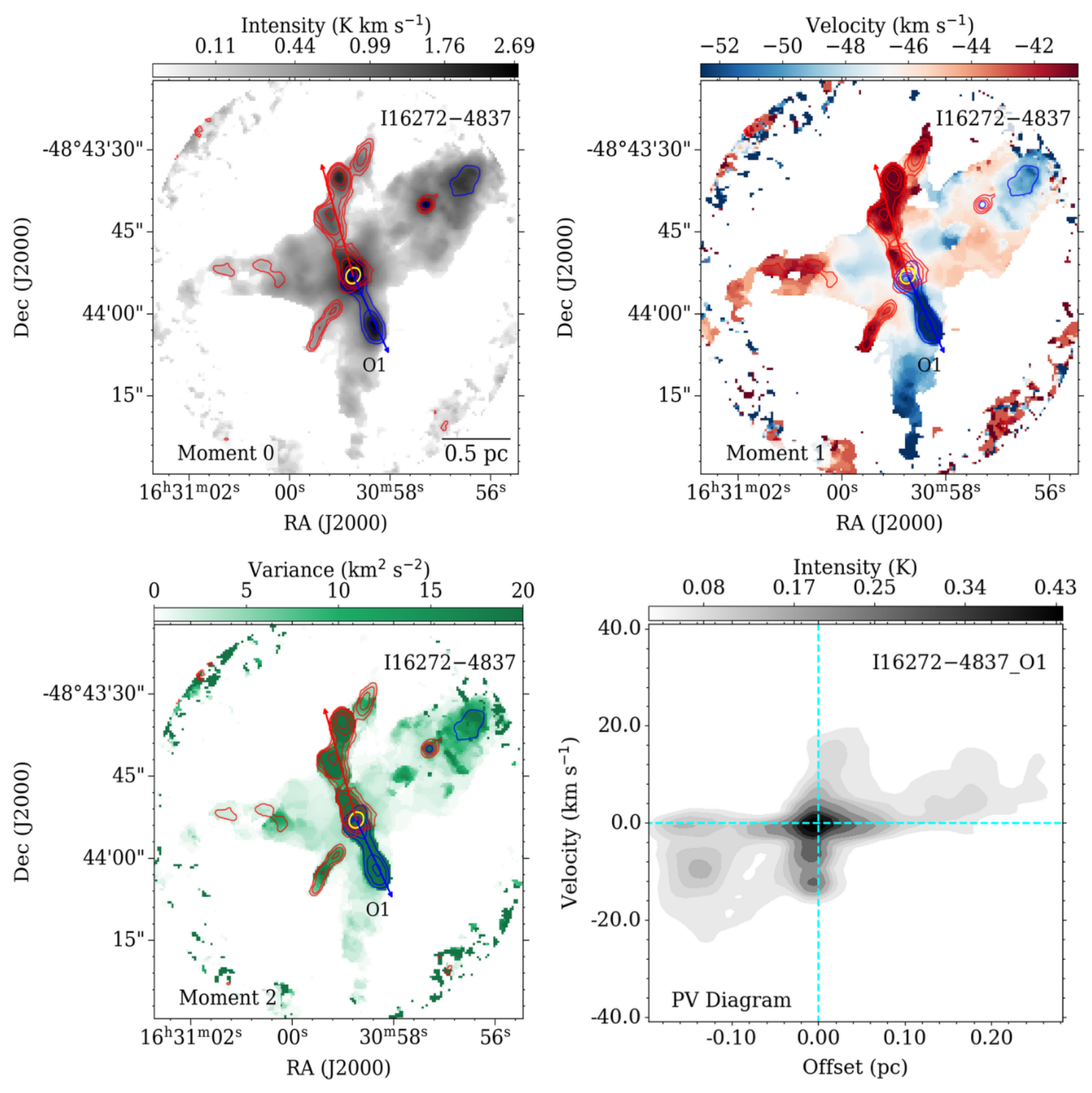}
\figsetgrpnote{ The top panels show the moment-0 map (left) and moment-1 map (right) of HC$_3$N line. The bottom panels show the moment-2 map (left, with a color bar above it) and the PV diagram along the outflow axis (right).}
\figsetgrpend

\figsetgrpstart
\figsetgrpnum{1.16}
\figsetgrptitle{Outflows in I16348-4654}
\figsetplot{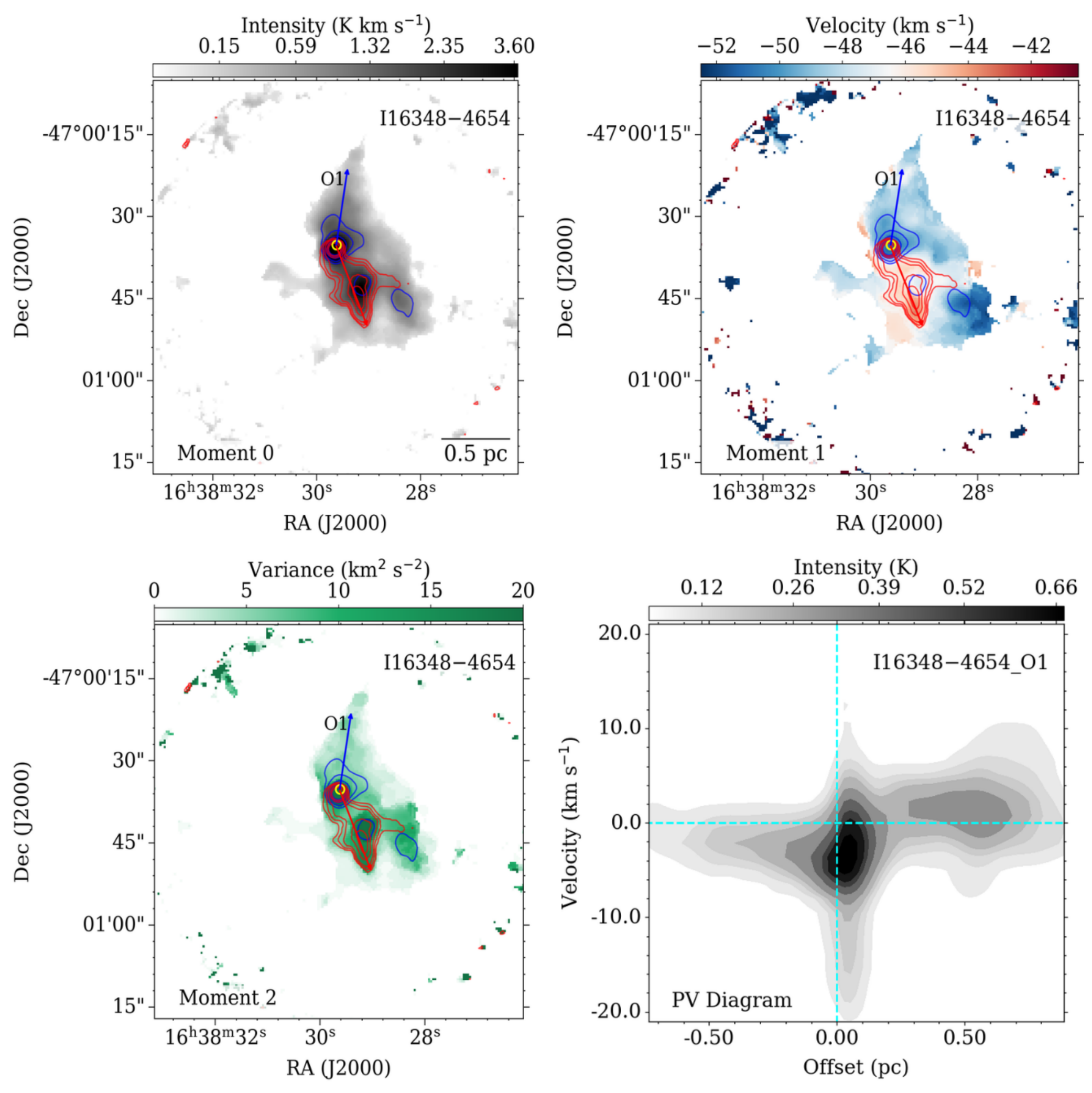}
\figsetgrpnote{ The top panels show the moment-0 map (left) and moment-1 map (right) of HC$_3$N line. The bottom panels show the moment-2 map (left, with a color bar above it) and the PV diagram along the outflow axis (right).}
\figsetgrpend

\figsetgrpstart
\figsetgrpnum{1.17}
\figsetgrptitle{Outflows in I16351-4722}
\figsetplot{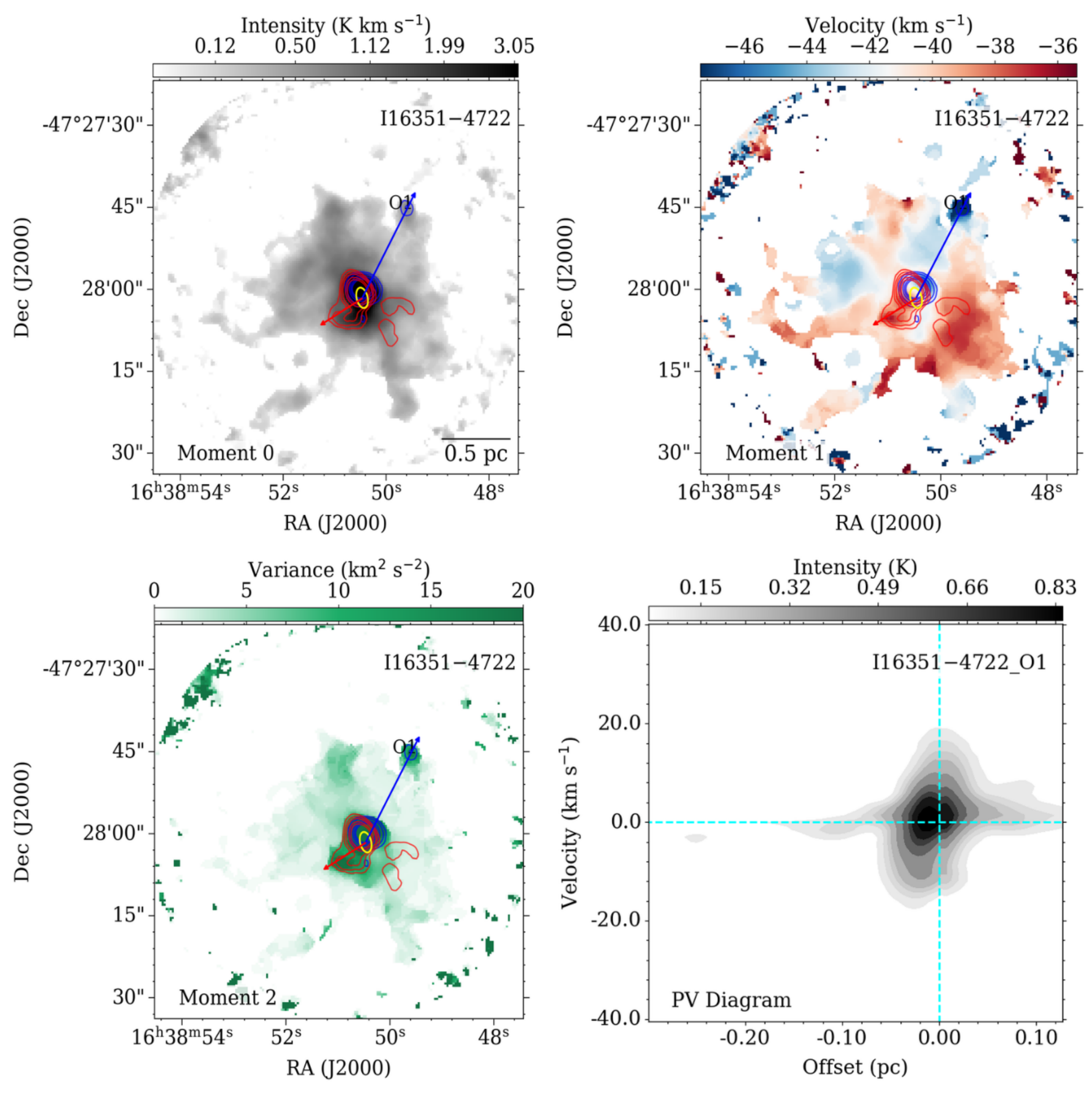}
\figsetgrpnote{ The top panels show the moment-0 map (left) and moment-1 map (right) of HC$_3$N line. The bottom panels show the moment-2 map (left, with a color bar above it) and the PV diagram along the outflow axis (right).}
\figsetgrpend

\figsetgrpstart
\figsetgrpnum{1.18}
\figsetgrptitle{Outflows in I16385-4619}
\figsetplot{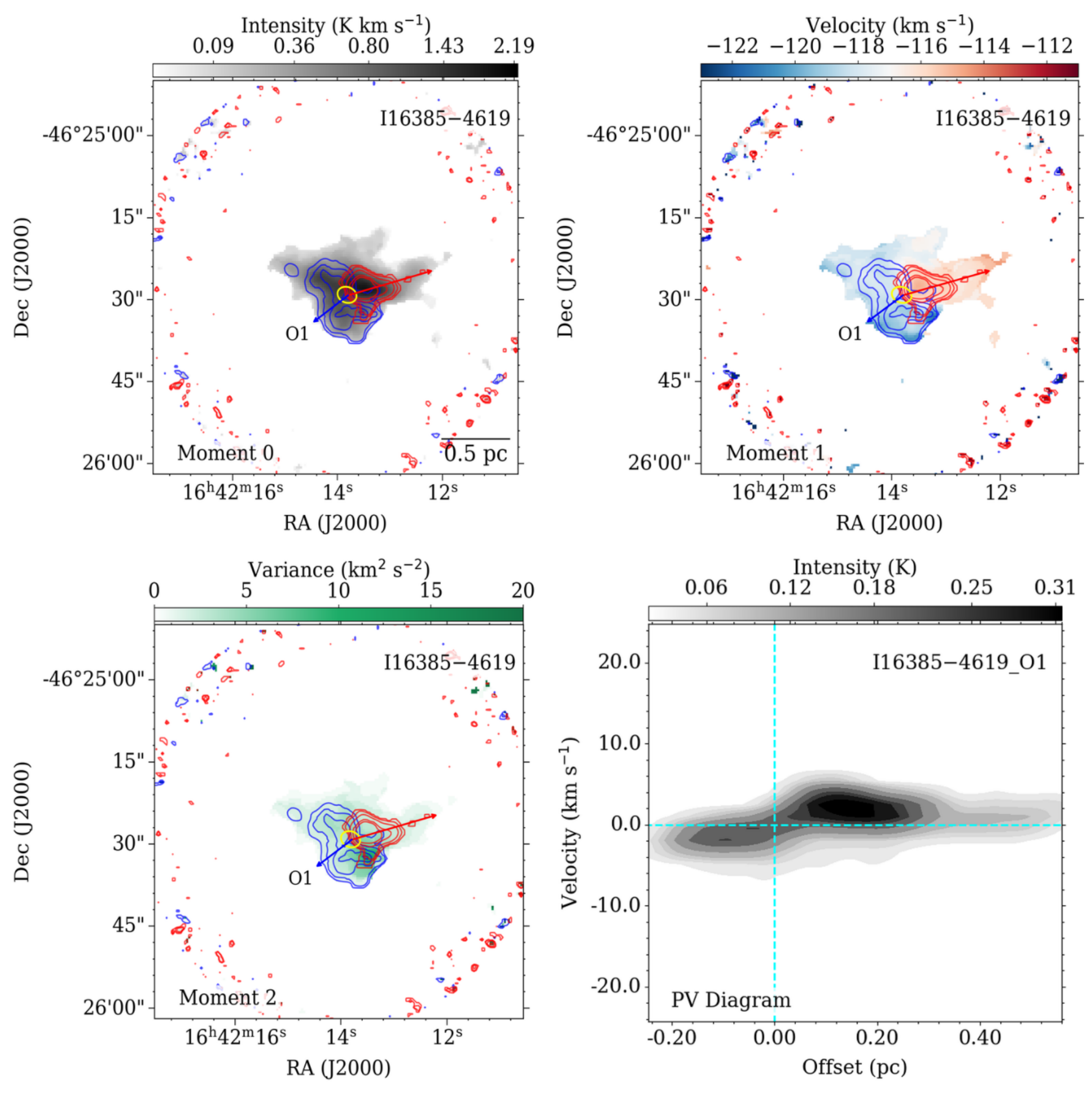}
\figsetgrpnote{ The top panels show the moment-0 map (left) and moment-1 map (right) of HC$_3$N line. The bottom panels show the moment-2 map (left, with a color bar above it) and the PV diagram along the outflow axis (right).}
\figsetgrpend

\figsetgrpstart
\figsetgrpnum{1.19}
\figsetgrptitle{Outflows in I16424-4531}
\figsetplot{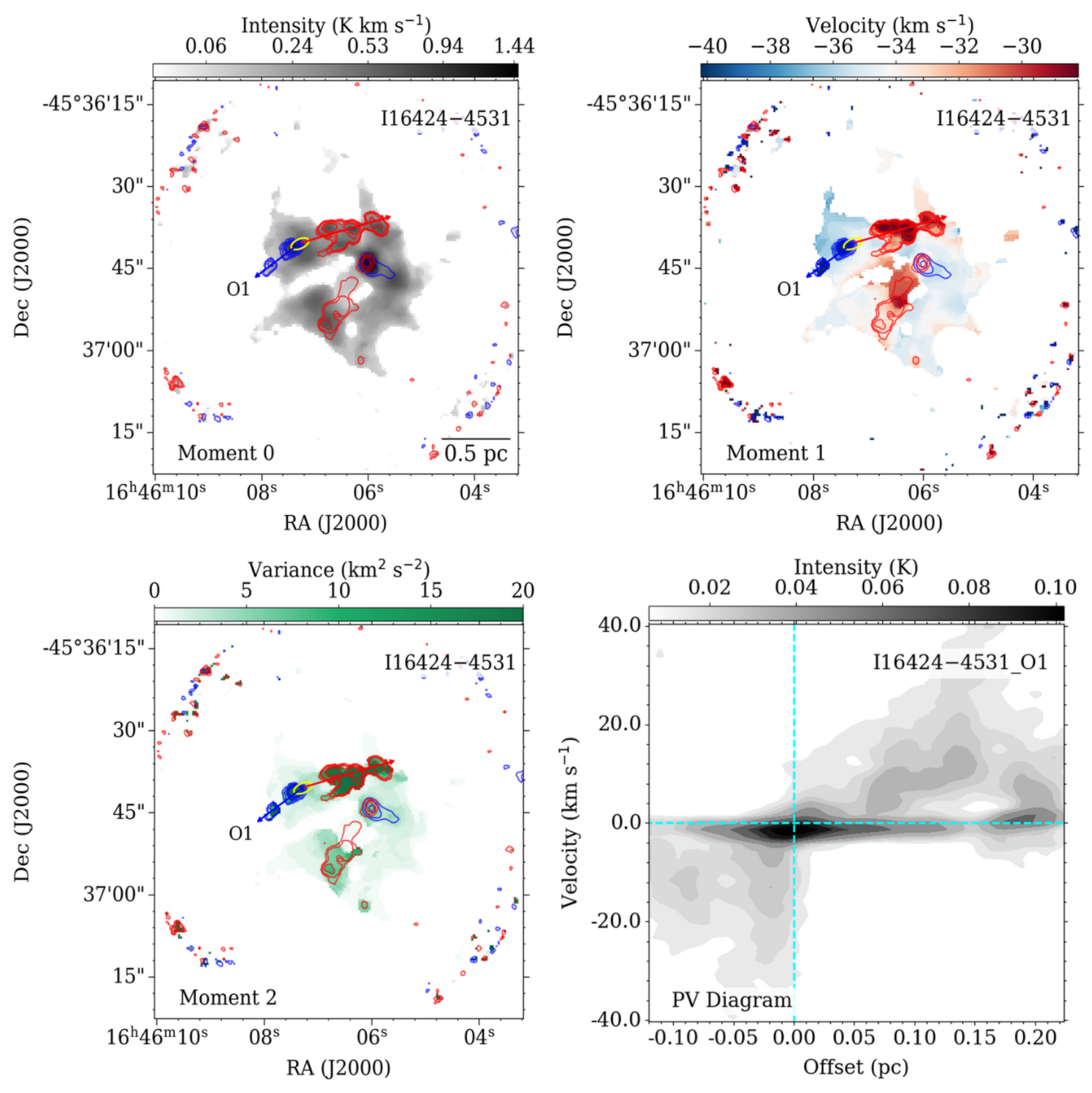}
\figsetgrpnote{ The top panels show the moment-0 map (left) and moment-1 map (right) of HC$_3$N line. The bottom panels show the moment-2 map (left, with a color bar above it) and the PV diagram along the outflow axis (right).}
\figsetgrpend

\figsetgrpstart
\figsetgrpnum{1.20}
\figsetgrptitle{Outflows in I16445-4459}
\figsetplot{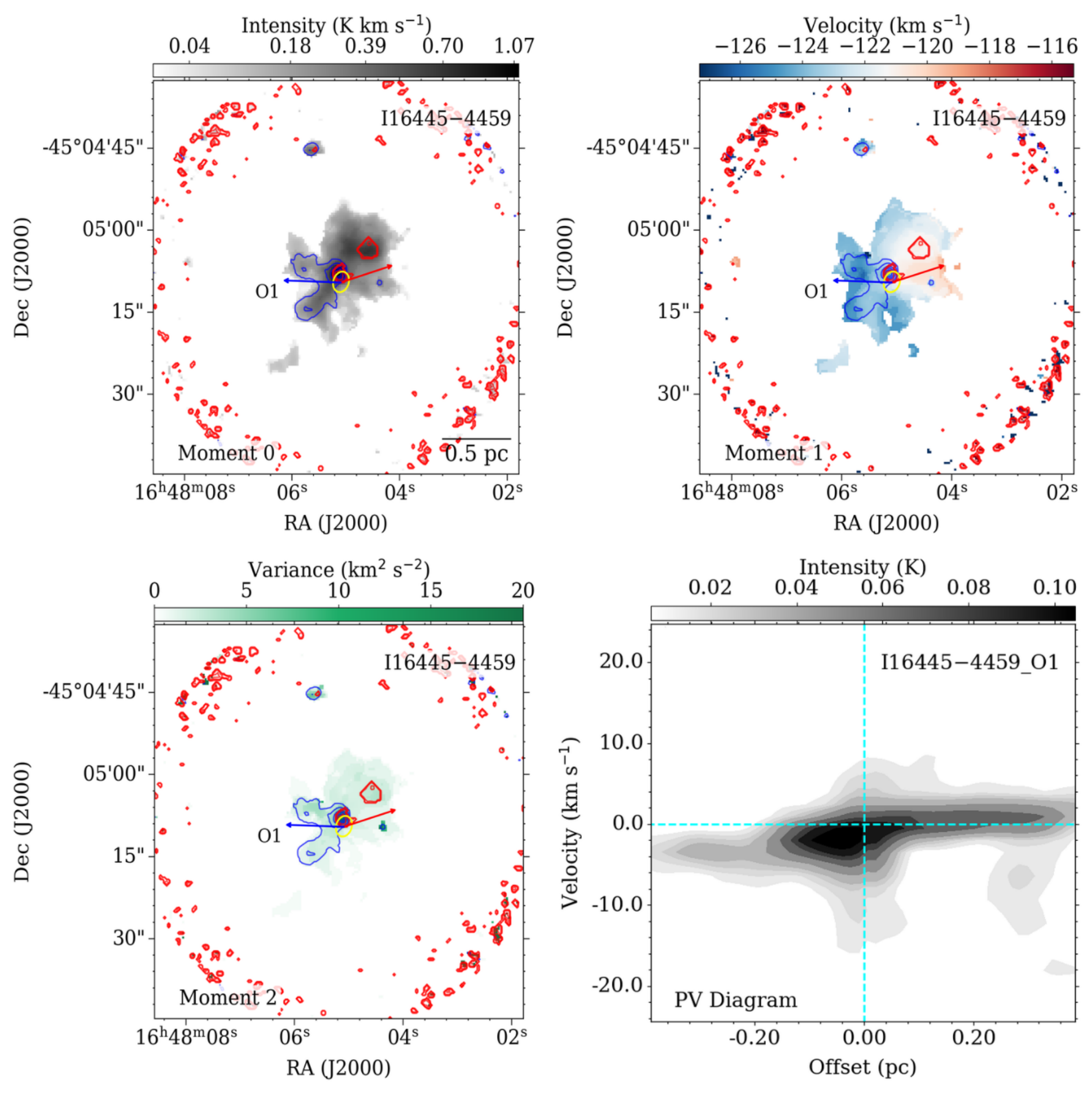}
\figsetgrpnote{ The top panels show the moment-0 map (left) and moment-1 map (right) of HC$_3$N line. The bottom panels show the moment-2 map (left, with a color bar above it) and the PV diagram along the outflow axis (right).}
\figsetgrpend

\figsetgrpstart
\figsetgrpnum{1.21}
\figsetgrptitle{Outflows in I16458-4512}
\figsetplot{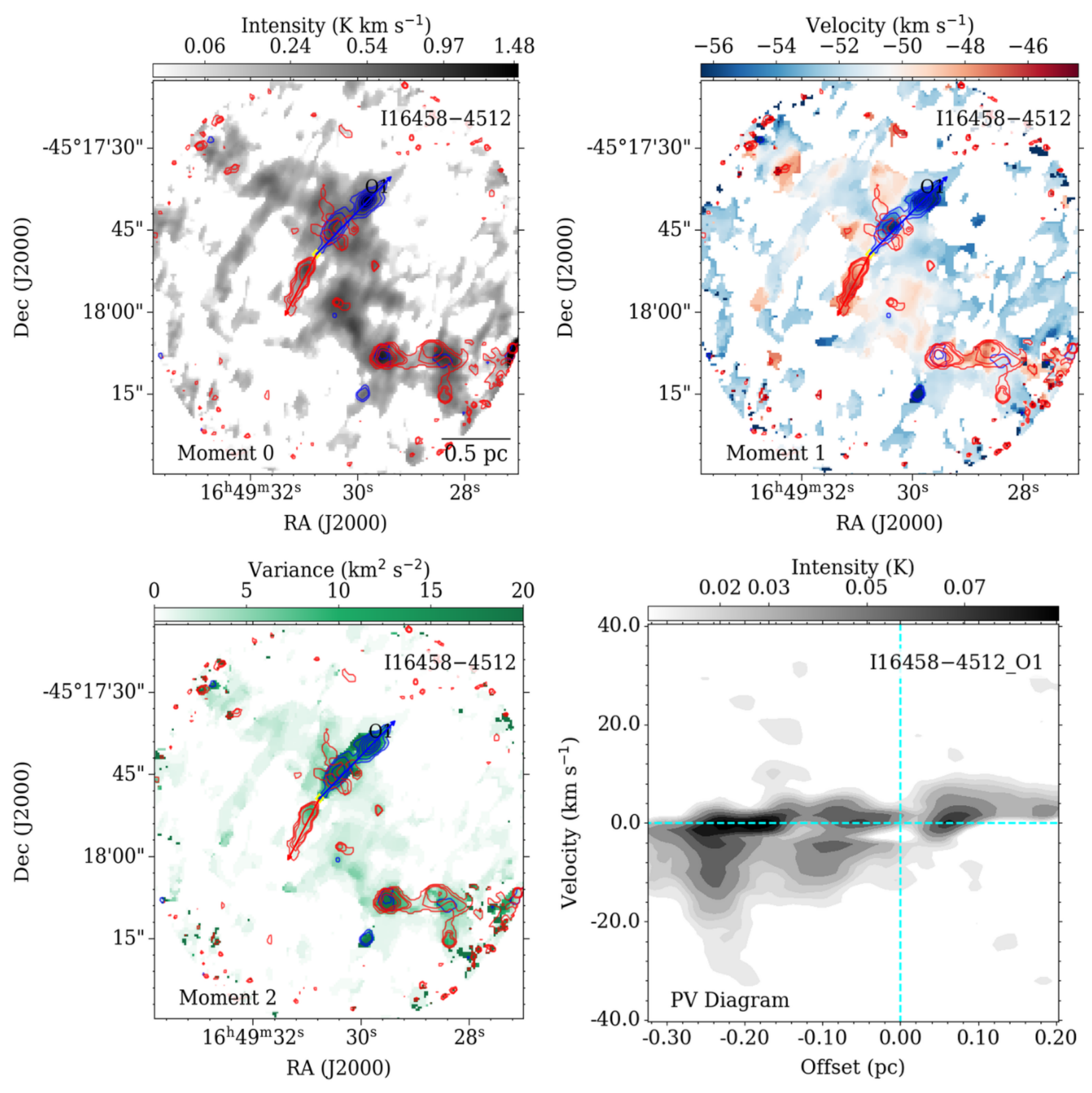}
\figsetgrpnote{ The top panels show the moment-0 map (left) and moment-1 map (right) of HC$_3$N line. The bottom panels show the moment-2 map (left, with a color bar above it) and the PV diagram along the outflow axis (right).}
\figsetgrpend

\figsetgrpstart
\figsetgrpnum{1.22}
\figsetgrptitle{Outflows in I16487-4423}
\figsetplot{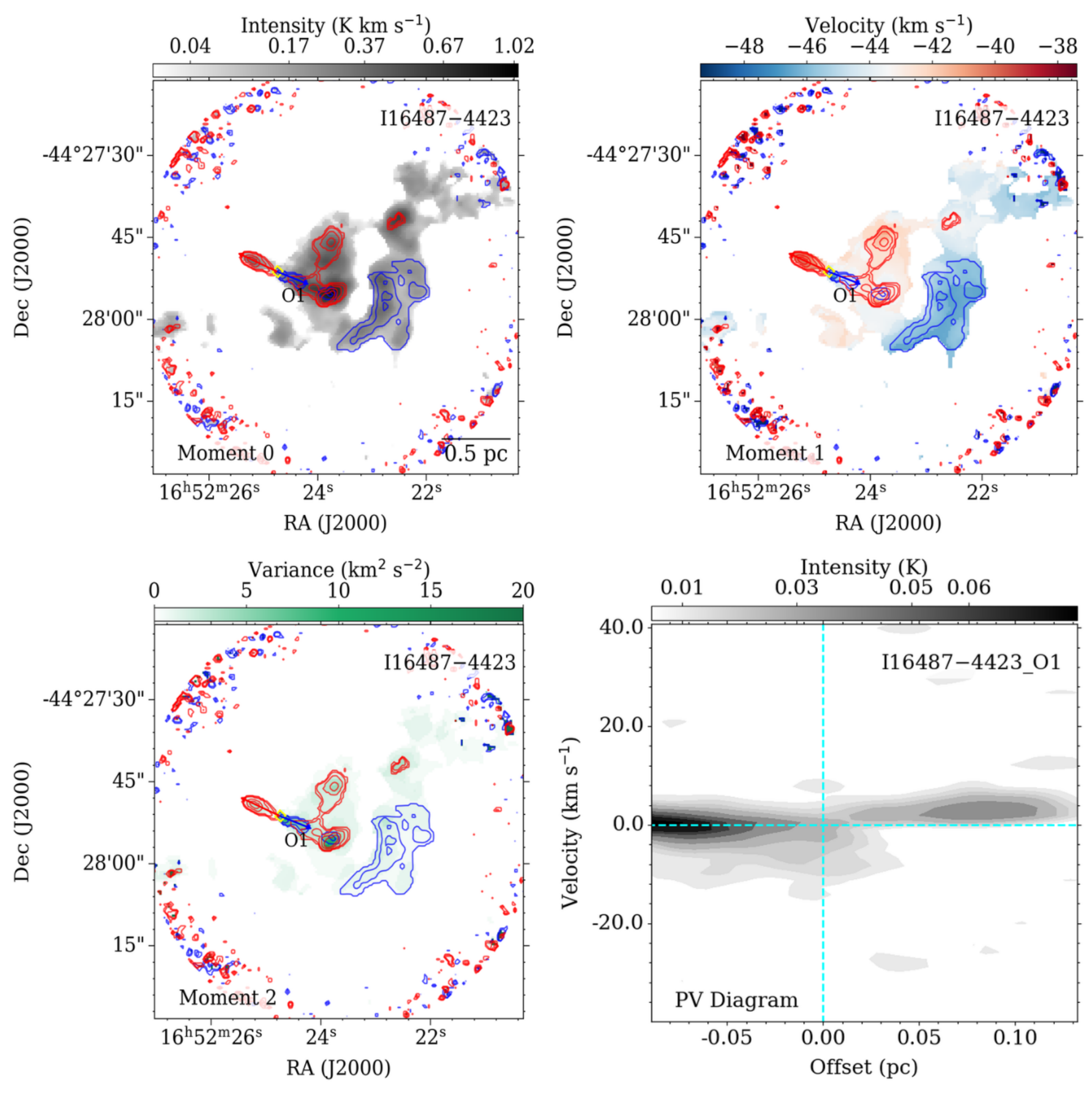}
\figsetgrpnote{ The top panels show the moment-0 map (left) and moment-1 map (right) of HC$_3$N line. The bottom panels show the moment-2 map (left, with a color bar above it) and the PV diagram along the outflow axis (right).}
\figsetgrpend

\figsetgrpstart
\figsetgrpnum{1.23}
\figsetgrptitle{Outflows in I16489-4431}
\figsetplot{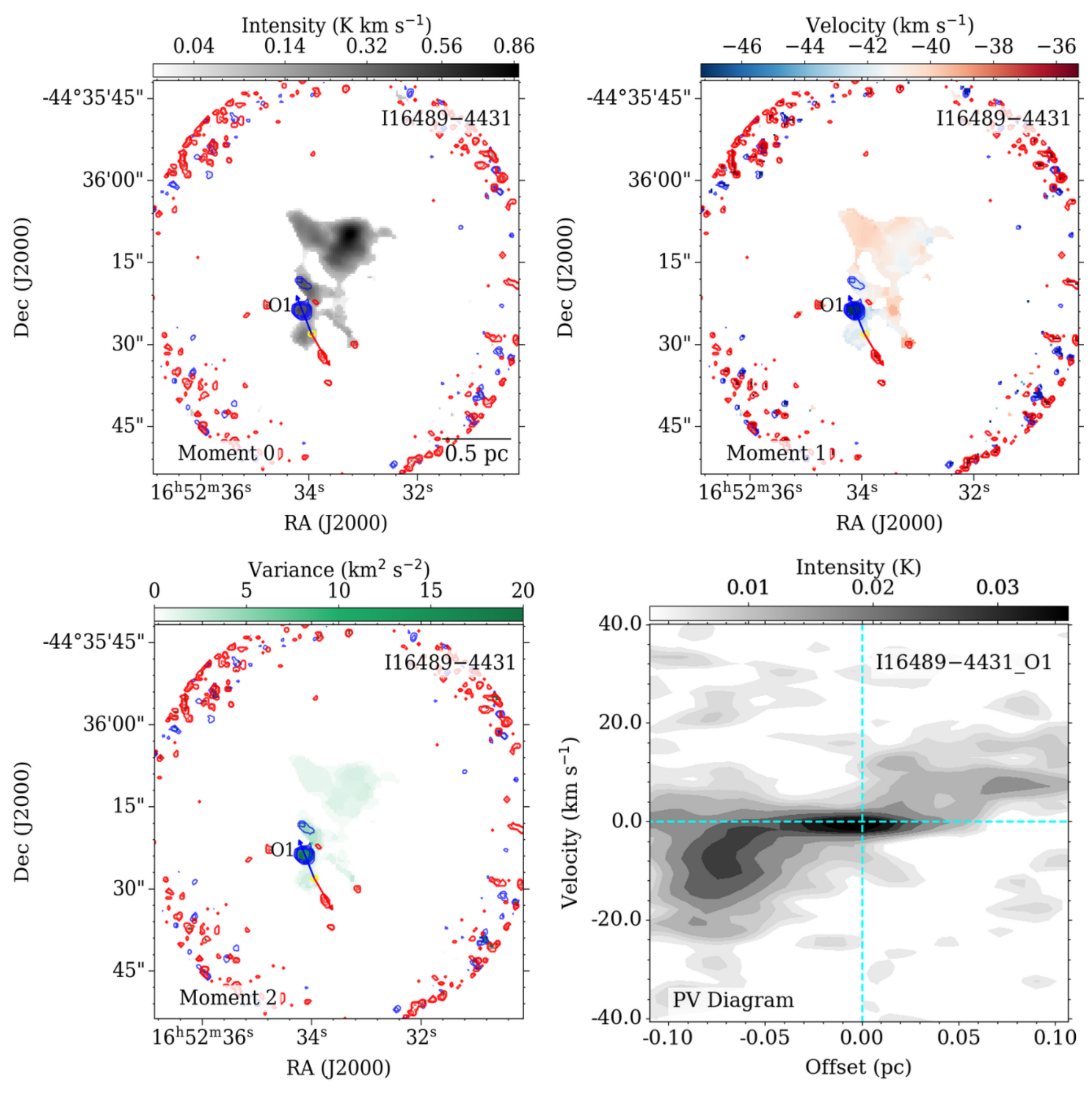}
\figsetgrpnote{ The top panels show the moment-0 map (left) and moment-1 map (right) of HC$_3$N line. The bottom panels show the moment-2 map (left, with a color bar above it) and the PV diagram along the outflow axis (right).}
\figsetgrpend

\figsetgrpstart
\figsetgrpnum{1.24}
\figsetgrptitle{Outflows in I16547-4247}
\figsetplot{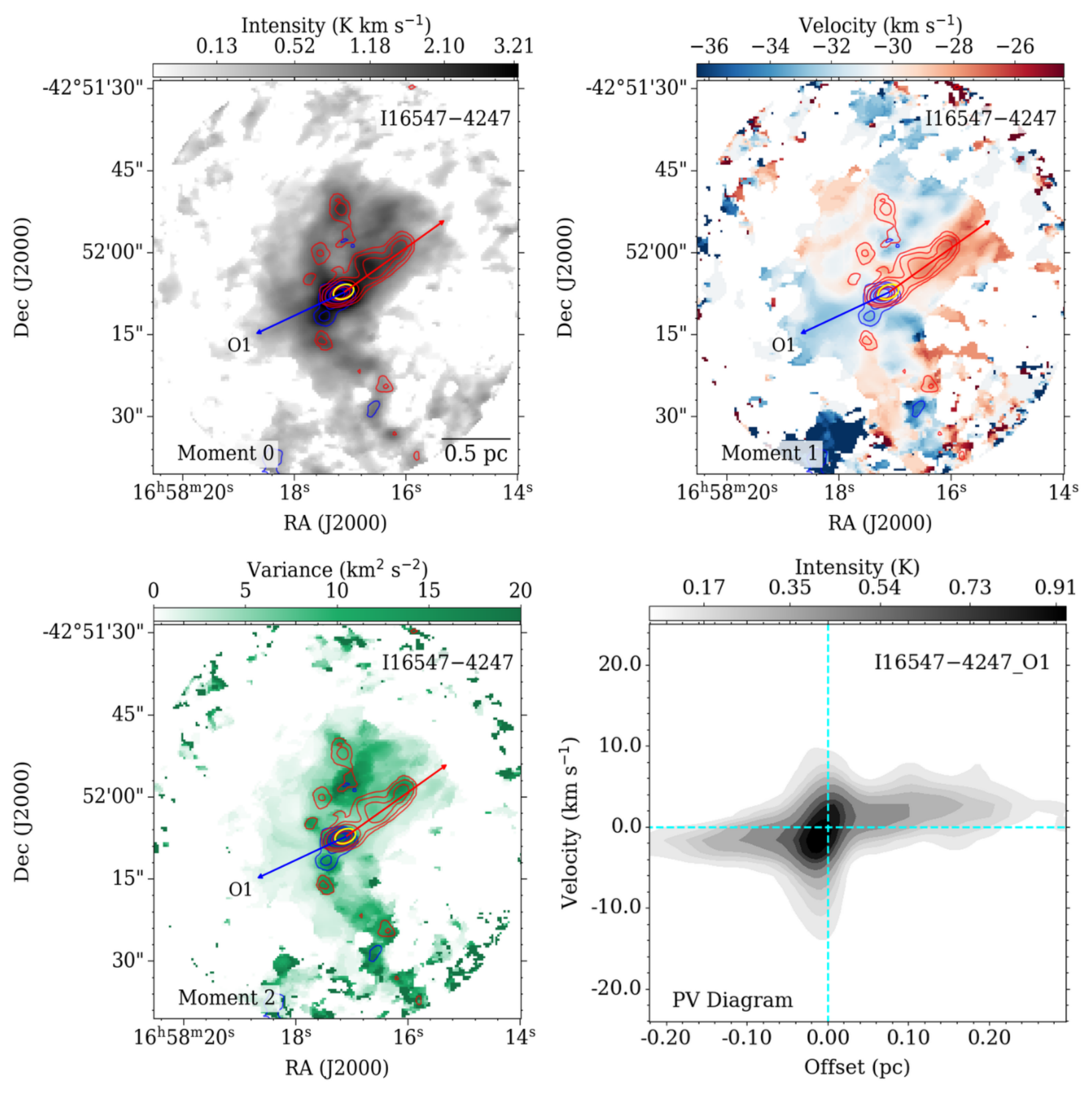}
\figsetgrpnote{ The top panels show the moment-0 map (left) and moment-1 map (right) of HC$_3$N line. The bottom panels show the moment-2 map (left, with a color bar above it) and the PV diagram along the outflow axis (right).}
\figsetgrpend

\figsetgrpstart
\figsetgrpnum{1.25}
\figsetgrptitle{Outflows in I16571-4029}
\figsetplot{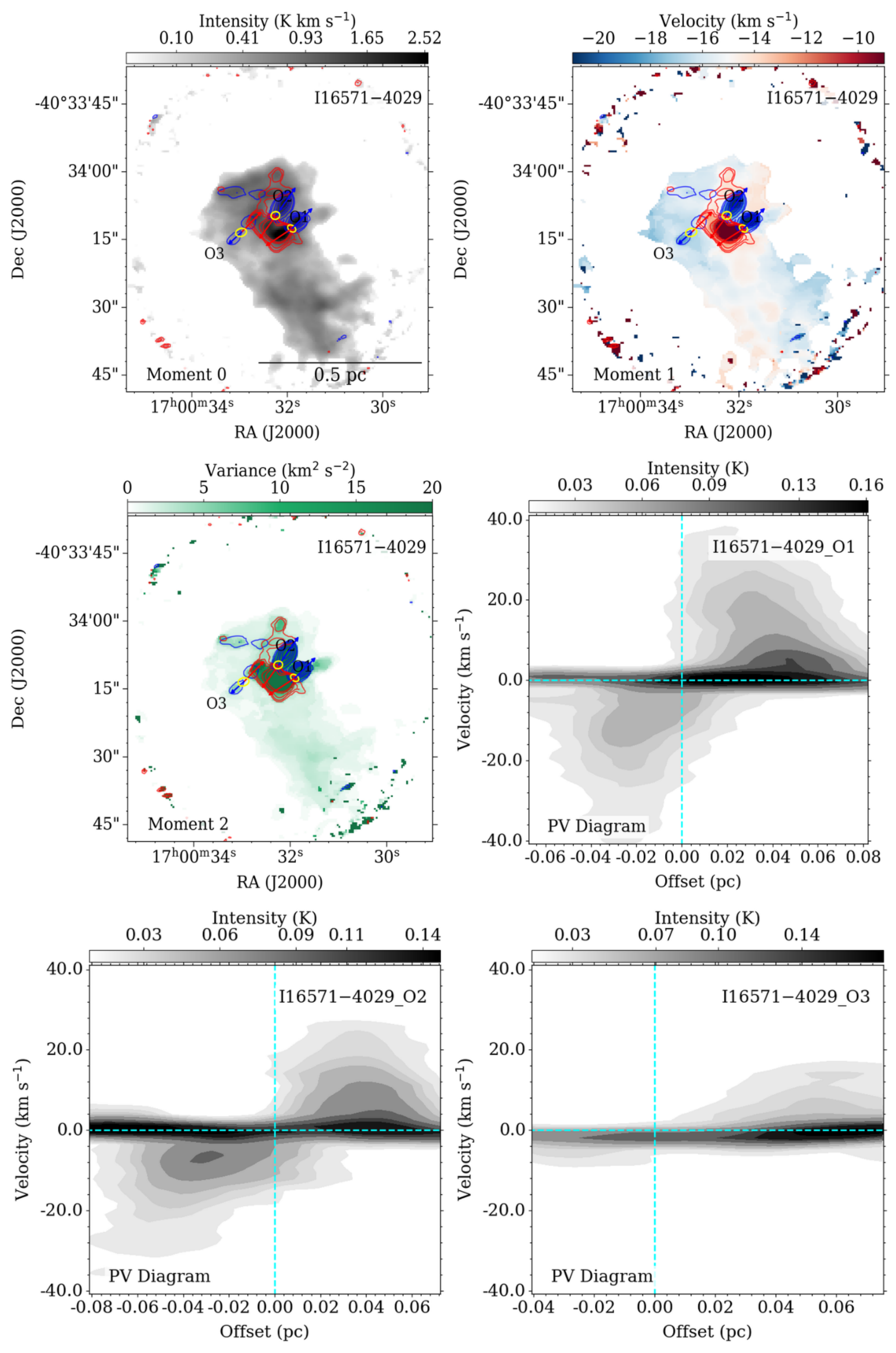}
\figsetgrpnote{ The top panels show the moment-0 map (left) and moment-1 map (right) of HC$_3$N line. The bottom panels show the moment-2 map (left, with a color bar above it) and the PV diagram along the outflow axis (right).}
\figsetgrpend

\figsetgrpstart
\figsetgrpnum{1.26}
\figsetgrptitle{Outflows in I17006-4215}
\figsetplot{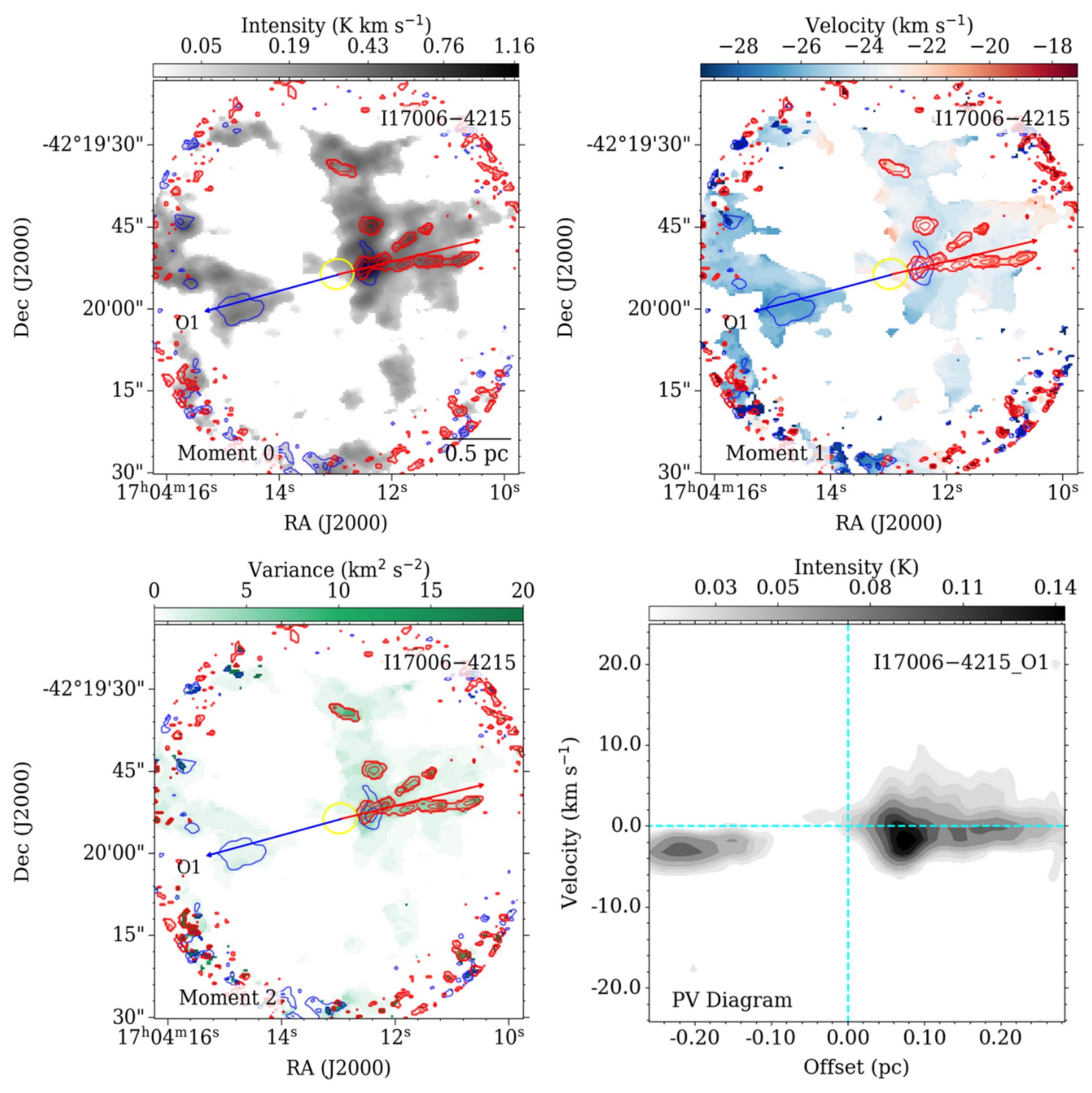}
\figsetgrpnote{ The top panels show the moment-0 map (left) and moment-1 map (right) of HC$_3$N line. The bottom panels show the moment-2 map (left, with a color bar above it) and the PV diagram along the outflow axis (right).}
\figsetgrpend

\figsetgrpstart
\figsetgrpnum{1.27}
\figsetgrptitle{Outflows in I17016-4124}
\figsetplot{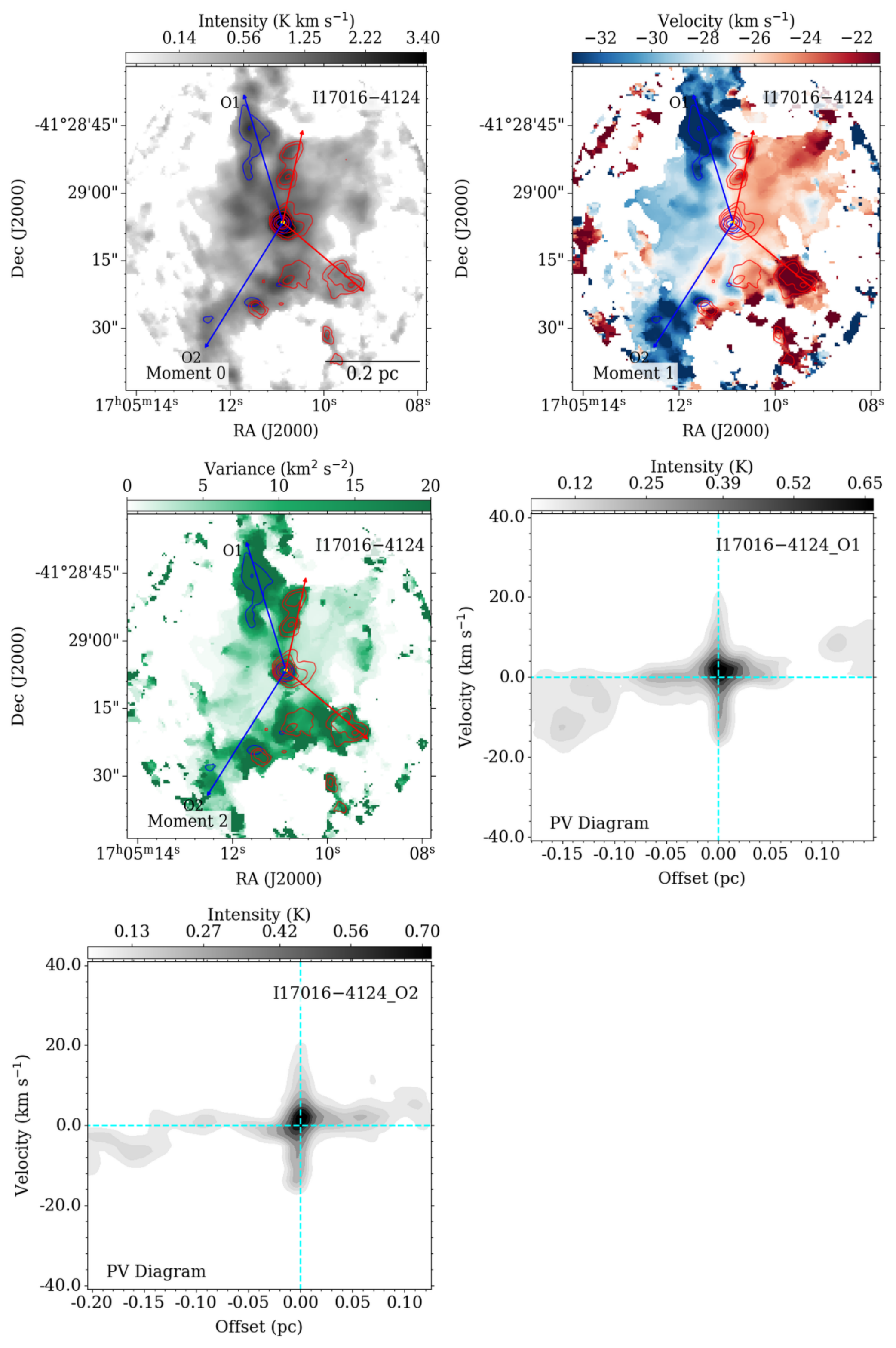}
\figsetgrpnote{ The top panels show the moment-0 map (left) and moment-1 map (right) of HC$_3$N line. The bottom panels show the moment-2 map (left, with a color bar above it) and the PV diagram along the outflow axis (right).}
\figsetgrpend

\figsetgrpstart
\figsetgrpnum{1.28}
\figsetgrptitle{Outflows in I17175-3544}
\figsetplot{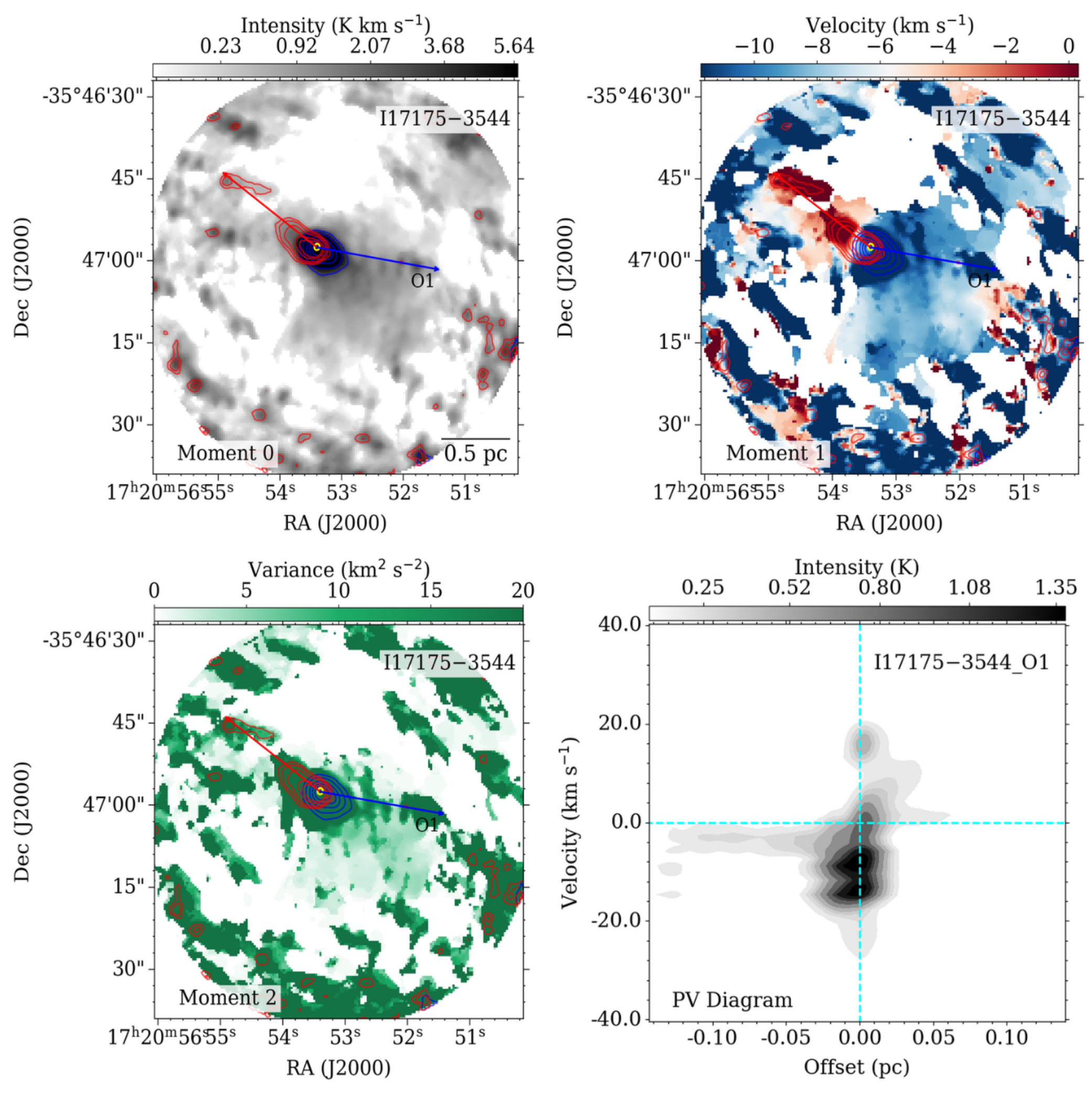}
\figsetgrpnote{ The top panels show the moment-0 map (left) and moment-1 map (right) of HC$_3$N line. The bottom panels show the moment-2 map (left, with a color bar above it) and the PV diagram along the outflow axis (right).}
\figsetgrpend

\figsetgrpstart
\figsetgrpnum{1.29}
\figsetgrptitle{Outflows in I17204-3636}
\figsetplot{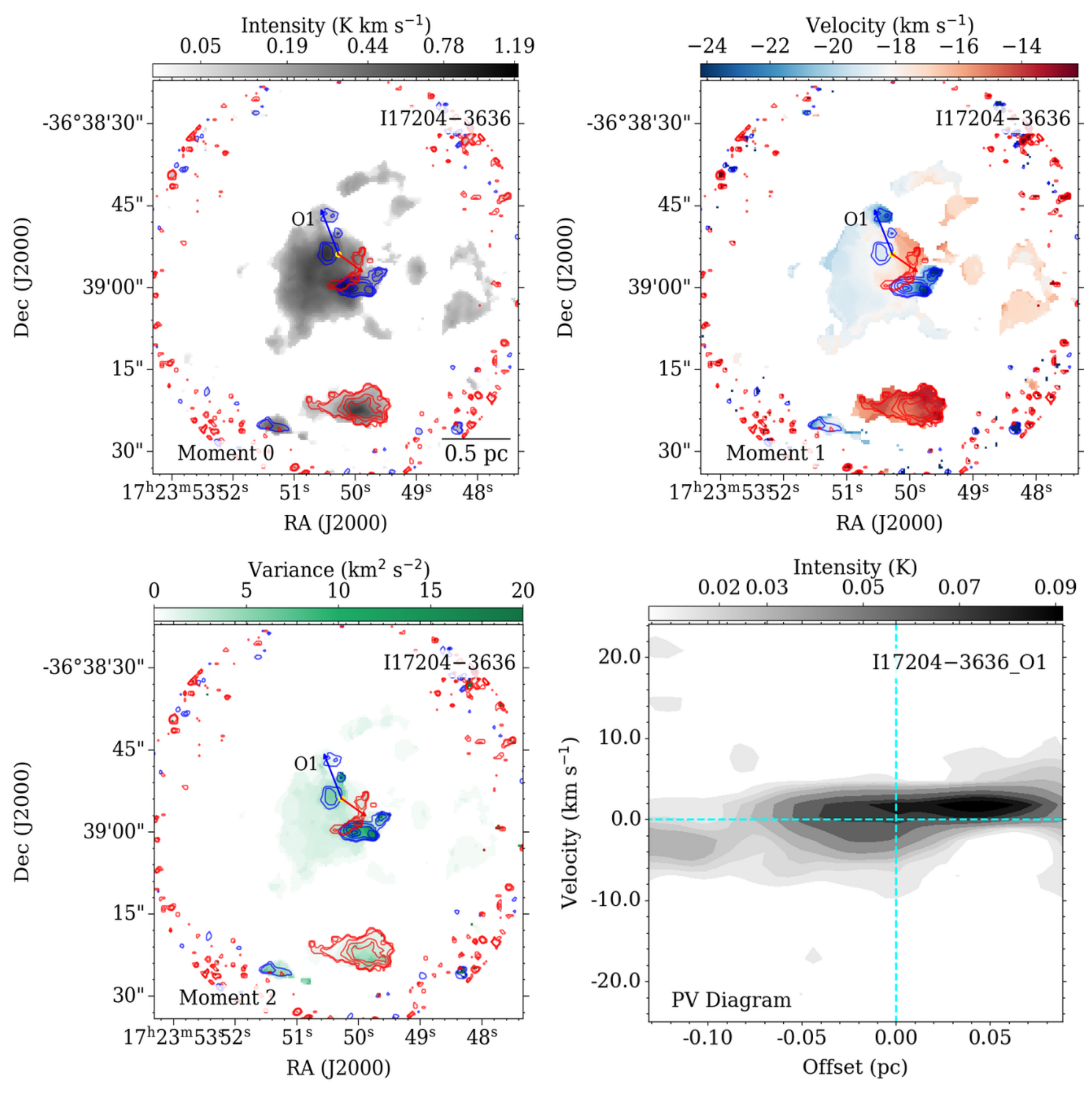}
\figsetgrpnote{ The top panels show the moment-0 map (left) and moment-1 map (right) of HC$_3$N line. The bottom panels show the moment-2 map (left, with a color bar above it) and the PV diagram along the outflow axis (right).}
\figsetgrpend

\figsetgrpstart
\figsetgrpnum{1.30}
\figsetgrptitle{Outflows in I17220-3609}
\figsetplot{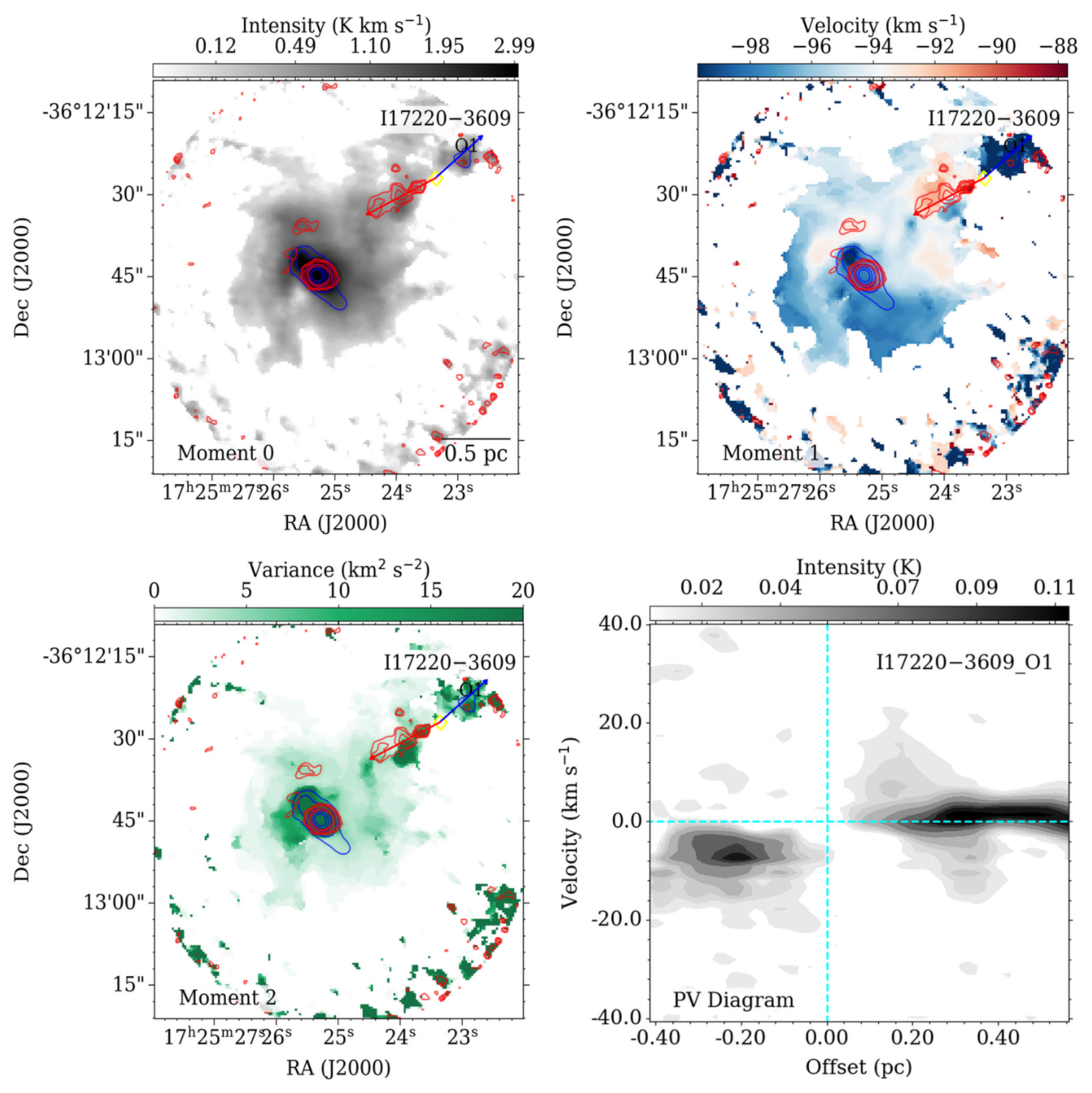}
\figsetgrpnote{ The top panels show the moment-0 map (left) and moment-1 map (right) of HC$_3$N line. The bottom panels show the moment-2 map (left, with a color bar above it) and the PV diagram along the outflow axis (right).}
\figsetgrpend

\figsetgrpstart
\figsetgrpnum{1.31}
\figsetgrptitle{Outflows in I17233-3606}
\figsetplot{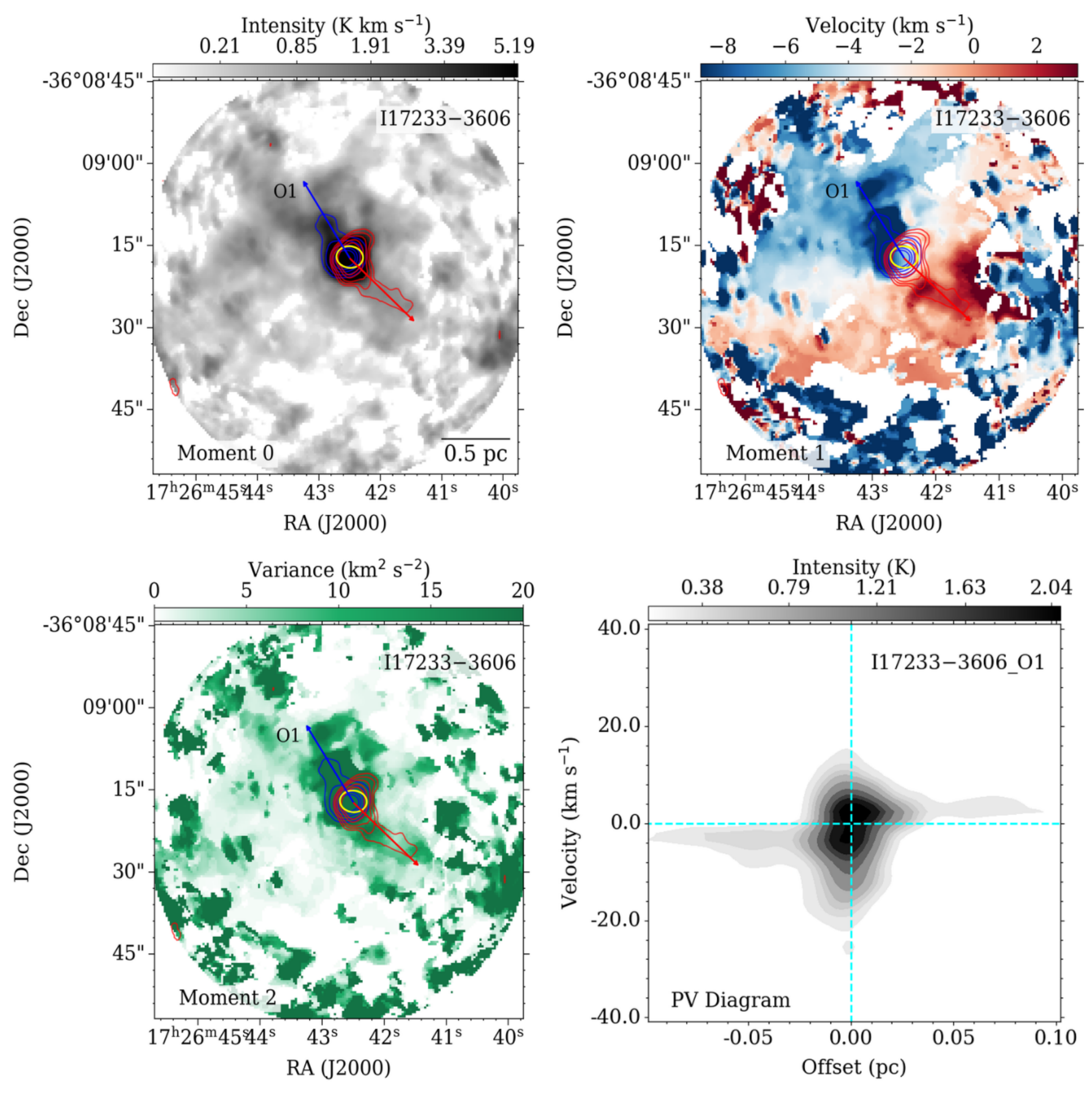}
\figsetgrpnote{ The top panels show the moment-0 map (left) and moment-1 map (right) of HC$_3$N line. The bottom panels show the moment-2 map (left, with a color bar above it) and the PV diagram along the outflow axis (right).}
\figsetgrpend

\figsetgrpstart
\figsetgrpnum{1.32}
\figsetgrptitle{Outflows in I17271-3439}
\figsetplot{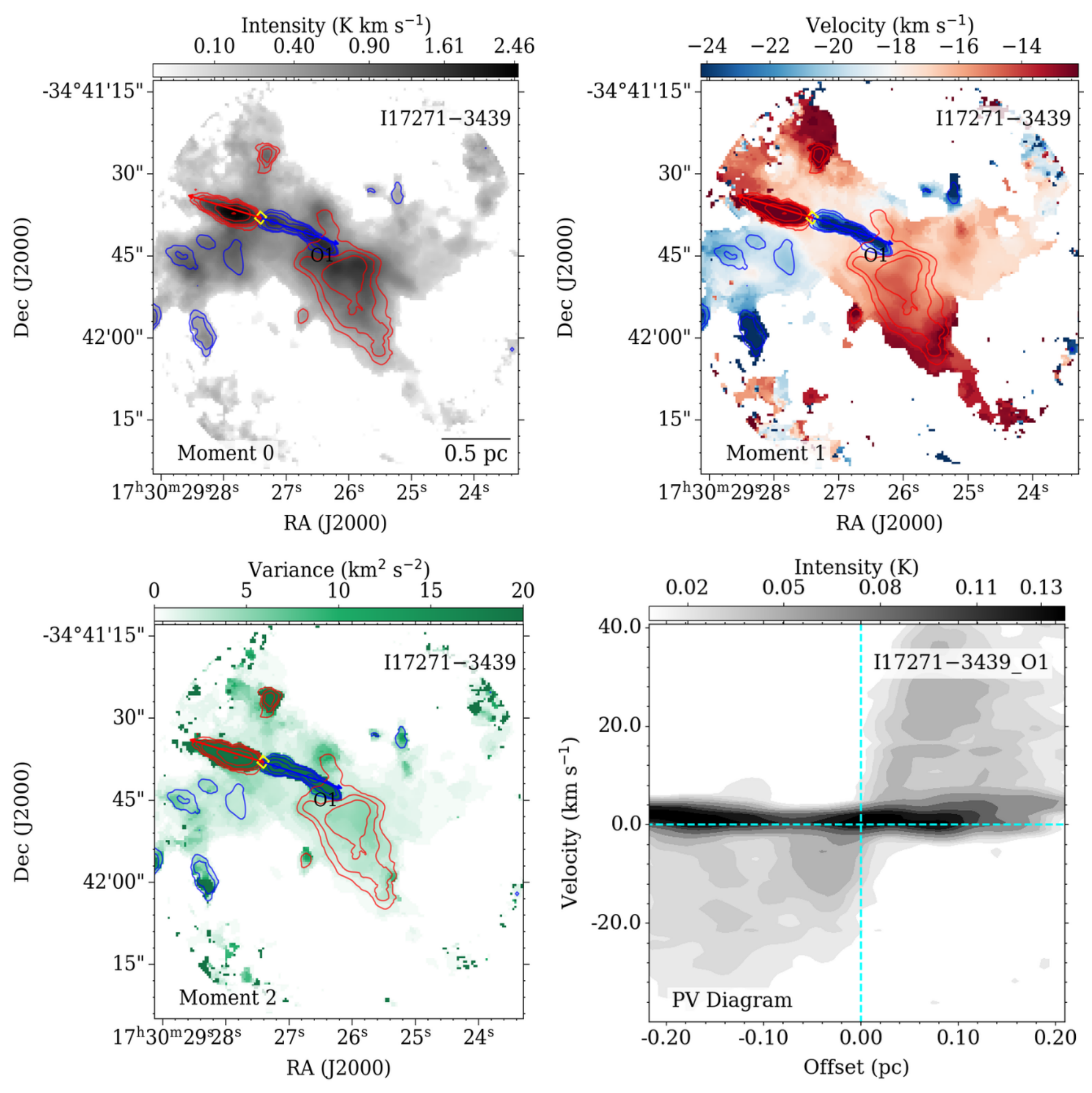}
\figsetgrpnote{ The top panels show the moment-0 map (left) and moment-1 map (right) of HC$_3$N line. The bottom panels show the moment-2 map (left, with a color bar above it) and the PV diagram along the outflow axis (right).}
\figsetgrpend

\figsetgrpstart
\figsetgrpnum{1.33}
\figsetgrptitle{Outflows in I18117-1753}
\figsetplot{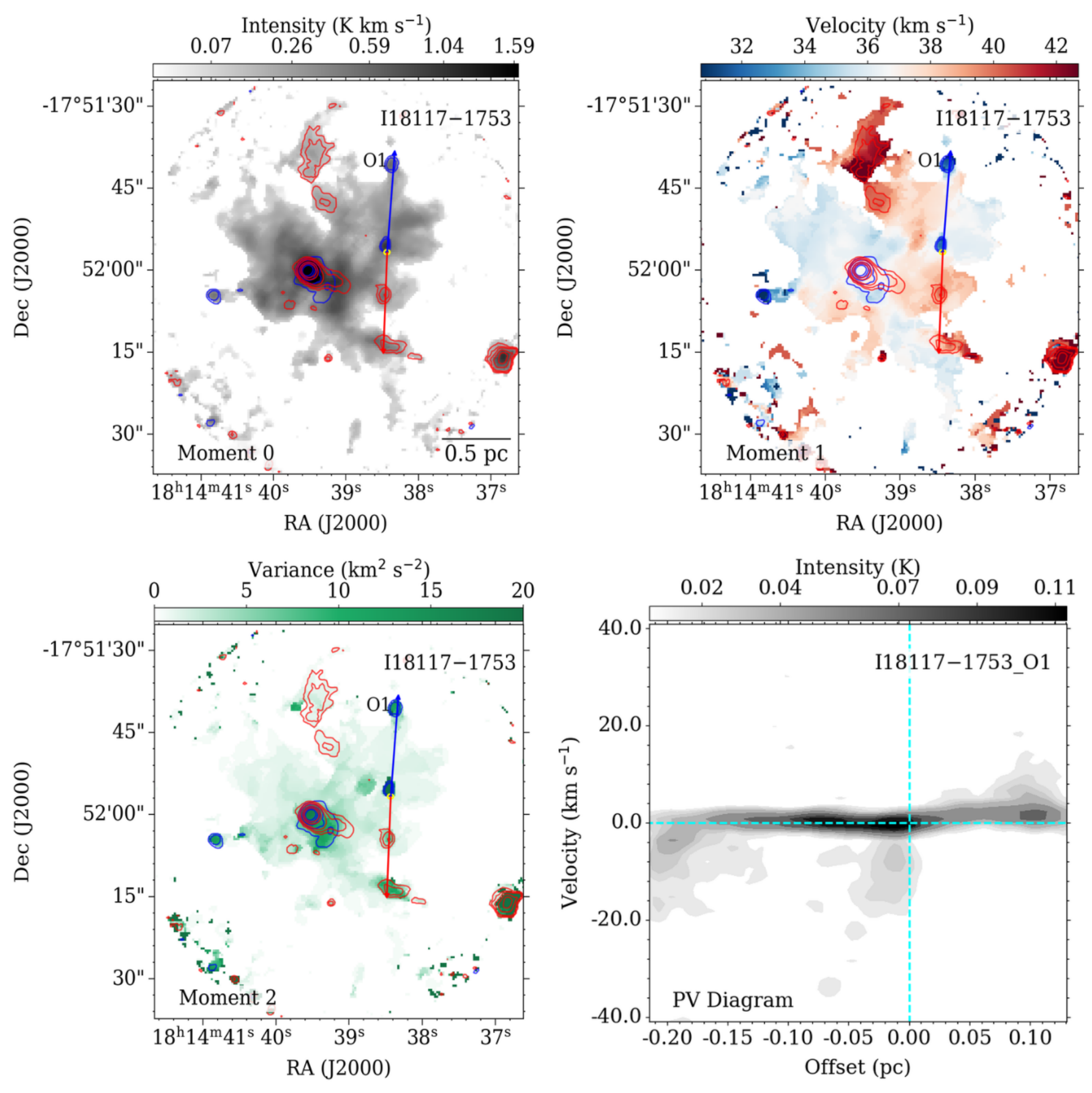}
\figsetgrpnote{ The top panels show the moment-0 map (left) and moment-1 map (right) of HC$_3$N line. The bottom panels show the moment-2 map (left, with a color bar above it) and the PV diagram along the outflow axis (right).}
\figsetgrpend

\figsetgrpstart
\figsetgrpnum{1.34}
\figsetgrptitle{Outflows in I18159-1648}
\figsetplot{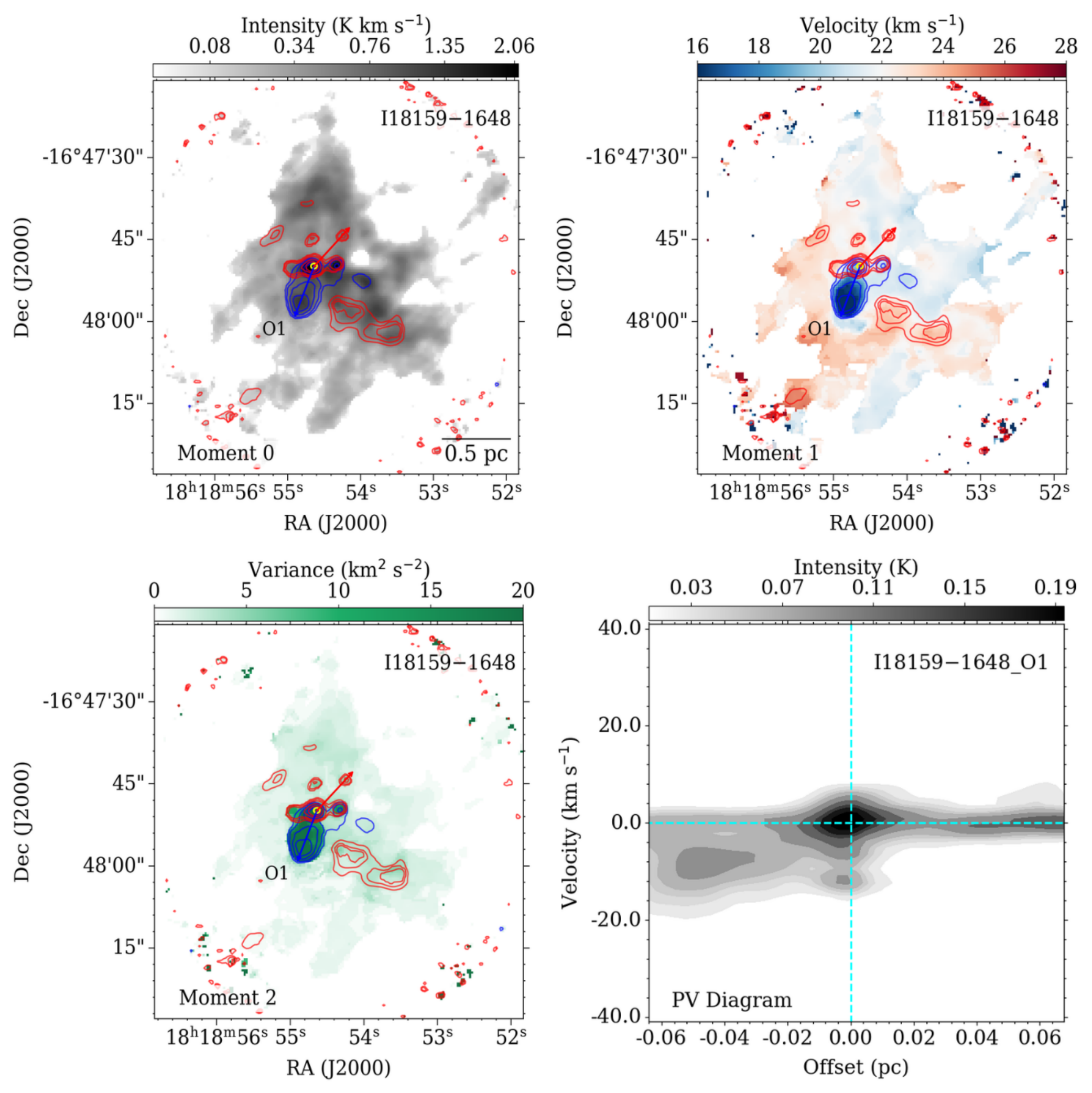}
\figsetgrpnote{ The top panels show the moment-0 map (left) and moment-1 map (right) of HC$_3$N line. The bottom panels show the moment-2 map (left, with a color bar above it) and the PV diagram along the outflow axis (right).}
\figsetgrpend

\figsetgrpstart
\figsetgrpnum{1.35}
\figsetgrptitle{Outflows in I18182-1433}
\figsetplot{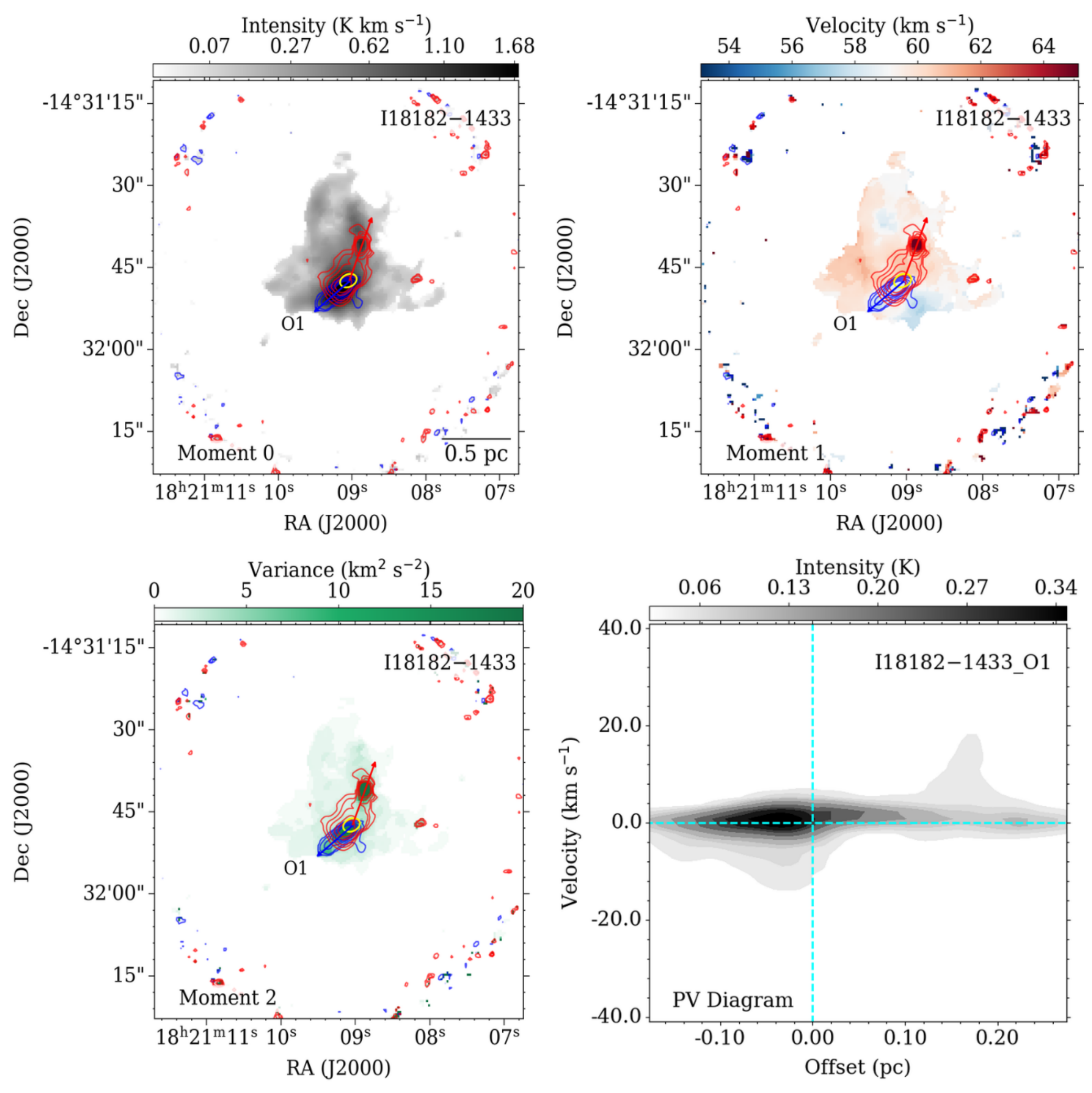}
\figsetgrpnote{ The top panels show the moment-0 map (left) and moment-1 map (right) of HC$_3$N line. The bottom panels show the moment-2 map (left, with a color bar above it) and the PV diagram along the outflow axis (right).}
\figsetgrpend

\figsetgrpstart
\figsetgrpnum{1.36}
\figsetgrptitle{Outflows in I18264-1152}
\figsetplot{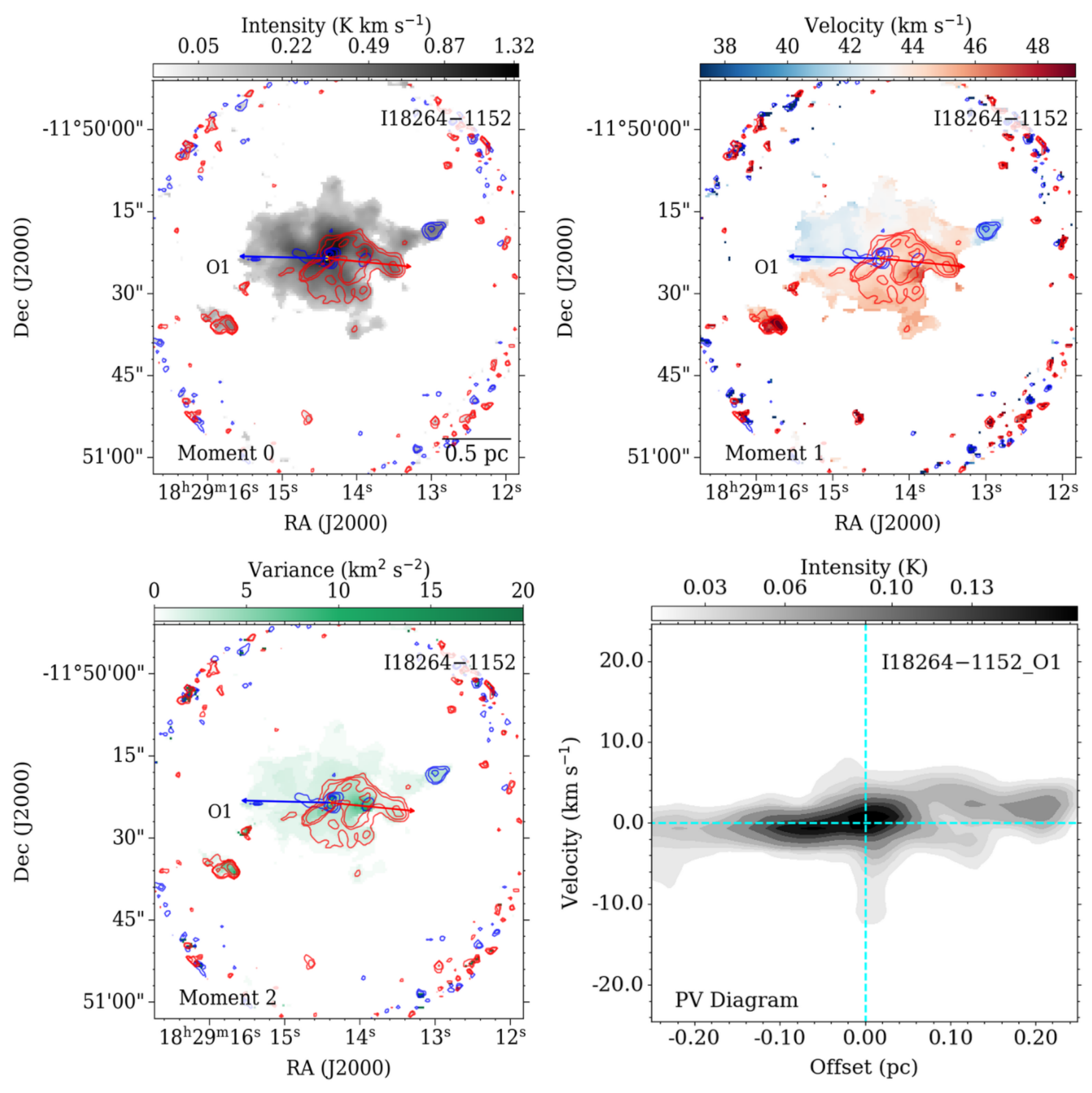}
\figsetgrpnote{ The top panels show the moment-0 map (left) and moment-1 map (right) of HC$_3$N line. The bottom panels show the moment-2 map (left, with a color bar above it) and the PV diagram along the outflow axis (right).}
\figsetgrpend

\figsetgrpstart
\figsetgrpnum{1.37}
\figsetgrptitle{Outflows in I18290-0924}
\figsetplot{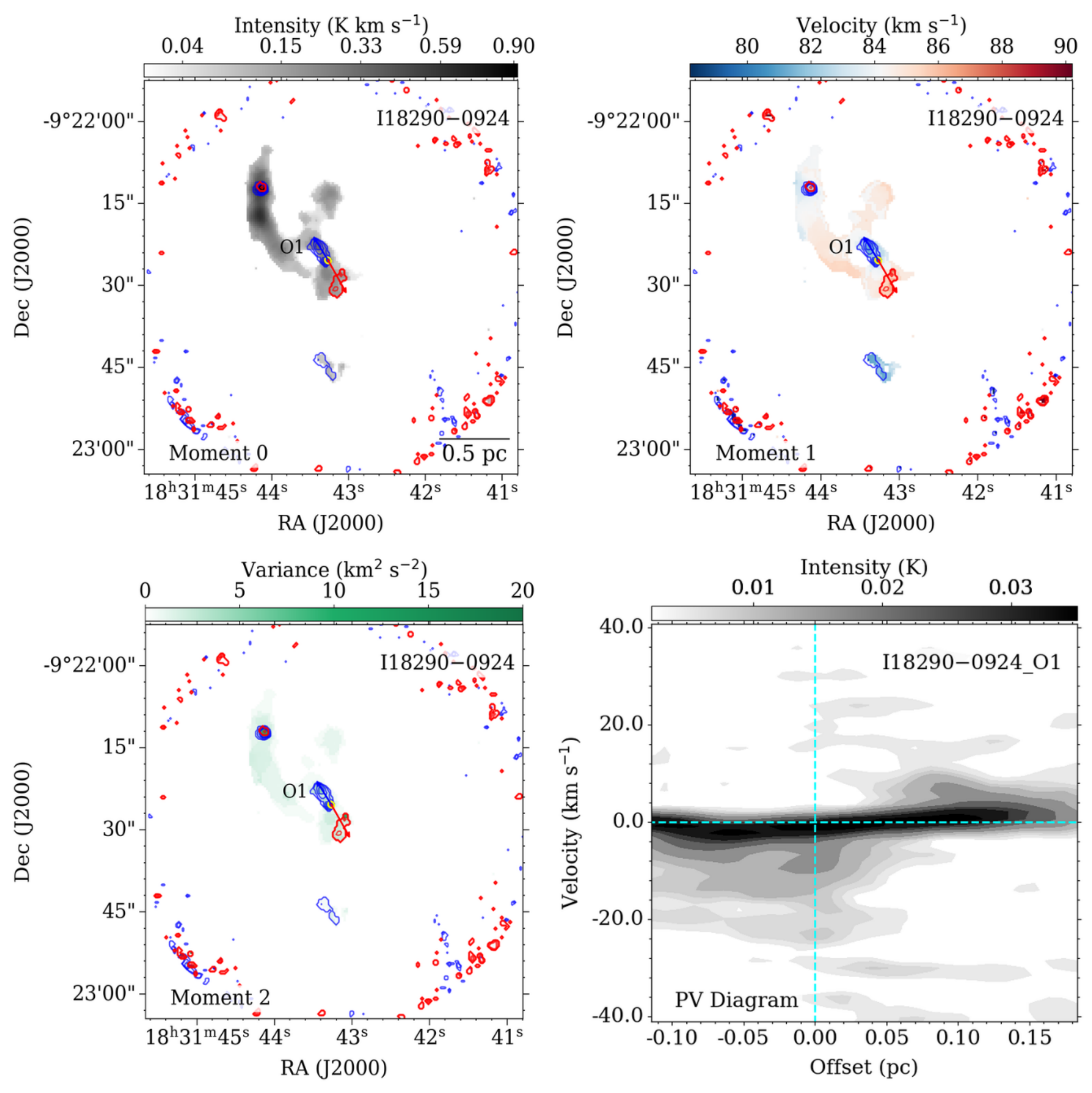}
\figsetgrpnote{ The top panels show the moment-0 map (left) and moment-1 map (right) of HC$_3$N line. The bottom panels show the moment-2 map (left, with a color bar above it) and the PV diagram along the outflow axis (right).}
\figsetgrpend

\figsetgrpstart
\figsetgrpnum{1.38}
\figsetgrptitle{Outflows in I18445-0222}
\figsetplot{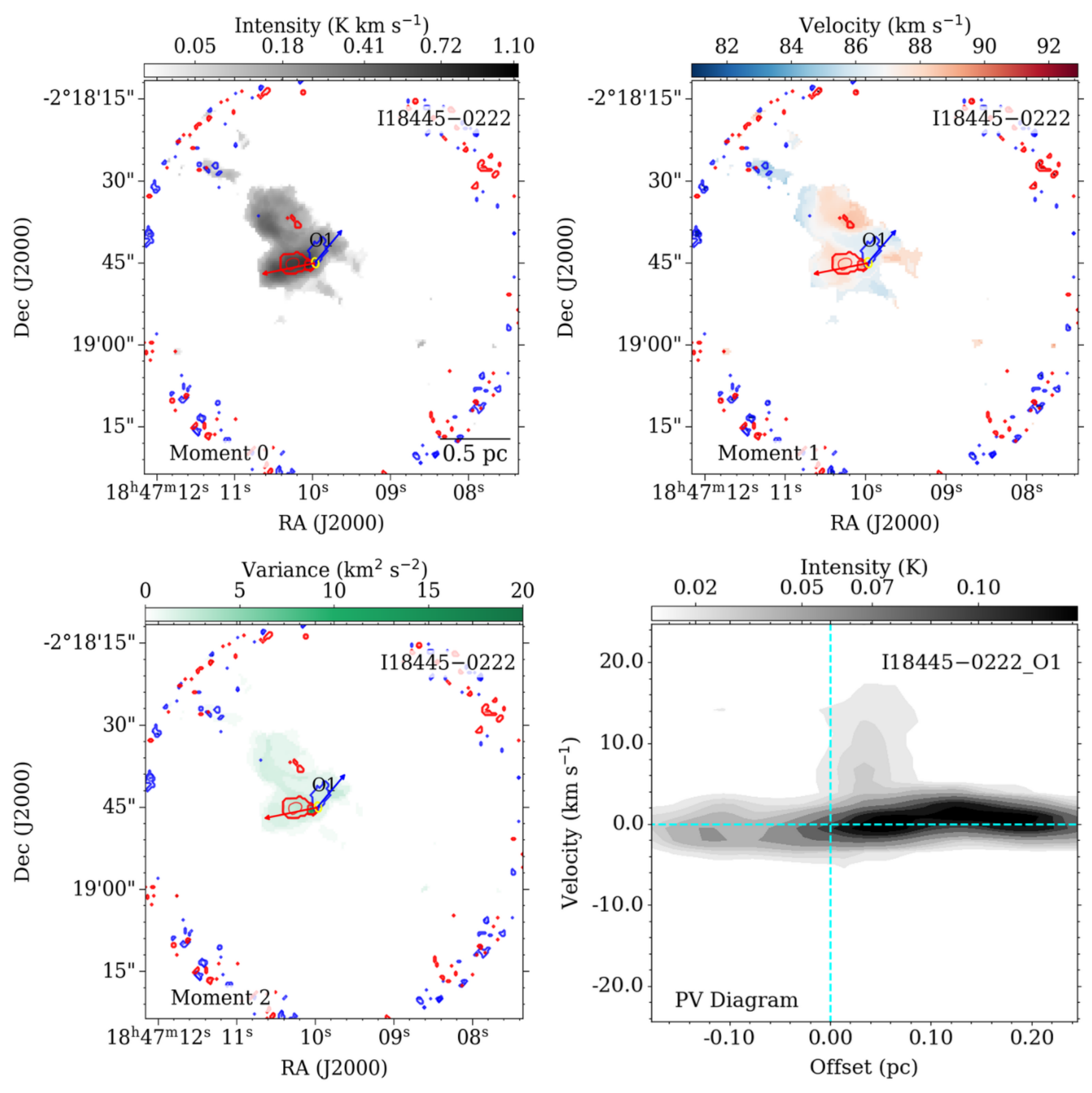}
\figsetgrpnote{ The top panels show the moment-0 map (left) and moment-1 map (right) of HC$_3$N line. The bottom panels show the moment-2 map (left, with a color bar above it) and the PV diagram along the outflow axis (right).}
\figsetgrpend

\figsetgrpstart
\figsetgrpnum{1.39}
\figsetgrptitle{Outflows in I18479-0005}
\figsetplot{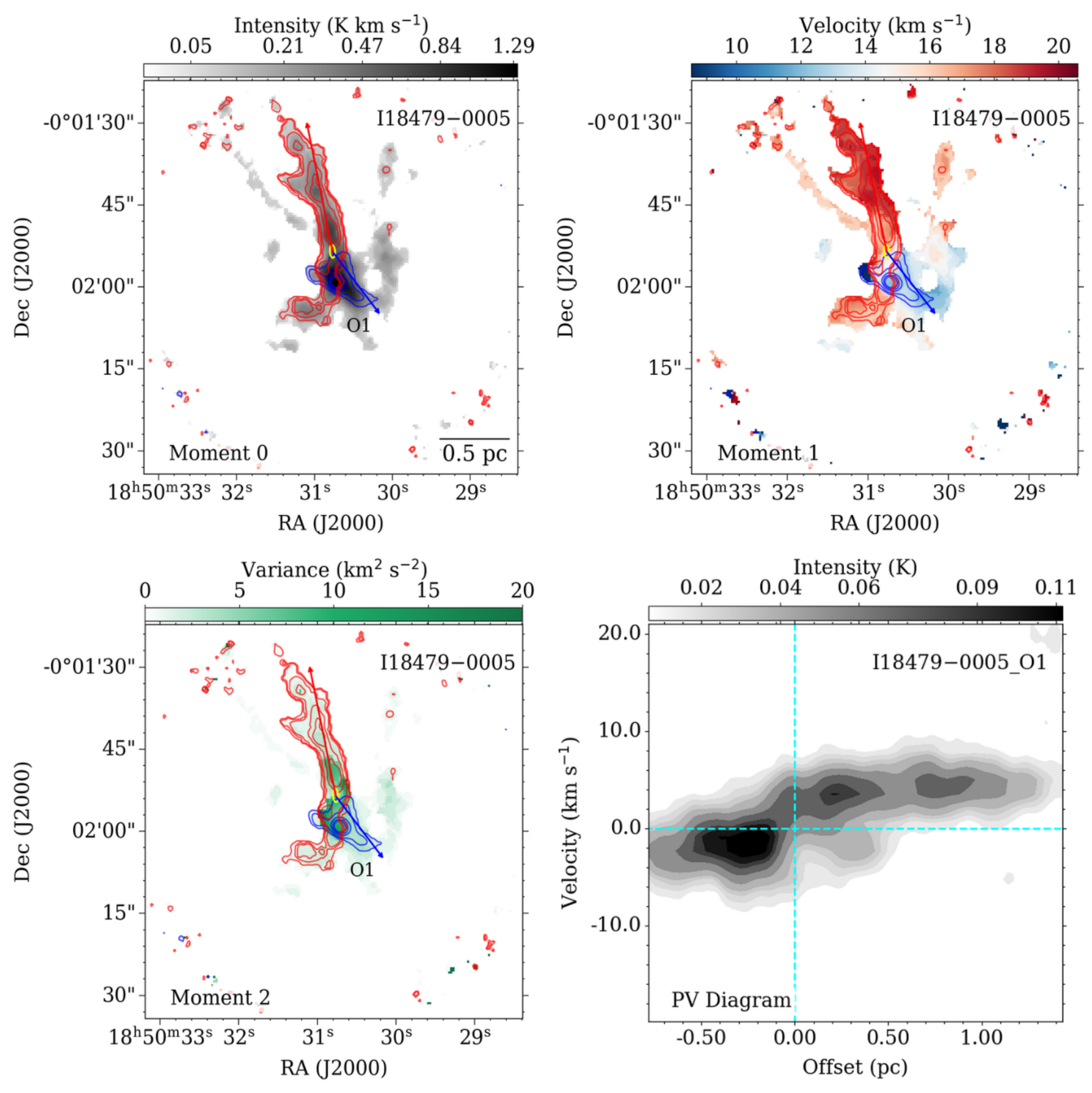}
\figsetgrpnote{ The top panels show the moment-0 map (left) and moment-1 map (right) of HC$_3$N line. The bottom panels show the moment-2 map (left, with a color bar above it) and the PV diagram along the outflow axis (right).}
\figsetgrpend

\figsetgrpstart
\figsetgrpnum{1.40}
\figsetgrptitle{Outflows in I18507+0121}
\figsetplot{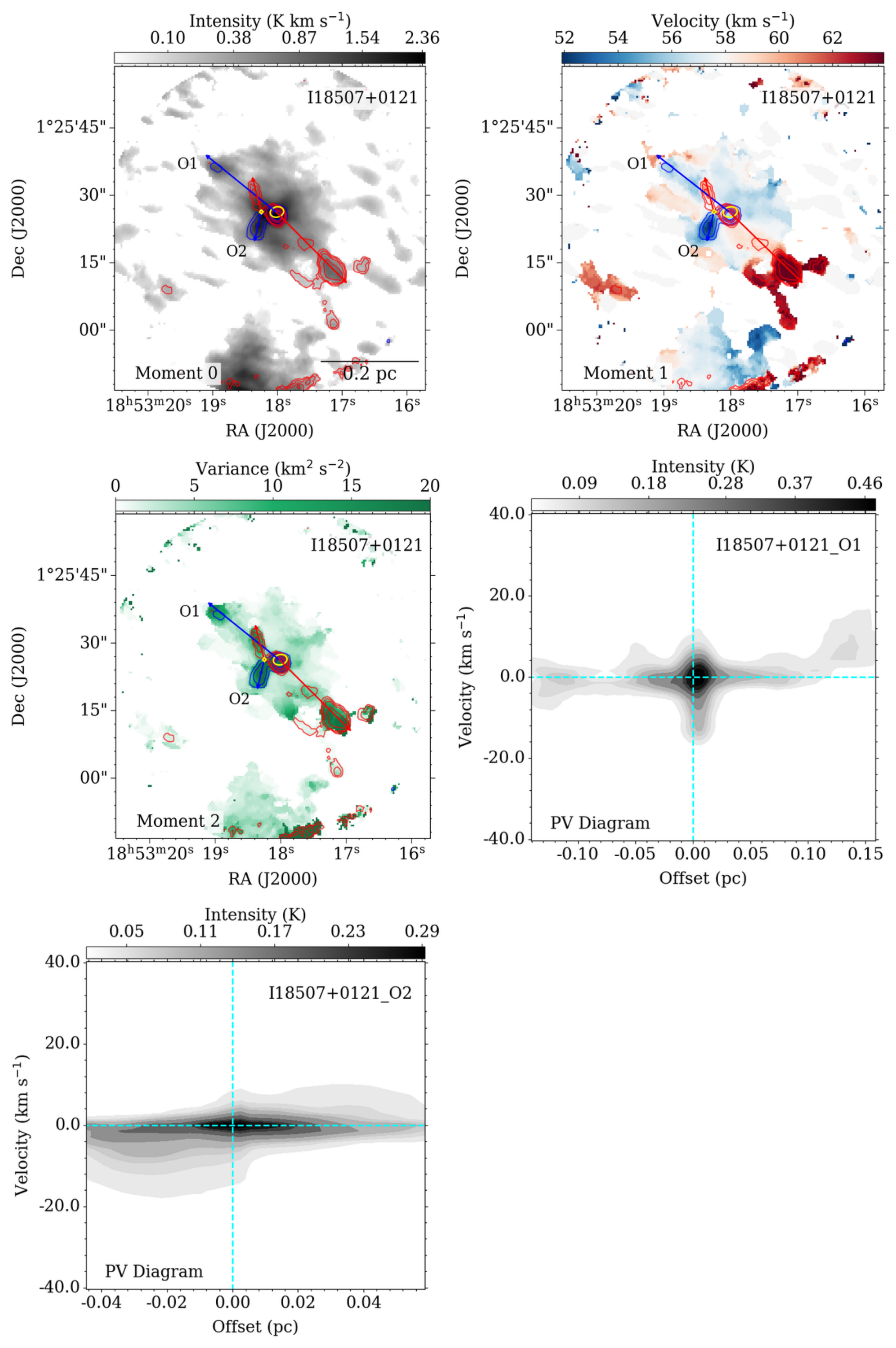}
\figsetgrpnote{ The top panels show the moment-0 map (left) and moment-1 map (right) of HC$_3$N line. The bottom panels show the moment-2 map (left, with a color bar above it) and the PV diagram along the outflow axis (right).}
\figsetgrpend

\figsetend

\begin{figure}[ht!]
\figurenum{A}
\plotone{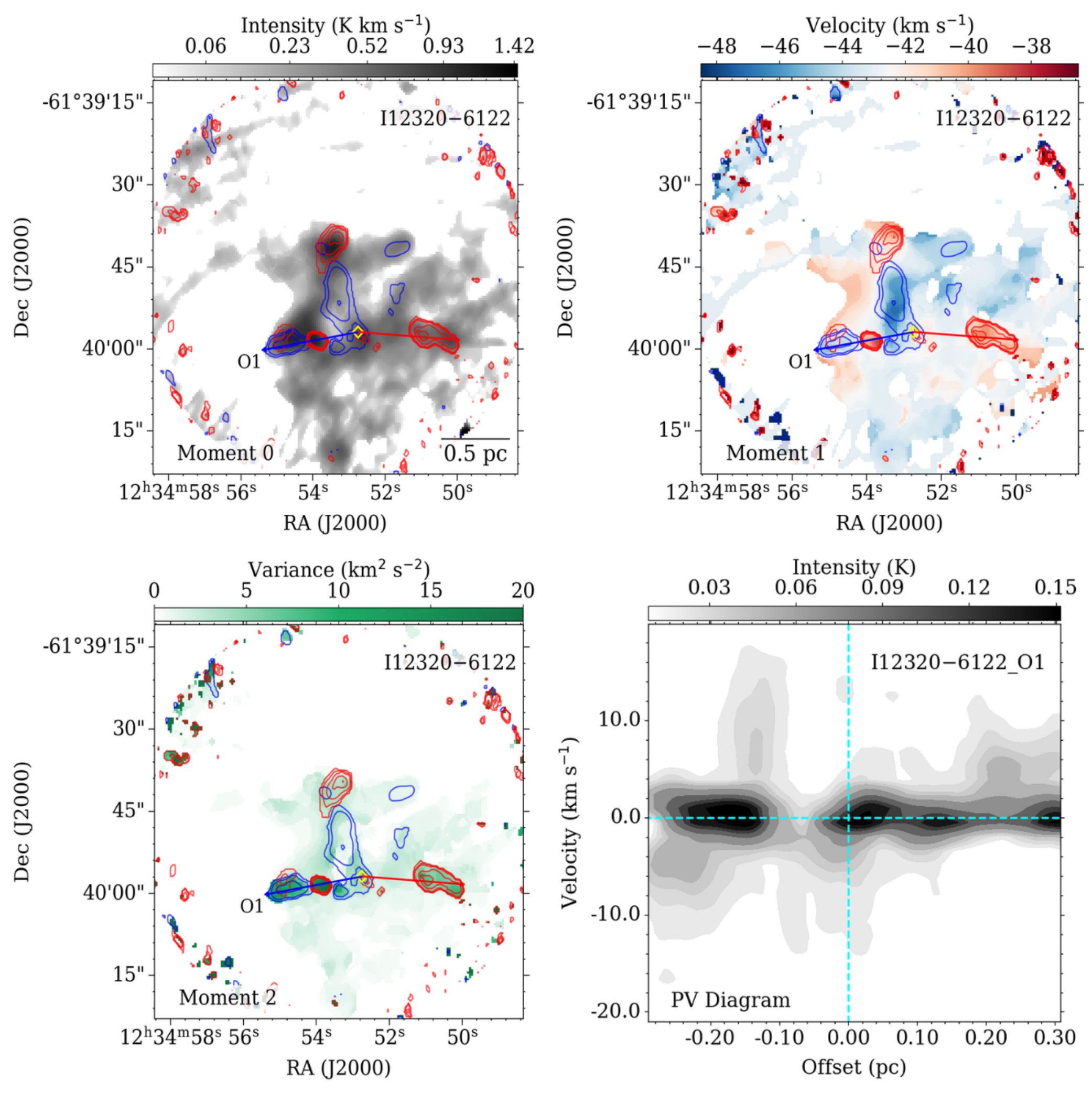}
\caption{ The top panels show the moment-0 map (left) and moment-1 map (right) of HC$_3$N line. The outflow axis is represented by thin blue (for blue-shifted lobe) and red arrows (for red-shifted lobe) in the moment maps, and the position of 3 mm dust continuum core (host core) is marked with a yellow ellipse. The red-shifted and blue-shifted components of the emission are also shown with red and blue contours, respectively. The bottom panels show the moment-2 map (left, with a color bar above it) and the PV diagram along the outflow axis (right). The vertical and horizontal cyan dashed lines represent the position and the V$_{\text{lsr}}$ of the host core, respectively. The complete figure set (40 figures) is available in the online journal.}
\end{figure}

\section{Basic Properties of the Clumps Associated with HC$_3$N Outflows}
\label{Appendix:Clumps}
In this section, we present the basic information of the clumps associated with HC$_3$N outflows in Table~\ref{tab:clump}. The clump information of all the sources from ATOMS survey can be found in \citet[][]{liu2020}.

\begin{deluxetable*}{ccccccccc}

\tabletypesize{\scriptsize}
\tablecaption{Basic Properties of the Clumps Associated with HC$_3$N Outflows  \label{tab:clump}}
\tablehead{
\colhead{Outflow Name} & \colhead{Associated Clump} & \colhead{$^{\dagger}$RA (J2000)} & \colhead{$^{\dagger}$DEC (J2000)}  & \colhead{$^{\dagger}$Vlsr} & \colhead{$^{\dagger}$Distance} & \colhead{$^{\dagger}$$T_{\text{dust}}$} & \colhead{$^{\dagger}$log($L_{\text{bol}}$)} &  \colhead{$^{\dagger}$log($M_{\text{clump}}$)}   \\
{} & {} &(h:m:s) & (d:m:s) & (km/s) & (kpc) & (K) & ($L_\odot$) & ($M_\odot$)}
\startdata
I12320-6122\_O1 & I12320-6122 & 12:34:53.38 & -61:39:46.9 & -42.5 & 3.43 & 44.6 & 5.6 & 3.0 \\
I13140-6226\_O1 & I13140-6226 & 13:17:15.70 &  -62:42:27.5 & -33.9 & 3.8 & 22.6 & 3.8 & 2.9 \\
I13291-6249\_O1 & I13291-6249 & 13:32:31.23 & -63:05:21.8 & -34.7 & 7.61 & 27.2 & 5.2 & 3.7 \\
I14164-6028\_O1 & I14164-6028 & 14:20:08.23 & -60:42:05.0 & -46.5 & 3.19 & 28.7 & 3.7 & 2.2 \\
I14453-5912\_O1 & I14453-5912 & 14:49:07.77 & -59:24:49.7 & -40.2 & 2.82 & 24.1 & 4.2 & 2.9 \\
I15394-5358\_O1  & I15394-5358 & 15:43:16.48 & -54:07:16.9 & -41.6 & 1.82 & 20.7 & 3.7 & 2.9 \\
I15394-5358\_O2 & I15394-5358 & 15:43:16.48 & -54:07:16.9 & -41.6 & 1.82 & 20.7 & 3.7 & 2.9 \\
I15520-5234\_O1 & I15520-5234 & 15:55:48.84 & -52:43:06.2 & -41.3 & 2.65 & 32.2 & 5.1 & 3.2 \\
I15520-5234\_O2 & I15520-5234 & 15:55:48.84 & -52:43:06.2 & -41.3 & 2.65 & 32.2 & 5.1 & 3.2 \\
I15520-5234\_O3 & I15520-5234 & 15:55:48.84 & -52:43:06.2 & -41.3 & 2.65 & 32.2 & 5.1 & 3.2 \\
I15520-5234\_O4 & I15520-5234 & 15:55:48.84 & -52:43:06.2 & -41.3 & 2.65 & 32.2 & 5.1 & 3.2 \\
I15520-5234\_O5 & I15520-5234 & 15:55:48.84 & -52:43:06.2 & -41.3 & 2.65 & 32.2 & 5.1 & 3.2 \\
I15520-5234\_O6 & I15520-5234 & 15:55:48.84 & -52:43:06.2 & -41.3 & 2.65 & 32.2 & 5.1 & 3.2 \\
I15520-5234\_O7 & I15520-5234 & 15:55:48.84 & -52:43:06.2 & -41.3 & 2.65 & 32.2 & 5.1 & 3.2 \\
I15557-5215\_O1 & I15557-5215 & 15:59:40.76 & -52:23:27.7 & -67.6 & 4.03 & 20.7 & 3.9 & 3.2 \\
I15584-5247\_O1 & I15584-5247 & 16:02:19.63 & -52:55:22.4 & -76.8 & 4.41 & 23.9 & 4.2 & 3.1 \\
I16060-5146\_O1 & I16060-5146 & 16:09:52.85 & -51:54:54.7 & -91.6 & 5.3 & 32.2 & 5.8 & 3.9 \\
I16071-5142\_O1 & I16071-5142 & 16:11:00.01 & -51:50:21.6 & -87 & 5.3 & 23.9 & 4.8 & 3.7 \\
I16119-5048\_O1 & I16119-5048 & 16:15:45.65 & -50:55:53.5 & -48.2 & 3.1 & 24.0 & 4.3 & 3.2 \\
I16158-5055\_O1 & I16158-5055 & 16:19:38.63 & -51:03:20.0 & -49.2 & 3.57 & 28.3 & 5.2 & 3.5 \\
I16164-5046\_O1 & I16164-5046 & 16:20:10.91 & -50:53:15.5 & -57.3 & 3.57 & 31.4 & 5.5 & 3.7 \\
I16272-4837\_O1 & I16272-4837 & 16:30:59.08 & -48:43:53.3 & -46.6 & 2.92 & 23.1 & 4.3 & 3.2 \\
I16348-4654\_O1 & I16348-4654 & 16:38:29.64 & -47:00:41.1 & -46.5 & 12.09 & 23.6 & 5.4 & 4.4 \\
I16351-4722\_O1 & I16351-4722 & 16:38:50.98 & -47:27:57.8 & -41.4 & 3.02 & 30.4 & 4.9 & 3.2 \\
I16385-4619\_O1 & I16385-4619 & 16:42:14.04 & -46:25:25.9 & -117.0 & 7.11 & 31.9 & 5.1 & 3.2 \\
I16424-4531\_O1 & I16424-4531 & 16:46:06.61 & -45:36:46.6 & -34.2 & 2.63 & 24.6 & 3.9 & 2.7 \\
I16445-4459\_O1 & I16445-4459 & 16:48:05.18 & -45:05:08.6 & -121.3 & 7.95 & 24.6 & 5.0 & 3.9 \\
I16458-4512\_O1 & I16458-4512 & 16:49:30.41 & -45:17:53.6 & -50.4 & 3.56 & 21.4 & 4.5 & 3.6 \\
I16487-4423\_O1 & I16487-4423 & 16:52:23.67 & -44:27:52.3 & -43.4 & 3.26 & 24.6 & 4.4 & 3.0 \\
I16489-4431\_O1 & I16489-4431 & 16:52:33.50 & -44:36:17.7 & -41.3 & 3.26 & 21.8 & 3.8 & 2.9 \\
I16547-4247\_O1 & I16547-4247 & 16:58:17.26 & -42:52:04.5 & -30.4 & 2.74 & 28.9 & 4.8 & 3.2 \\
I16571-4029\_O1 & I16571-4029 & 17:00:32.21 & -40:34:12.7 & -15 & 2.38 & 27.0 & 4.3 & 2.9 \\
I16571-4029\_O2 & I16571-4029 & 17:00:32.21 & -40:34:12.7 & -15 & 2.38 & 27.0 & 4.3 & 2.9 \\
I16571-4029\_O3 & I16571-4029 & 17:00:32.21 & -40:34:12.7 & -15 & 2.38 & 27.0 & 4.3 & 2.9 \\
I17006-4215\_O1 & I17006-4215 & 17:04:12.99 & -42:19:54.2 & -23.2 & 2.21 & 27.7 & 4.4 & 2.8 \\
I17016-4124\_O1 & I17016-4124 & 17:05:11.02 & -41:29:07.8 & -27.1 & 1.37 & 32.0 & 5.3 & 3.8 \\
I17016-4124\_O2 & I17016-4124 & 17:05:11.02 & -41:29:07.8 & -27.1 & 1.37 & 32.0 & 5.3 & 3.8 \\
I17175-3544\_O1 & I17175-3544 & 17:20:53.10 & -35:47:03.0 & -5.7 & 1.34 & 30.6 & 4.8 & 3.1 \\
I17204-3636\_O1 & I17204-3636 & 17:23:50.32 & -36:38:58.1 & -18.2 & 3.32 & 25.8 & 4.2 & 2.9 \\
I17220-3609\_O1 & I17220-3609 & 17:25:24.99 & -36:12:45.1 & -93.7 & 8.01 & 25.4 & 5.7 & 4.3 \\
I17233-3606\_O1 & I17233-3606 & 17:26:42.73 & -36:09:20.8 & -2.7 & 1.34 & 29.9 & 4.6 & 3.0 \\
I17271-3439\_O1 & I17271-3439 & 17:30:26.21 & -34:41:48.9 & -18.2 & 3.1 & 35.0 & 5.6 & 4.0 \\
I18117-1753\_O1 & I18117-1753 & 18:14:39.14 & -17:52:01.3 & 36.7 & 2.57 & 23.6 & 4.6 & 3.5 \\
I18159-1648\_O1 & I18159-1648 & 18:18:54.34 & -16:47:51.9 & 22 & 1.48 & 21.6 & 3.8 & 2.8 \\
I18182-1433\_O1 & I18182-1433 & 18:21:09.22 & -14:31:46.8 & 59.1 & 4.71 & 24.7 & 4.3 & 3.1 \\
I18264-1152\_O1 & I18264-1152 & 18:29:14.28 & -11:50:27.0 & 43.2 & 3.33 & 20.3 & 3.9 & 3.2 \\
I18290-0924\_O1 & I18290-0924 & 18:31:43.23 & -09:22:28.5 & 84.2 & 5.34 & 22.1 & 4.0 & 3.2 \\
I18445-0222\_O1 & I18445-0222 & 18:47:09.76 & -02:18:47.6 & 86.9 & 5.16 & 27.0 & 4.6 & 3.4 \\
I18479-0005\_O1 & I18479-0005 & 18:50:30.79 & -00:01:58.2 & 14.6 & 12.96 & 34.2 & 6.1 & 4.2 \\
I18507+0121\_O1 & I18507+0121 & 18:53:18.12 & 01:25:22.7 & 57.9 & 1.56 & 22.7 & 3.5 & 2.6 \\
I18507+0121\_O2 & I18507+0121 & 18:53:18.12 & 01:25:22.7 & 57.9 & 1.56 & 22.7 & 3.5 & 2.6
\enddata
\tablenotetext{\dagger
}{Reference:\citet[][]{liu2020}}
\end{deluxetable*}

\section{Position Velocity diagrams across the Outflow Lobes}
\label{sec:transverse}
As discussed in Section~\ref{sec:narrow jets}, we explored the possibility of HC$_3$N to trace fast narrow jets as well as slow wide outflow components within the same outflow source. We identified 10 of such sources among 45 outflows. In this section, we present the position-velocity diagram across the outflow lobes of the 10 sources  (See Fig. Set 2).

\figsetstart
\figsetnum{2}
\figsettitle{PV Diagram Across Outflow Lobes}

\figsetgrpstart
\figsetgrpnum{2.1}
\figsetgrptitle{I15394-5358}
\figsetplot{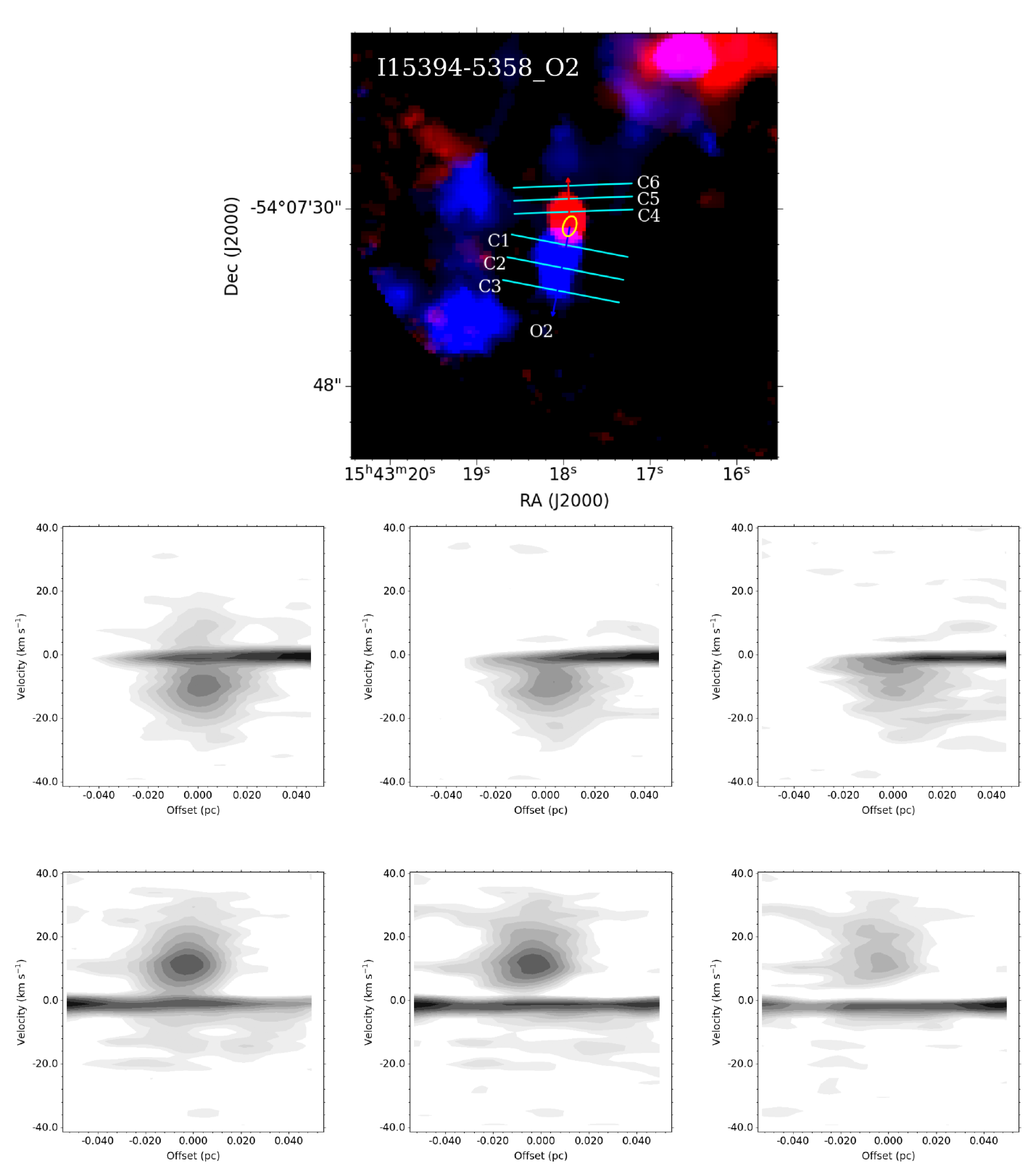}
\figsetgrpnote{The top panel shows the color composite image of the outflow lobes overlaid with six cuts (in Cyan; C1--C6) along which PV diagrams are generated. The middle and bottom panels represent the PV diagrams of C1--C3 and C4--C6 cuts, respectively.}
\figsetgrpend

\figsetgrpstart
\figsetgrpnum{2.2}
\figsetgrptitle{I16071-5142}
\figsetplot{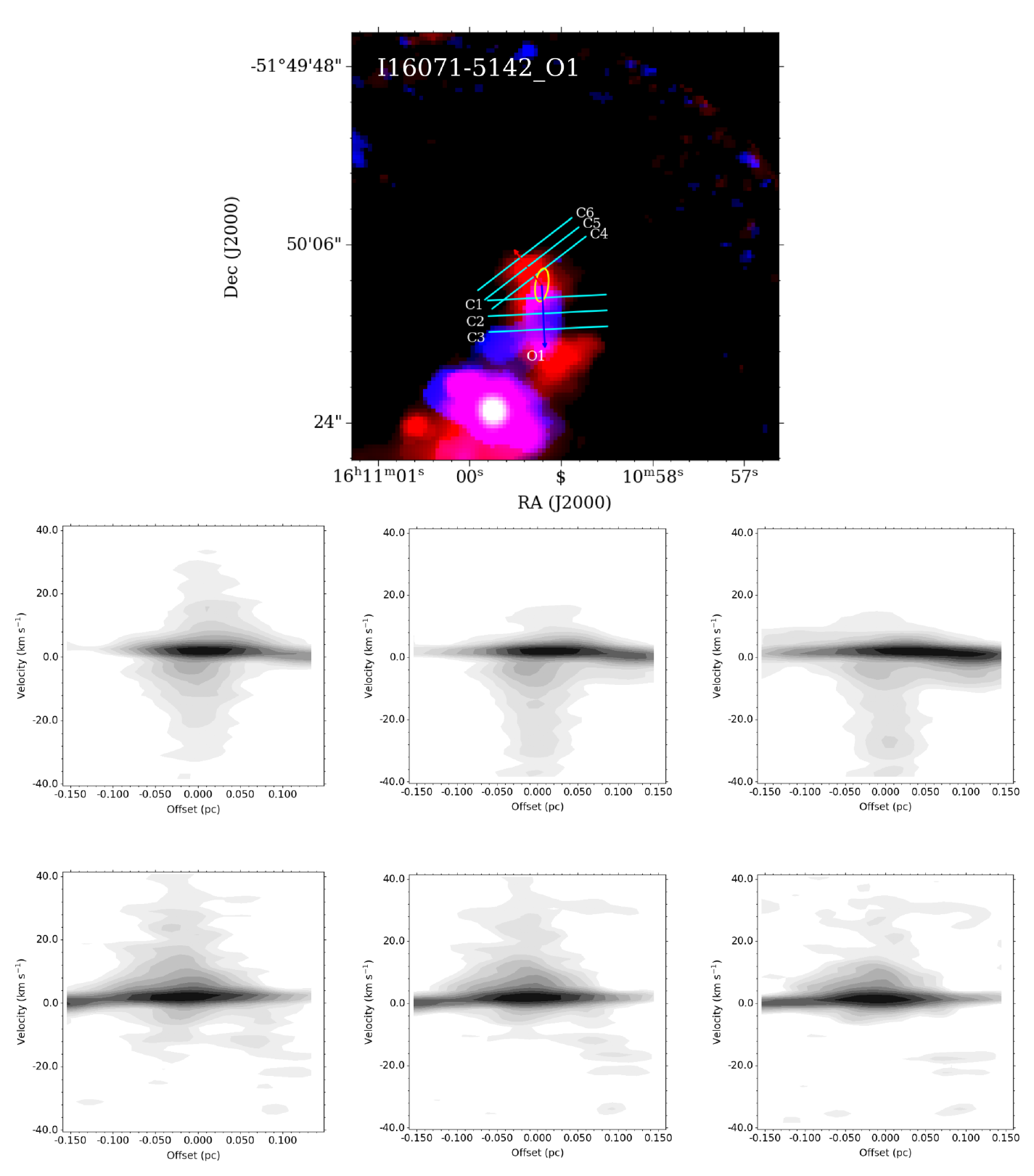}
\figsetgrpnote{The top panel shows the color composite image of the outflow lobes overlaid with six cuts (in Cyan; C1--C6) along which PV diagrams are generated. The middle and bottom panels represent the PV diagrams of C1--C3 and C4--C6 cuts, respectively.}
\figsetgrpend

\figsetgrpstart
\figsetgrpnum{2.3}
\figsetgrptitle{I16272-4837}
\figsetplot{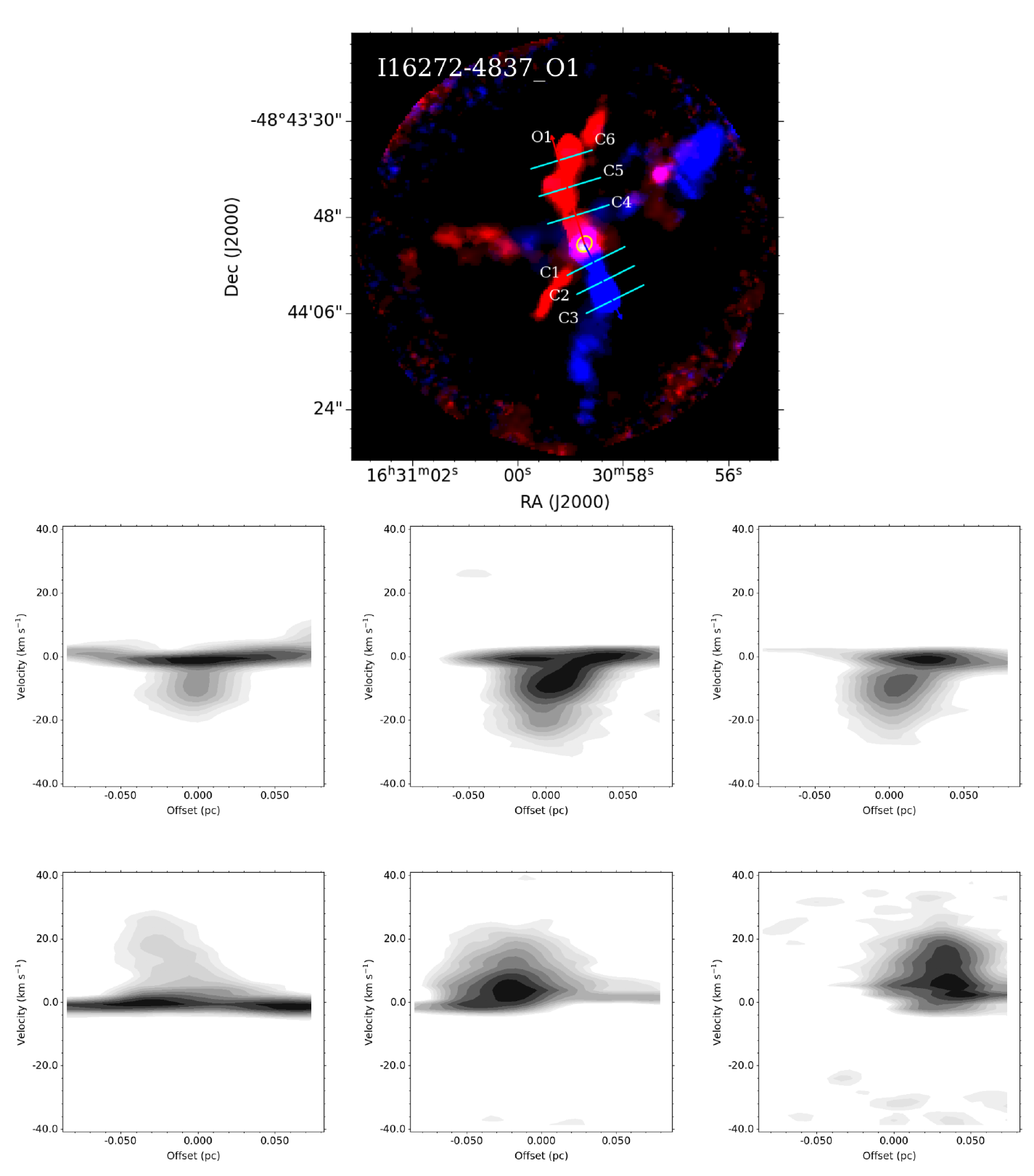}
\figsetgrpnote{The top panel shows the color composite image of the outflow lobes overlaid with six cuts (in Cyan; C1--C6) along which PV diagrams are generated. The middle and bottom panels represent the PV diagrams of C1--C3 and C4--C6 cuts, respectively.}
\figsetgrpend

\figsetgrpstart
\figsetgrpnum{2.4}
\figsetgrptitle{I16351-4722}
\figsetplot{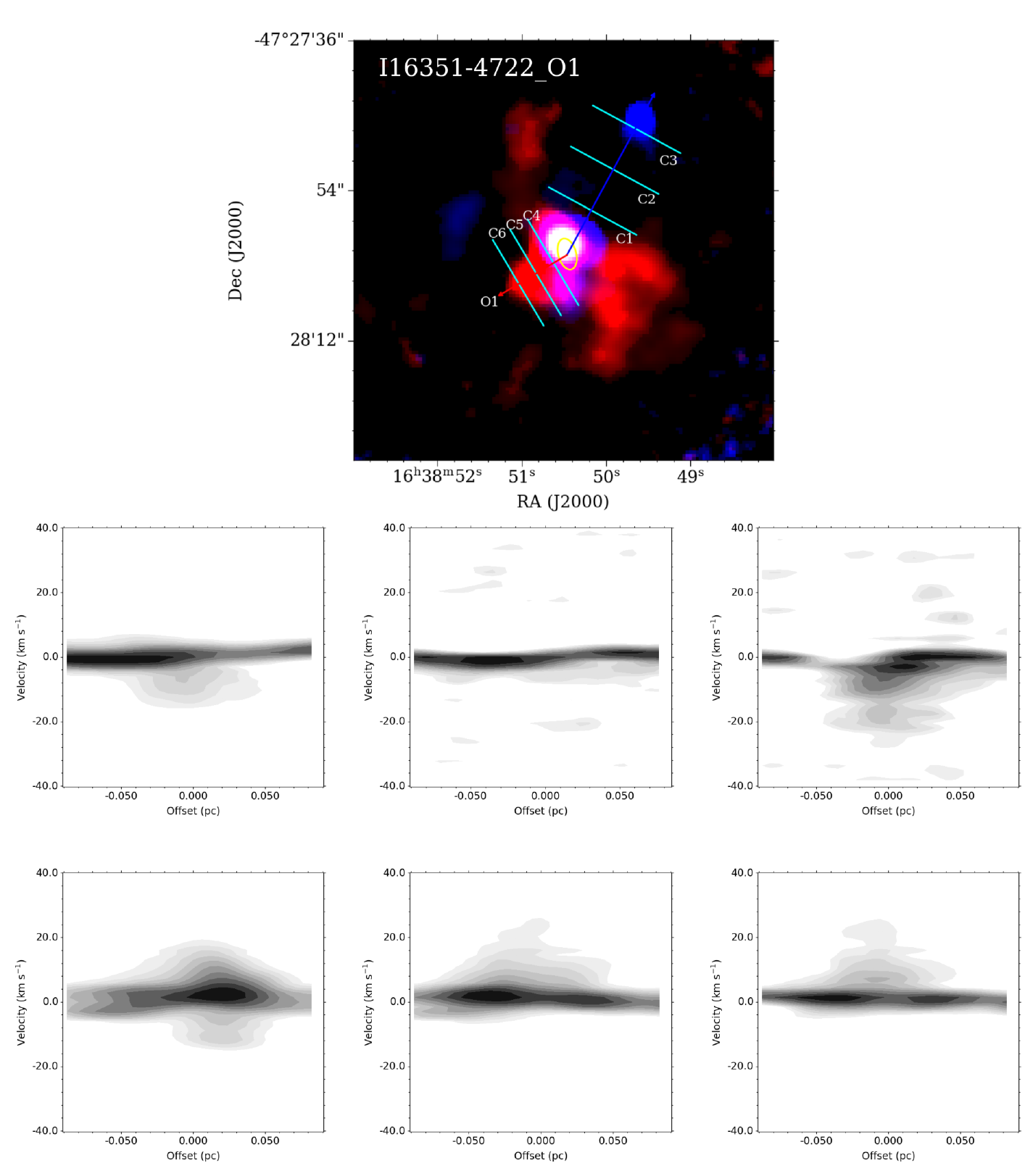}
\figsetgrpnote{The top panel shows the color composite image of the outflow lobes overlaid with six cuts (in Cyan; C1--C6) along which PV diagrams are generated. The middle and bottom panels represent the PV diagrams of C1--C3 and C4--C6 cuts, respectively.}
\figsetgrpend

\figsetgrpstart
\figsetgrpnum{2.5}
\figsetgrptitle{I16424-4531}
\figsetplot{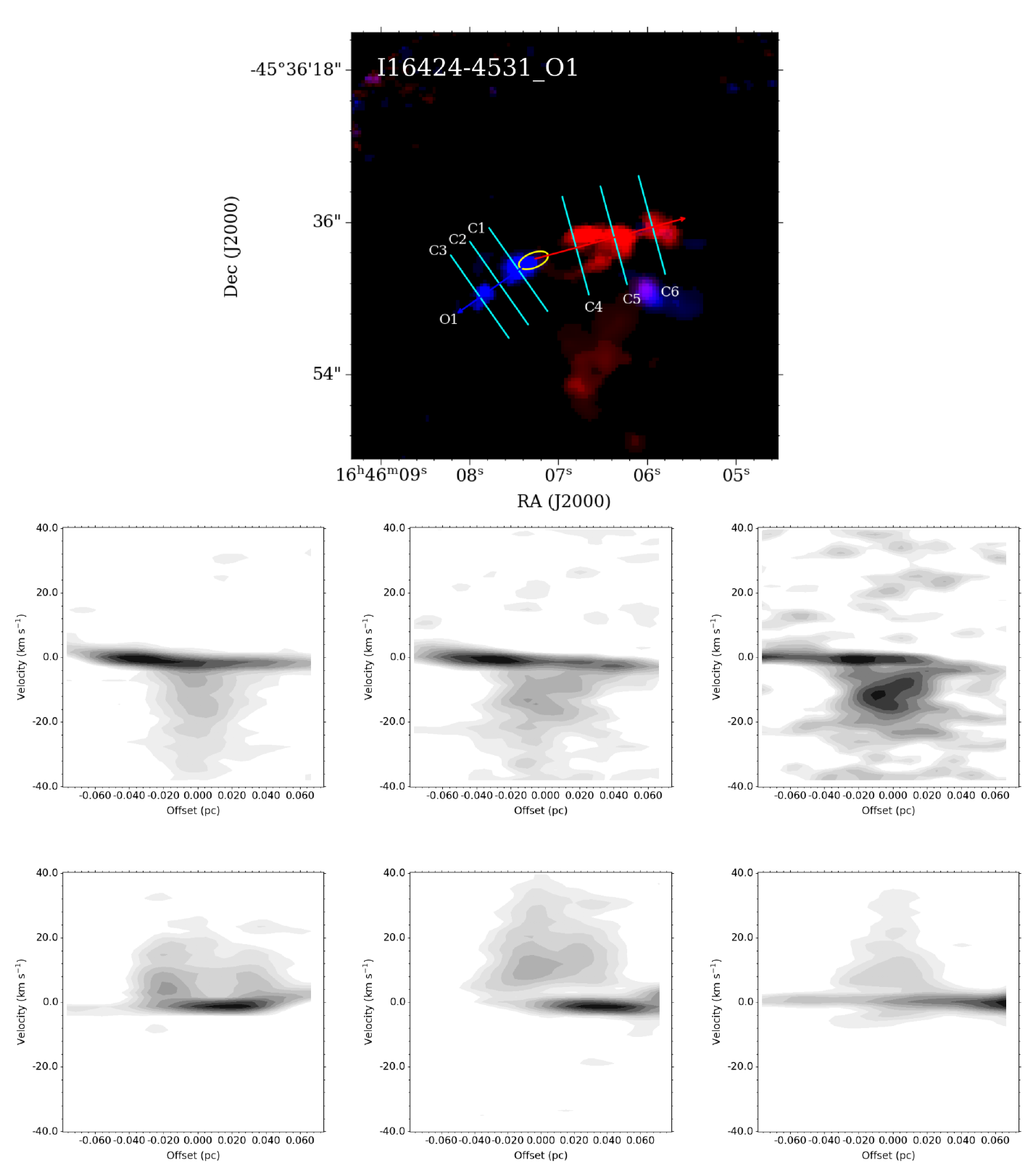}
\figsetgrpnote{The top panel shows the color composite image of the outflow lobes overlaid with six cuts (in Cyan; C1--C6) along which PV diagrams are generated. The middle and bottom panels represent the PV diagrams of C1--C3 and C4--C6 cuts, respectively.}
\figsetgrpend

\figsetgrpstart
\figsetgrpnum{2.6}
\figsetgrptitle{I16489-4431}
\figsetplot{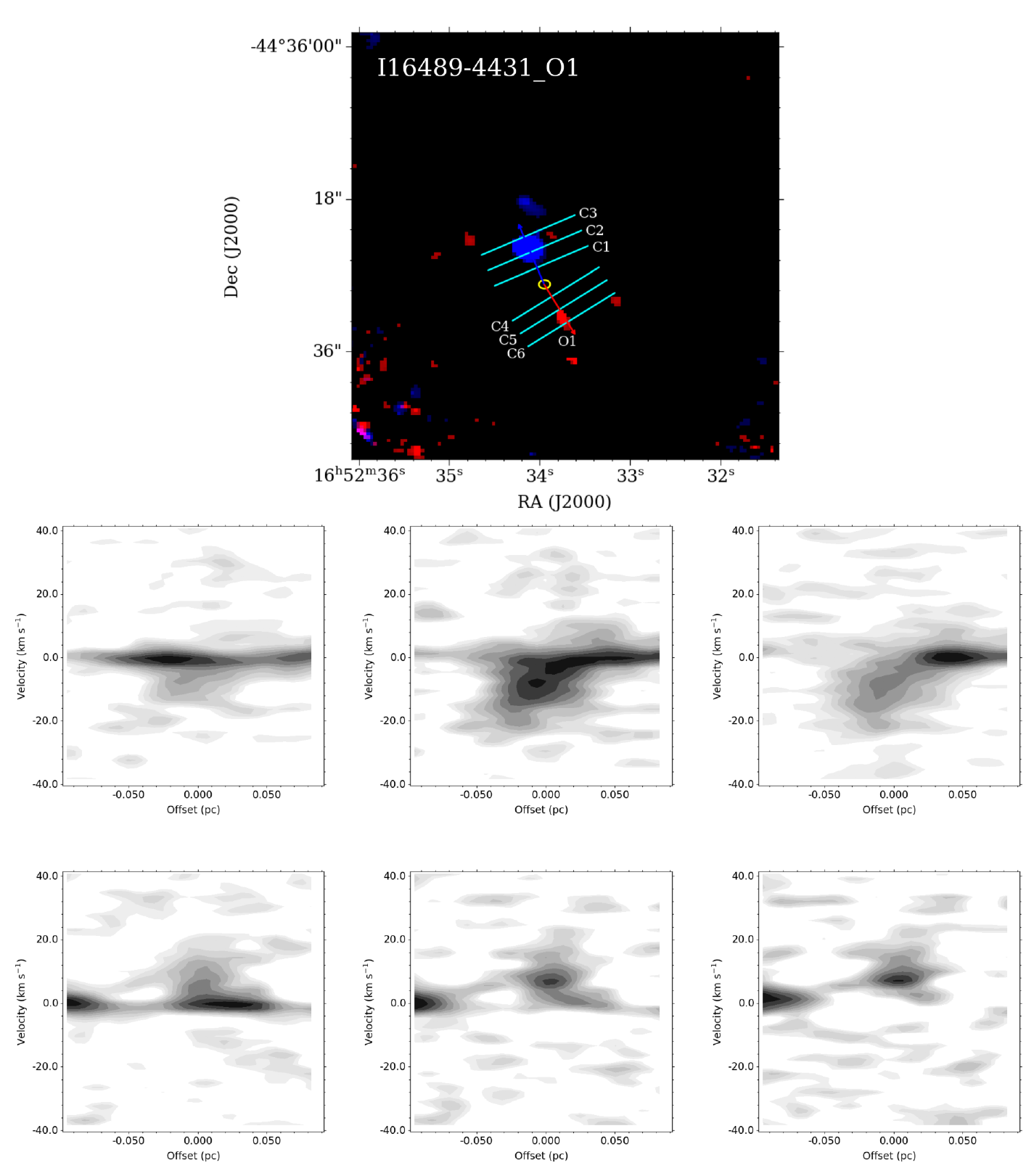}
\figsetgrpnote{The top panel shows the color composite image of the outflow lobes overlaid with six cuts (in Cyan; C1--C6) along which PV diagrams are generated. The middle and bottom panels represent the PV diagrams of C1--C3 and C4--C6 cuts, respectively.}
\figsetgrpend

\figsetgrpstart
\figsetgrpnum{2.7}
\figsetgrptitle{I17175-3544}
\figsetplot{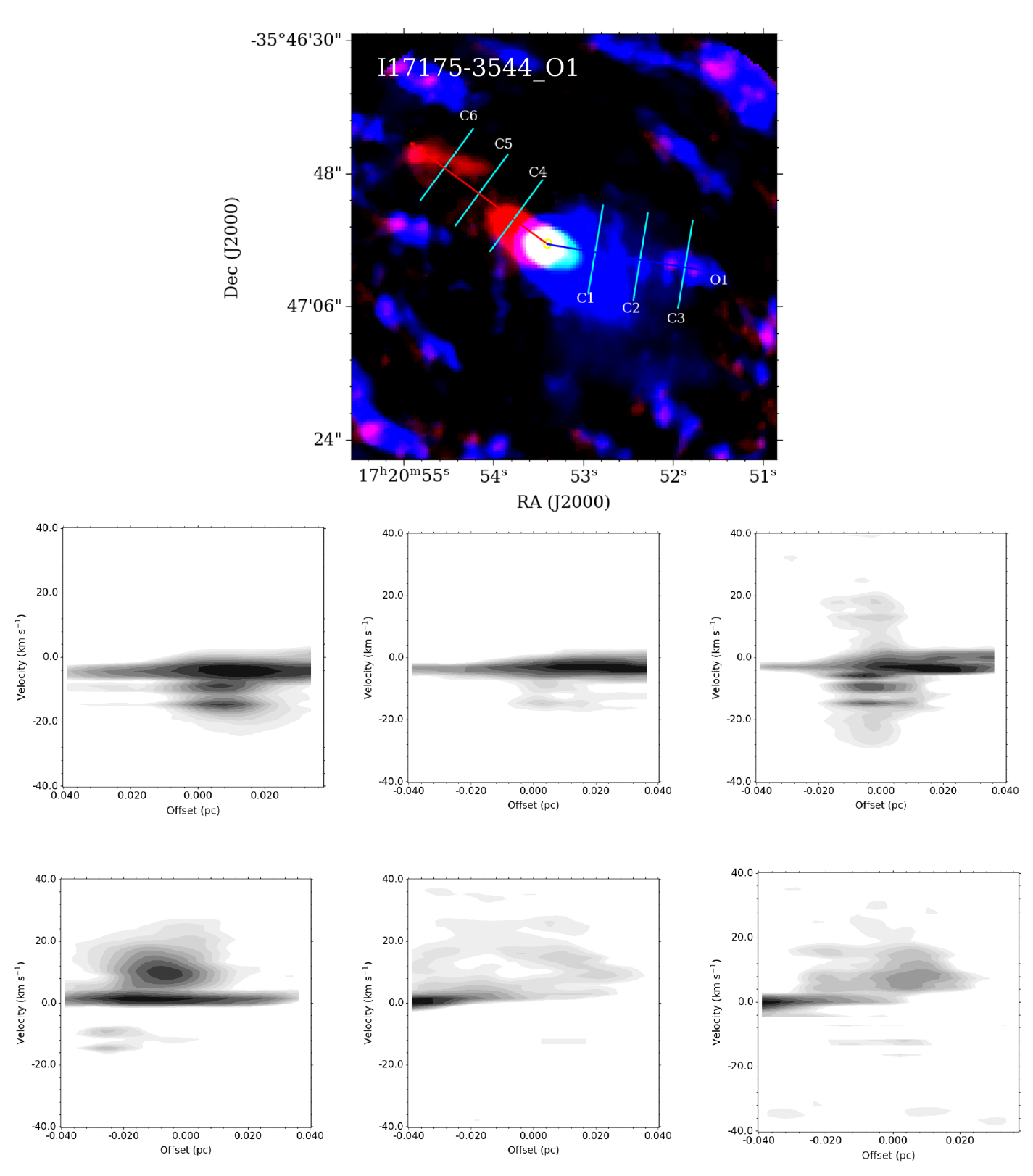}
\figsetgrpnote{The top panel shows the color composite image of the outflow lobes overlaid with six cuts (in Cyan; C1--C6) along which PV diagrams are generated. The middle and bottom panels represent the PV diagrams of C1--C3 and C4--C6 cuts, respectively.}
\figsetgrpend

\figsetgrpstart
\figsetgrpnum{2.8}
\figsetgrptitle{I17271-3439}
\figsetplot{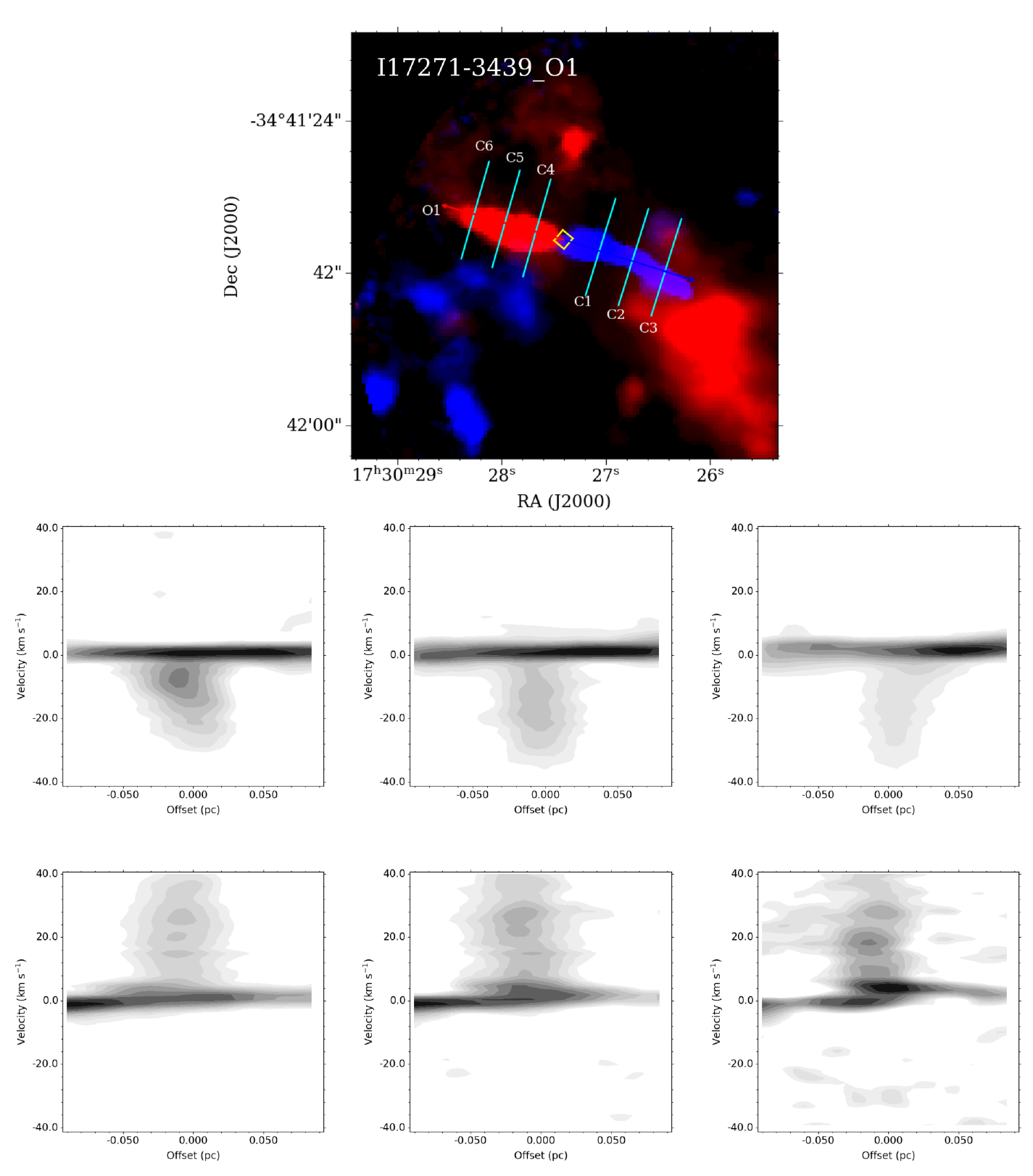}
\figsetgrpnote{The top panel shows the color composite image of the outflow lobes overlaid with six cuts (in Cyan; C1--C6) along which PV diagrams are generated. The middle and bottom panels represent the PV diagrams of C1--C3 and C4--C6 cuts, respectively.}
\figsetgrpend

\figsetgrpstart
\figsetgrpnum{2.9}
\figsetgrptitle{I18182-1433}
\figsetplot{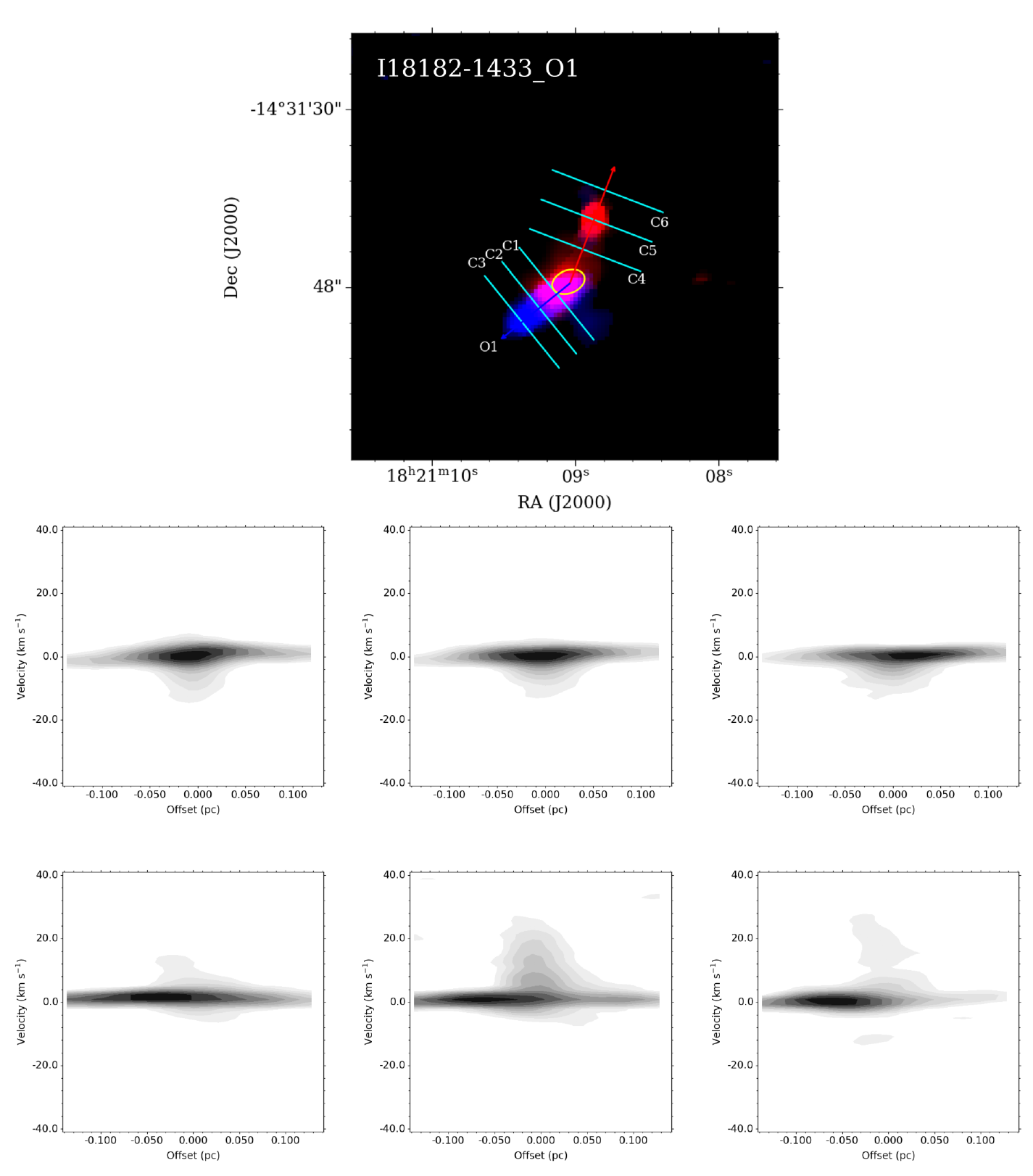}
\figsetgrpnote{The top panel shows the color composite image of the outflow lobes overlaid with six cuts (in Cyan; C1--C6) along which PV diagrams are generated. The middle and bottom panels represent the PV diagrams of C1--C3 and C4--C6 cuts, respectively.}
\figsetgrpend

\figsetgrpstart
\figsetgrpnum{2.10}
\figsetgrptitle{I18290-0924}
\figsetplot{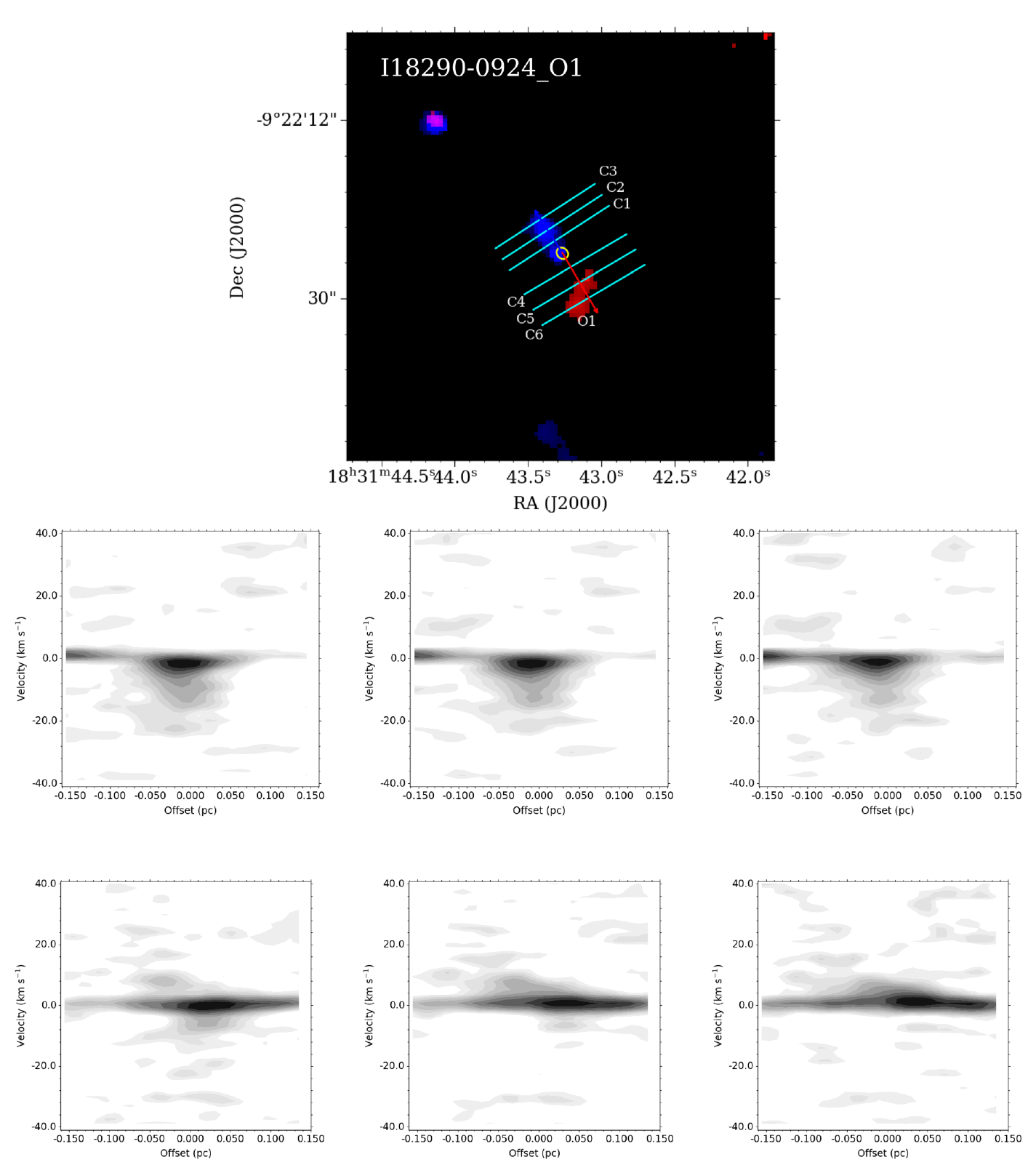}
\figsetgrpnote{The top panel shows the color composite image of the outflow lobes overlaid with six cuts (in Cyan; C1--C6) along which PV diagrams are generated. The middle and bottom panels represent the PV diagrams of C1--C3 and C4--C6 cuts, respectively.}
\figsetgrpend

\figsetend

\begin{figure}[ht!]
\figurenum{B}
\plotone{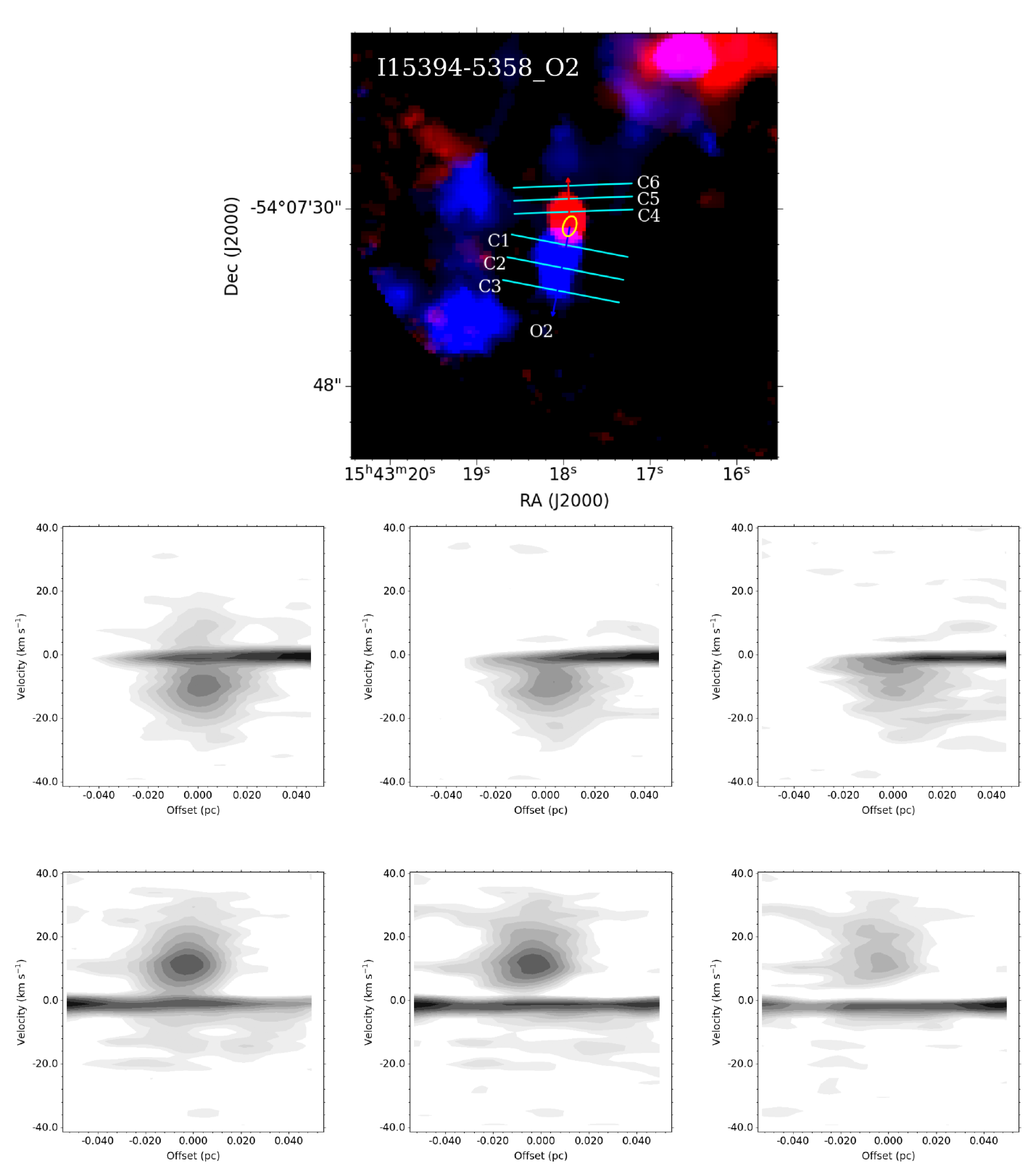}
\caption{The top panel shows the color composite image of the outflow lobes overlaid with six cuts (in Cyan; C1--C6) along which PV diagrams are generated. The middle and bottom panels represent the PV diagrams of C1--C3 and C4--C6 cuts, respectively. The complete figure set (10 figures) is available in the online journal.}
\end{figure}
\section{Detection of HC$_3$N Outflows with Clump Parameters}
\begin{figure}[ht!]
\figurenum{C}
    \centering
    \includegraphics[width=0.48\textwidth]{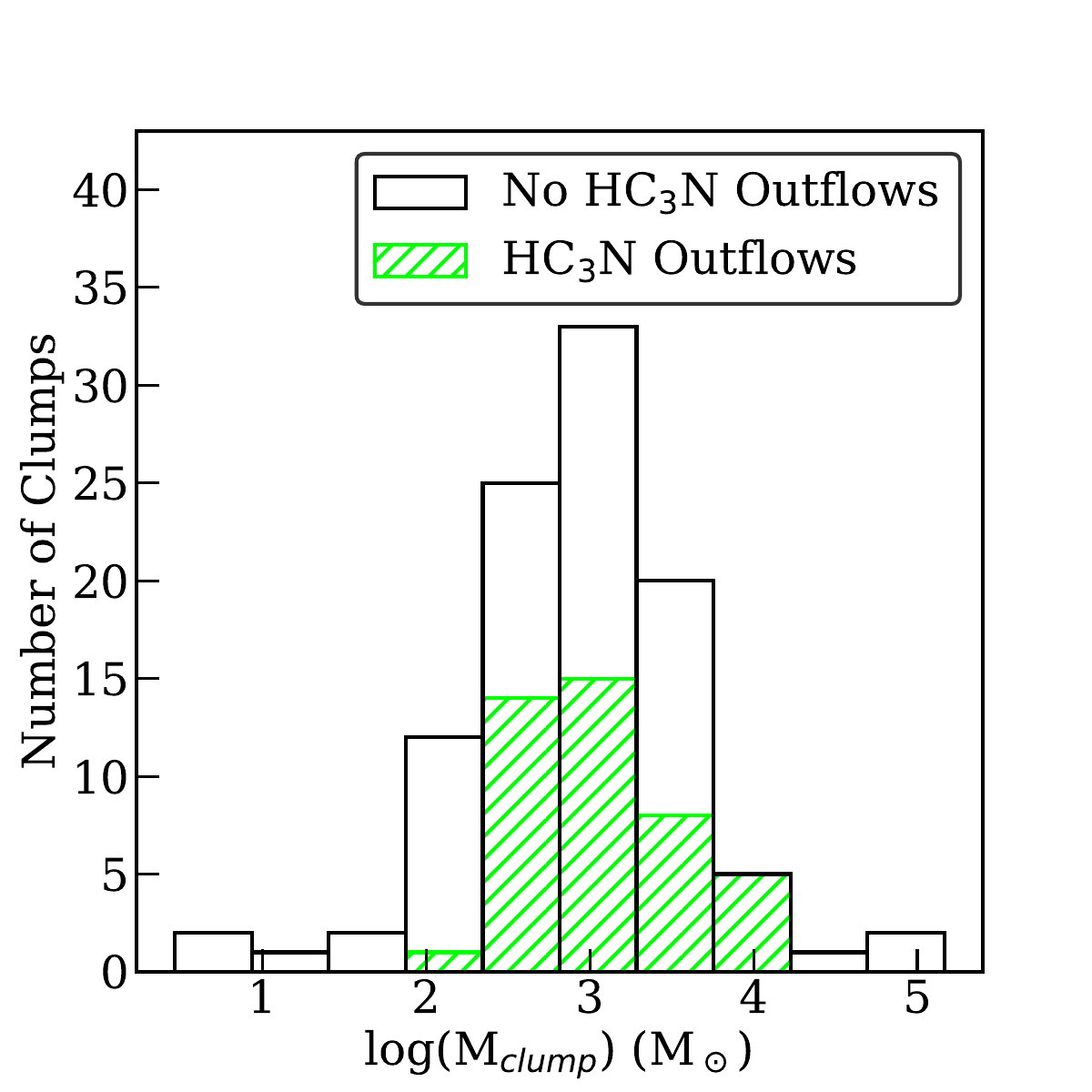}
    \includegraphics[width=0.48\textwidth]{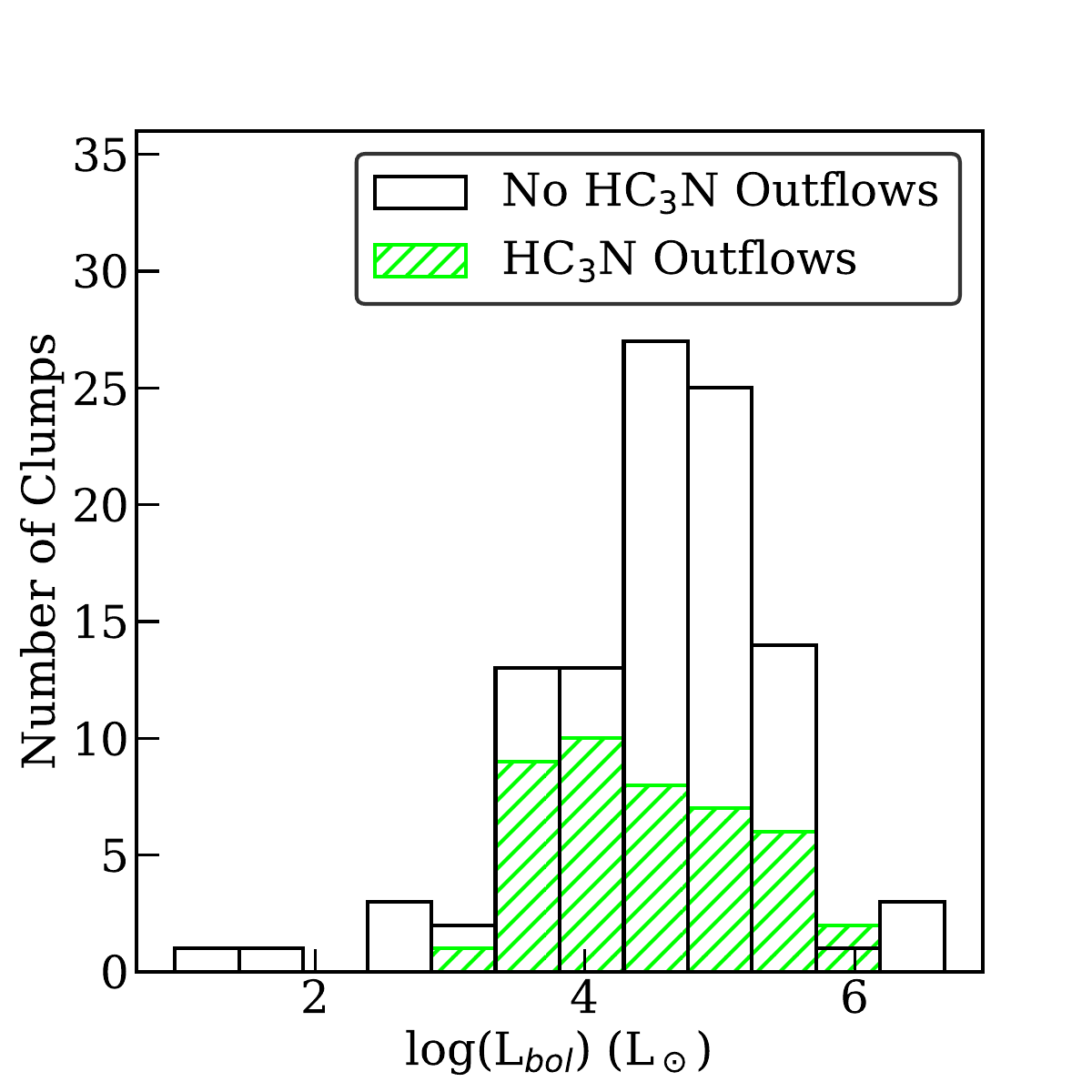}\\
    \includegraphics[width=0.48\textwidth]{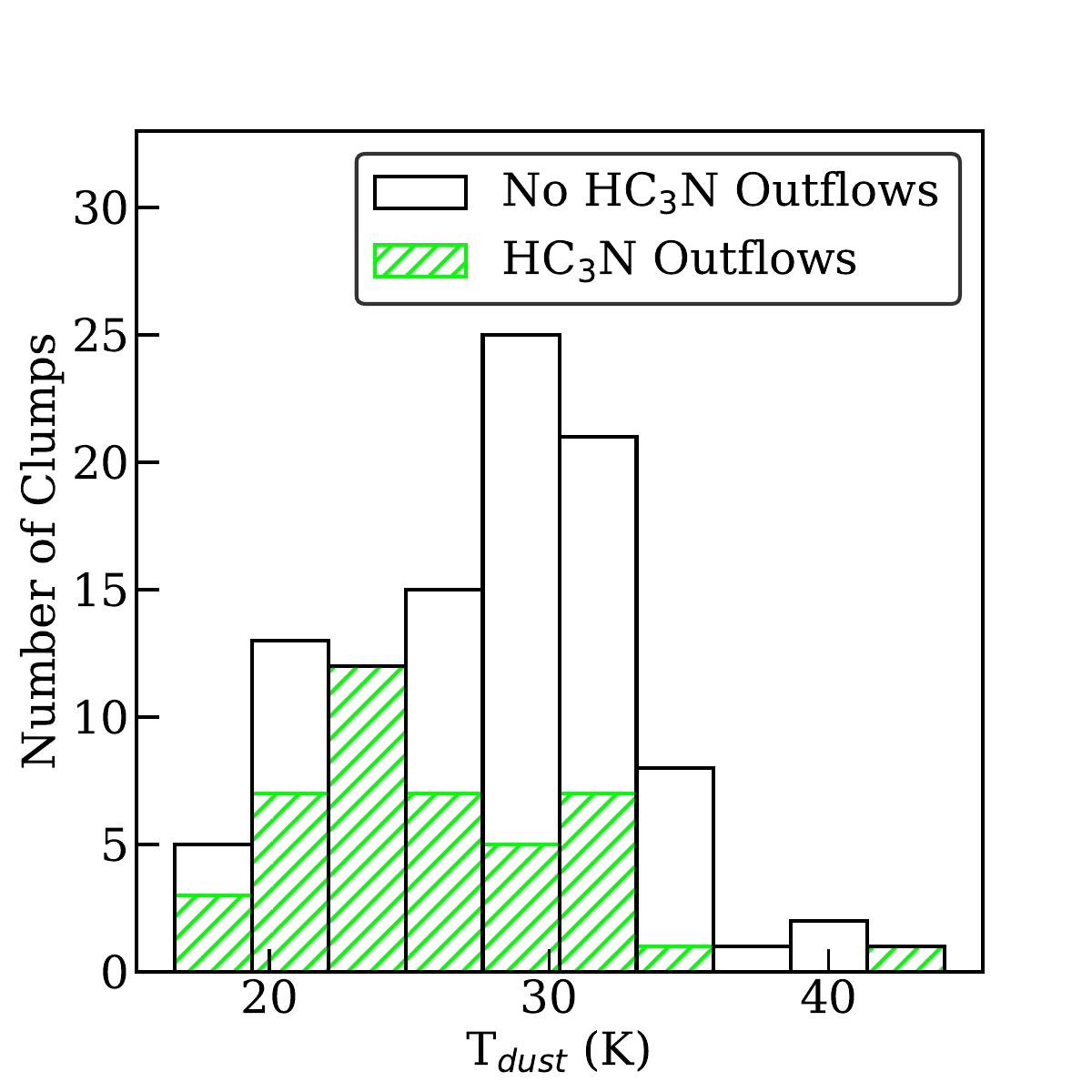}
    \includegraphics[width=0.48\textwidth]{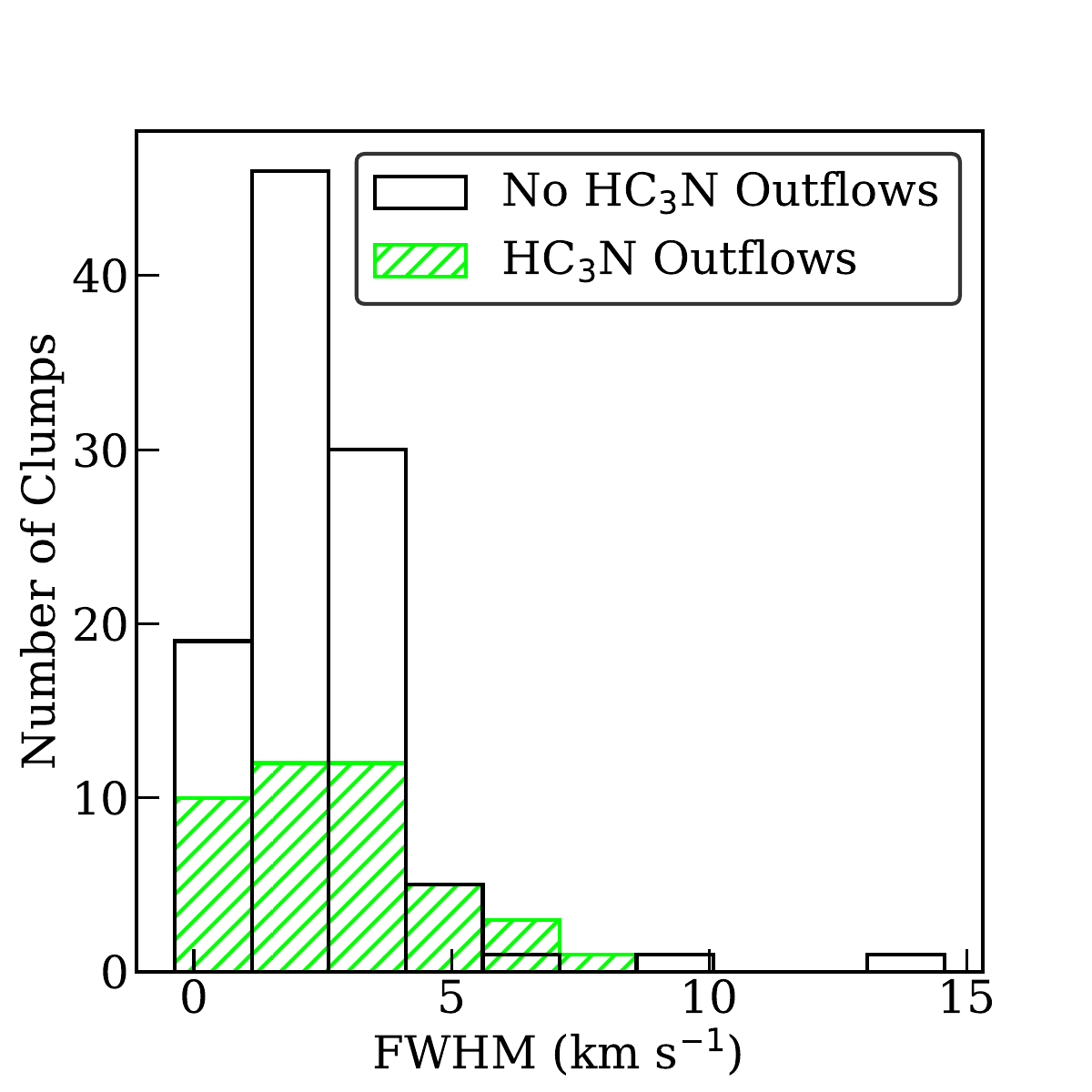}
    \caption{(Top) Distribution of HC$_3$N outflows with associated clump mass (Left) and luminosity (Right), (Bottom) Distribution of HC$_3$N outflows with the average dust temperature (Left) and the velocity FWHM of the clumps (Right).}
    \label{fig:detection}
\end{figure}

In this section, we present the distribution of HC$_3$N outflows with clump mass, luminosity, average dust temperature and velocity FWHM. Figure~\ref{fig:detection} shows the histogram of clumps associated with HC$_3$N outflows and those not associated, in terms of clump parameters.

\label{sec:detection}
\end{document}